\documentclass[11pt,dvipsnames]{article}
\pdfoutput=1

\usepackage{graphicx}
\usepackage{amssymb}
\usepackage{textcomp}
\usepackage{amsmath}
\usepackage{geometry}                
\usepackage{graphicx}
\usepackage{amssymb}
\usepackage{epstopdf}
\usepackage{pgfplots}
\usepackage{tikz-3dplot}
 
\tdplotsetmaincoords{60}{115}
\pgfplotsset{compat=newest}
\usepackage{mathrsfs}

\makeatletter
\renewcommand*\env@matrix[1][\arraystretch]{%
  \edef\arraystretch{#1}%
  \hskip -\arraycolsep
  \let\@ifnextchar\new@ifnextchar
  \array{*\c@MaxMatrixCols c}}
\makeatother

\usepackage{geometry}                
\usepackage{graphicx}
\usepackage{amssymb}
\usepackage{epstopdf}
\usepackage{amsmath}

\usepackage{bm}
\usepackage{times}
\usepackage{epsfig}
\usepackage{color}
\usepackage{array,multirow}
\usepackage{jheppub}

\newcommand{\trix}[1]{\left(\begin{array}{#1}}
\newcommand{\notrix}{\end{array}\right)}
\newcommand{\comment}[1]{}

\def\beq{\begin{equation}}
\def\eeq{\end{equation}}
\def\bea{\begin{eqnarray}}
\def\eea{\end{eqnarray}}

\usepackage[normalem]{ulem}

\usepackage[T1]{fontenc}

\usepackage[utf8]{inputenc}

\usepackage{mathtools} 

\usepackage{amsfonts}

\usepackage{ amssymb }

\usepackage[toc,page,title]{appendix}

\usepackage{xcolor}

\usepackage[compat=1.1.0]{tikz-feynman}

\usepackage{graphicx}

\usepackage{subcaption}

\usepackage{float}

\usepackage{caption}

\usepackage{slashed}

\usepackage{breqn}

\usepackage{amsmath}
\usetikzlibrary{arrows.meta,angles,quotes,calc}

\title{\Large  {{\bf{Moduli and Hidden Matter\\ in Heterotic M-Theory\\ with an Anomalous $U(1)$ Hidden Sector}}}}
\author{{Sebastian Dumitru$^{1}$ and Burt A.~Ovrut$^{1}$} \\[2mm
]
   {${}$\it Department of Physics, University of Pennsylvania} \\[.1cm]
   {\it Philadelphia, PA 19104, USA}

}
\date{\today}

{\footnotetext{\mbox{sdumitru@sas.upenn.edu, ~ovrut@elcapitan.hep.upenn.edu} }}
\abstract{
This paper discusses the dilaton, K\"ahler moduli and hidden sector matter chiral superfields of heterotic $M$-theory vacua in which the hidden sector gauge bundle is chosen to be a line bundle with an anomalous U(1) structure group.  For simplicity of notation, the theory is compactified on a Calabi-Yau threefold with $h^{1,1}=1$, although all methods and results apply to more general heterotic compactifications. After introducing a non-perturbative $F$-term potential and coupling to supergravity, the canonically normalized scalar and fermion mass eigenstates, evaluated around a fixed supersymmetry breaking vacuum, are computed and the explicit expressions for their masses presented. In addition, the relevant couplings of these eigenstates to themselves and to chiral matter in the observable sector are evaluated. The decay rates of generic observable sector scalars into both moduli and hidden sector matter scalars and fermions are then presented. This opens the door to explicit calculations of the decay of an observable sector cosmological inflaton into moduli and hidden sector dark matter candidates. Finally, an explicit flux and gaugino condensate induced non-perturbative superpotential is introduced which is shown to stabilize three of the four real components of the moduli fields.} 
\begin{document}

\maketitle

\newpage

\section{Introduction}

Heterotic $M$-theory is eleven-dimensional Horava-Witten theory
 \cite{Horava:1995qa,Horava:1996ma} dimensionally reduced to five-dimensions by compactifying on a Calabi-Yau (CY) complex
threefold. It was first introduced in \cite{Lukas:1997fg} and discussed in detail in \cite{Lukas:1998yy, Lukas:1998tt}. The five-dimensional heterotic $M$-theory consists of two four-dimensional orbifold planes separated by a finite fifth-dimension.
The two orbifold planes, each
with an $E_{8}$ gauge group, are called the observable and hidden sectors respectively~\cite{Lukas:1997fg, Lukas:1998yy, Lukas:1998tt, Donagi:1998xe,Ovrut:2000bi}. By choosing a suitable CY threefold, as well as an
appropriate holomorphic vector bundle~\cite{Donagi:1999gc} on the CY compactification at the observable sector, one can find realistic low energy $N=1$ supersymmetric particle physics models. A number of such realistic observable sector theories
have been constructed. See, for example,~\cite{Braun:2005nv,Braun:2005bw,Braun:2005ux,Bouchard:2005ag,Anderson:2009mh,Braun:2011ni,Anderson:2011ns,Anderson:2012yf,Anderson:2013xka,Nibbelink:2015ixa,Nibbelink:2015vha,Braun:2006ae,Blaszczyk:2010db,Andreas:1999ty,Curio:2004pf}. In \cite{Braun:2013wr,Ovrut:2018qog,Ashmore:2020ocb}, it was shown that for a heterotic vacuum to be phenomenogically viable, the hidden sector must be
consistent with a series of constraints: 1) allowing for five-branes in the $S^1/\mathbb{Z}_2$ orbifold interval, the
entire theory must be anomaly-free~\cite{Lukas:1999nh,Donagi:1998xe}; 2) the unified gauge coupling
parameters must be positive in both the observable and hidden sectors; 3) the regions of K\"ahler moduli space where both the observable and hidden sector bundles are  slope-stable must overlap and 4) prior to a specified supersymmetry breaking mechanism being introduced, $N = 1$ SUSY must be preserved at the compactification scale. Early attempts~\cite{Braun:2013wr} to build such a hidden sector were valid only in the weakly coupled heterotic string regime in which, however, one cannot obtain reasonable values for the observable sector unification scale and unified gauge coupling~\cite{Witten:1996mz,Banks:1996ss,Banks:1996rr}.

This problem was rectified in \cite{Ashmore:2020ocb}, where we proposed that the hidden sector gauge bundle be a rank-two Whitney sum $L\oplus L^{-1}$  constructed from a line bundle $L$.  The associated $U(1)$ structure group is then embedded into the $E_{8}$ gauge group via the mapping  $U(1) \rightarrow SU(2)\subset E_8$. 
Importantly, within a substantial region of K\"ahler moduli
space, the genus-one corrected slope of the bundle $L$ vanishes, yielding a poly-stable $L\oplus L^{-1}$ bundle. For a given set of line bundles $L$, this vacuum was shown to satisfy all of the above constraints within the context of  \emph{strongly coupled} heterotic $M$-theory vacua. The coupling parameter was found to be large enough
to yield the correct value for the observable sector $SO(10)$ unification mass and
gauge coupling. This work was extended in \cite{Ashmore:2021xdm} to hidden sector gauge bundles consisting of Whitney sums of multiple line bundles, with similar results.  Importantly, these admissible hidden sectors have gauge bundles which contain so-called ``anomalous'' $U(1)$ factors \cite{Dine:1987xk,Dine:1987gj,Anastasopoulos:2006cz,Blumenhagen:2005ga}. 

In \cite{Ashmore:2020wwv}, we analyzed a possible SUSY-breaking mechanism for the vacua found in \cite{Ashmore:2020ocb}, namely
gaugino condensation in the hidden sector. The condensate induces non-zero F-terms in the 4D
effective theory which break SUSY globally~\cite{Choi:1997cm,Kaplunovsky:1993rd,Horava:1996vs,Lukas:1997rb,Nilles:1998sx,Binetruy:1996xja,Antoniadis:1997xk,Minasian:2017eur,Gray:2007qy,Lukas:1999kt,Font:1990nt}. This effect is mediated by gravity and induces calculable
moduli-dependent soft SUSY-breaking terms in the observable sector. We found a subregion inside the K\"ahler cone that leads to realistic four-dimensional physics, satisfying all known phenomenological constraints in the observable sector of the theory. However, we did not compute the canonically normalized mass eigenstates involving the moduli and the hidden sector matter scalars and fermions. Nor did we discuss their possible interactions or their interactions with the observable sector fields.

This was partially accomplished in the analysis given in \cite{Dumitru:2021jlh}. In that paper, we reviewed the general mathematical formalism for computing the inhomogeneous transformations of the dilaton and K\"ahler moduli axions in the presence of an ``anomalous'' $U(1)$ in the hidden sector. Along with matter multiplets, which transform homogeneously under $U(1)$, an important, but restricted, part of the $U(1)$ invariant low energy hidden sector Lagrangian was presented and analyzed. This analysis, however, did not include non-perturbative effects and, hence, the vacua were ``D-flat'' and preserved $N=1$ supersymmetry. Within this context, a detailed mathematical formalism was given for rotating these fields to a new basis of chiral superfields with normalized kinetic energy and a diagonal mass matrix. Two explicit examples were presented, the first with vanishing and the second with non-zero Fayet-Iliopoulos (FI) term. Other studies of the properties of heterotic vacua with anomalous $U(1)$ exist in the literature, but they are usually within the context of the observable sector \cite{Ibanez:2001nd,Aldazabal:2000dg}, or relatively specific contexts that are not directly hidden sectors in a realistic heterotic $M$-theory vacuum \cite{Blumenhagen:2005ga,Blumenhagen:2006ux,Weigand:2006yj,Lukas:1999nh,Anderson:2009nt,Anderson:2010mh,Binetruy:1996uv}

The main goal of the present paper is to extend the analysis described above and compute the low energy Lagrangian for the moduli and hidden sector matter fields after non-perturbative effects spontaneously break the 4D $N=1$ supersymmetry. We will also discuss, in detail, the effects of including some relevant non-gauge interactions and coupling the theory to supergravity. Doing this allows us to explicitly compute the masses of the canonically normalized scalars and fermions, as well as to calculate physically relevant interactions of these fields with themselves and with observable sector fields. Although our previous work in \cite{Dumitru:2021jlh} was carried out within the context of a specific Calabi-Yau three-fold with $h^{1,1}=3$, in this paper, for simplicity, we choose a simpler CY threefold for which $h^{1,1}=1$. Hence, there is only one K\"ahler modulus present in the theory, which greatly simplifies our notation. Furthermore, we will assume that only a limited number of matter supermultiplets are present on the hidden sector. We make no attempt to build a realistic, phenomonologically viable model in this $h^{1,1}=1$ context; this choice serves solely to reduce the degrees of freedom in the system and, hence, to significantly simplify the process of computing the final mass eigenstates after supersymmetry is broken. These mass eigenstates mix different types of moduli fields and matter fields. We think that our method, as well as our conclusions, become more clear in this reduced set-up. Furthermore, once this analysis is well understood, it is straightforward to extend it to more complicated heterotic vacua constructions, such as those studied in \cite{Ashmore:2020ocb}.

Specifically, we do the following. In Section \ref{sec:Effective theory}, we present the matter, moduli and gauge field content of our model. In Section \ref{sec:D-term_s}, we discuss the D-term stabilization mechanism in vacua with an anomalous $U(1)$ present in the 4D theory. We identify two distinct types of D-flat vacua, depending on whether the genus-one corrected FI term vanishes or not. Most of this section is a review of our work in \cite{Dumitru:2021jlh}, but now applied to the simpler $h^{1,1}=1$ model. In Section \ref{mass_spect_sec}, we compute the full matter spectrum after $N=1$ supersymmetry is broken by non-perturbative effects in the hidden sector. The analysis is general, and applies to any particular method of supersymmetry breaking. We compute the mass matrices and the mass eigenstates, in both types of D-flat vacua. In Section \ref{sec:DM}, we show how the massive moduli and hidden matter field states couple to the observable sector. We also discuss some interesting dark matter candidates and propose a mechanism of producing these states during reheating. In Section \ref{Sec:Mod_Stab}, we give explicit examples of non-perturbative effects that can break $N=1$ supersymmetry at low-energy. We discuss the possibility of stabilizing the moduli in these non-supersymmetric vacua. We also compute the moduli mass spectrum in each of these examples, applying the results of Section \ref{mass_spect_sec}. Mathematical details used in the computation of both scalar and fermion masses are presented in the Appendix.

\section{4D Effective Theory}\label{sec:Effective theory}

Consider heterotic $M$-theory vacua compactified on a Calabi-Yau (CY) threefold $X$. In our previous work \cite{Dumitru:2021jlh} we analyzed the D-term stabilization mechanism and chose $h^{1,1}=3$ to be consistent with various realistic heterotic $M$-theory vacua and, specifically, the $B-L$ MSSM~\cite{Ambroso:2009jd,Marshall:2014kea,Marshall:2014cwa,Ovrut:2012wg,Ovrut:2014rba,Barger:2008wn,FileviezPerez:2009gr}.  In the present work, in the context of such D-flat vacua, we will condider non-perturbative effects such as gaugino condensation to spontaneously break the $N=1$ supersymmetry. This introduces a moduli-dependent superpotential into the effective Lagrangian and, as a result, greatly complicates all relevant calculations. For this reason, in the present paper, we will consider heterotic vacua compactified on a CY threefold with $h^{1,1}=1$, greatly simplifying the formalism. It follows that the $D=4$ low energy effective theory contains, in addition to the universal dilaton chiral multiplet $\tilde{S}=(S, \psi_{S})$, a single K\"ahler modulus chiral superfield $\tilde{T}=(T, \psi_{T})$.  Furthermore, we will assume that the observable sector contains a phenomenologically realistic $N=1$ supersymmetric particle physics model; that is, the MSSM or some viable extension thereorf. We will, for simplicity, refer to any matter chiral supermultiplet in the observable sector theory simply as $\tilde{C}_{(o)}^{\cal{I}}=(C_{(o)}^{\cal{I}}, \psi_{(o)}^{\cal{I}})$, where ${\cal{I}}=1, \dots,{\cal{N}}_{(o)}$. We assume that the hidden sector gauge bundle consists of a single line bundle $L={\cal{O}}_{X}(l)$, where $l$ is an integer, with structure group $U(1)$ appropriately embedded in the hidden sector $E_{8}$ gauge group. In addition, we assume that the $U(1)$ is ``anomalous''. We denote the $U(1)$ gauge connection and its Weyl spinor gaugino by $A_{2 \mu}$ and $\lambda_{2}$ respectively. 
\
The low energy matter spectrum of this hidden sector generically contains chiral multiplets that transform under $U(1)$ but are singlets under the associated commutant subgroup. Generically, there can be many such ``singlet'' matter superfields. However, as we will show below, it is sufficient to assume, for simplicity, that there are only two such matter multiplets. The extension to more than two multiplets is trivial. We denote these two singlet matter chiral multiplets as $\tilde{C}^{L}=(C^{L}, \psi^{L}), L=1,2$.
In our analysis, the chiral matter multiplet transforming non-trivially under the associated commutant subgroup do not play any role and have therefore been left out.

In a series
of papers~\cite{Anderson:2010mh, Anderson:2011cza,Anderson:2011ty}, it has been shown that in particular examples (i.e. compactifying on a CY with a point-like
sub-locus in complex structure moduli space where the gauge bundle is holomorphic, such that $h^{1,2}_{\text{hol}}=0$) one could be
able to fix all the complex structure moduli at the compactification scale. We will assume this is the case in our analysis and henceforth, we will neglect the contribution of the complex structure moduli. The work in~\cite{Cicoli:2013rwa}, however, offers a more general discussion on the topic of stabilizing the complex structure moduli. Furthermore, we disregard any effects of the five-brane in the fifth dimensional bulk space--other than its role in canceling the anomaly. 

The properties of heterotic vacua with the above properties and assumptions can be completely determined using the formalism and definitions presented in~\cite{Lukas:1997fg,Lukas:1998tt} (see also~\cite{Brandle:2003uya} for the $h^{1,1}=3$ case). Specifically, one finds the following. In the absence of five-branes, the K\"ahler potential of the system is
\begin{equation}
K=K_S+K_T+K_{\text{matter}}^{\text{(hid)}}+K_{\text{matter}}^{\text{(obs)}}\ ,
\end{equation}
where
\begin{align}
K_S&=-\kappa_4^{-2}\ln(S+\bar S)\ ,\\
K_T&=-3\kappa_4^{-2}\ln(T+\bar T)\ ,\label{bl1} \\
K_{\text{matter}}^{\text{(hid)}}&=e^{\kappa_4^2K_T/3}C^1\bar C^1+e^{\kappa_4^2K_T/3}C^2\bar C^2\ ,\\
K_{\text{matter}}^{\text{(obs)}}&=e^{\kappa_4^2K_T/3 } \mathcal{G}_{{\cal{I}}{\bar{\cal{J}}} } C_{(o)}^{\cal{I}}\bar{C}_{(o)}^{\bar {\cal{J}}}\ .
\end{align}
Note that, for simplicity, we have chosen the internal metric in $K_{\text{matter}}^{\text{(hid)}}$ to be $\mathcal{G}_{L{\bar{M}}}=\delta_{L{\bar{M}}}$.
The complex scalar components of the moduli superfields decompose as
\begin{equation}
\begin{split}
\label{eq:def_scalar_intro}
& S=s+i\sigma\ ,\\
&T=t+i\chi\ ,\\
\end{split}
\end{equation}
where $\sigma$ and $\chi$ are the dilaton axion and the K\"ahler axion respectively.

The moduli and the hidden matter scalars transform under the anomalous $U(1)$ as~\cite{Anderson:2010mh,Dumitru:2021jlh}
\begin{equation}
\begin{split}
\label{eq:Killing_vects}
\delta_\theta S&=2i\pi a\epsilon_S^2\epsilon_R^2 \beta l \theta ~ \equiv k_S\theta\ ,\\
\delta_\theta T& =-2i a\epsilon_S\epsilon_R^2 l\theta  ~\equiv k_T\theta\ ,\\
\delta_\theta C^1&=-iQ_1C^1\theta   ~\equiv k_1\theta,\\
\delta_\theta C^2&=-iQ_2C^2\theta  ~\equiv k_2\theta  \ ,\\
\end{split}
\end{equation}
where the parameter $a$ depends on the line bundle embedding into the hidden $E_8$, while $\epsilon_{S}$ and $\epsilon_{R}$ are expansion parameters in the strong coupling regime.
The total scalar superpotential is given by
\begin{equation}
\begin{split}
\label{eq:superpotential}
W=W^{\text{(obs)}}+W^{\text{(mod)}}+W^{\text{(hid)}}\ ,
\end{split}
\end{equation}
where
\begin{equation}
\label{shoe1}
W^{\text{(obs)}}=\mu_{\cal{I} \cal{J}} C_{(o)}^{\cal{I}}C_{(o)}^{\cal{J}}+Y_{\cal{I}\cal{J}\cal{K}}C_{(o)}^{\cal{I}}C_{(o)}^{\cal{J}}C_{(o)}^{\cal{K}}\ 
\end{equation}
is the matter field superpotential on the observable sector,
\begin{equation}
W^{\text{(hid)}}= m_{LM}C^{L}C^{M}+\lambda_{KLM}C^KC^LC^M 
\end{equation}
is the matter field superpotential on the hidden sector, while 
\begin{equation}
W^{\text{(mod)}}=\hat W_{np}(S,T)
\end{equation}
is the moduli superpotential generated by non-perurbative effects, such as gaugino condensation on the hidden sector or five-brane instantons. 
Finally, to order $\kappa^{2/3}_4$ gauge threshold corrections~\cite{Lukas:1997fg}, the gauge kinetic functions on the observable and the hidden sectors are given by
\begin{equation}
f_1=f_2=S\ .
\end{equation}
 These gauge kinetic functions determine the values of the gauge couplings $g_{1}$ and $g_{2}$ on the observable and hidden sector respectively. That is,
\begin{equation}
{g_{1}^{2}}=\frac{\pi \hat{\alpha}_{\text{GUT}}}{a\text{Re} f_{1}} , \qquad  {g_{2}^{2}}=\frac{ \pi \hat{\alpha}_{\text{GUT}}}{a\text{Re} f_{2}}  \ .
\label{umb1}
\end{equation}

In this paper, we focus on the effective theory in the moduli and the hidden sector. Similarly, in this paper, as shown in an explicit example in \cite{Dumitru:2021jlh}, we use the fact that the matter scalars $C^1, C^2$ in the hidden sector generically cannot form a gauge invariant superpotential. Hence, we can also ignore  $W^{\text{(hid)}}$. However, since the main focus of the present work is to discuss the effect of gaugino condensation on the hidden sector, the non-perturbative superpotential $\hat W_{np}(S,T)$ is central to our analysis and will be introduced and discussed in detail below. However, before proceeding to this analysis, let us briefly summarize the results of \cite{Dumitru:2021jlh}--that is, no gaugino condensation and, hence, $\hat W_{np}(S,T)=0$--within the simplified context  used in this paper. 

\section{D-term Stabilization}\label{sec:D-term_s}

\subsection{Effective Potential}

The low-energy gauge group arising in the hidden sector from a line bundle $L$ necessarily includes an
“anomalous” $U(1)$ factor in the 4D low-energy gauge group. Associated with the anomalous $U(1)$
is a moduli dependent D-term, whose form is well-known \cite{Freedman:2012zz}.
Specifically, the D-term potential energy
\begin{equation}
\label{late1}
V_D=\frac{1}{2\text{Re}f_2}{\mathcal{P}^2}\ ,
\end{equation}
is generated perturbatively on the hidden sector after compactification. 
The moment map $\mathcal{P}$ depends on the first derivatives of the Kähler potential with respect to the scalar
fields which are charged under the anomalous $U(1)$. For the field content presented in the previous
section, it has the form

\begin{equation}
\label{last1}
\begin{split}
\mathcal{P}&=ik_S\partial_S K+ik_T\partial_T K+ik_1\partial_{C^1}K+ik_2\partial_{C^2}K\ .
\end{split}
\end{equation}
The first derivatives of the K\"ahler potential with respects to the scalar field components are 
\begin{equation}
\begin{split}
\label{eq:first_der_K}
\partial_SK&=-\kappa_4^{-2}\frac{1}{S+\bar S}=-\kappa_4^{-2}\frac{1}{2s}\ ,\\
\partial_TK&=-3\kappa_4^{-2}\frac{1}{T+\bar T}+\frac{\kappa_4^2}{3}\frac{\partial K_T}{\partial T}e^{\kappa_4^2K_T/3}C^1\bar C^1+\frac{\kappa_4^2}{3}\frac{\partial K_T}{\partial T}e^{\kappa_4^2K_T/3}C^2\bar C^2\ ,\\
&=-3\kappa_4^{-2}\frac{1}{2t}-\frac{1}{4t^2}C^1\bar C^1-\frac{1}{4t^2}C^2\bar C^2\ ,\\
\partial_{C^1}K&=e^{\kappa_4^2K_T/3}\bar C^1=\frac{1}{2t}\bar C^1\ ,\\
\partial_{C^2}K&=e^{\kappa_4^2K_T/3}\bar C^2=\frac{1}{2t}\bar C^2\ .
\end{split}
\end{equation}
Substituting the expressions for the Killing vectors defined in eq. \eqref{eq:Killing_vects} and for the first derivatives of the K\"ahler potential calculated in eq. \eqref{eq:first_der_K} into \eqref{last1}, we find
\begin{equation}
\label{burt1}
\begin{split}
\mathcal{P}=&-\frac{ a\epsilon_S\epsilon_R^2}{\kappa^{2}_{4}}
\left( -\frac{1}{s}\pi\beta \epsilon_Sl+\frac{3l}{t} \right)-e^{\kappa_4^2K_T/3}\tilde Q_1C^1\bar C^1-e^{\kappa_4^2K_T/3}\tilde Q_2C^2\bar C^2\ ,
\end{split}
\end{equation}
where $\beta$ is the hidden sector ``charge'' defined in \cite{Ashmore:2020ocb} and we have defined the moduli-dependent charges
\begin{equation}
\begin{split}
\tilde Q_1&=Q_1-2a\epsilon_S\epsilon_R^2 l\frac{\kappa_4^2}{3}\frac{\partial K_T}{\partial T}\ ,\\
\tilde Q_2&=Q_2-2a\epsilon_S\epsilon_R^2 l\frac{\kappa_4^2}{3}\frac{\partial K_T}{\partial T}\ .
\end{split}
\end{equation}
Note that the first term in \eqref{burt1} depends only on $s={\rm Re}S$ and $t={\rm Re}T$ and is independent of the matter scalar fields. Since Re$f_{2}$ also only depends on Re$S$ and Re$T$, it follows that the ``axion'' components $\sigma$ and $\chi$ of $S$ and $T$ respectively do not enter the scalar potential $V_{D}$. Therefore, in the context of minimizing the potential, they are free to take any values.

Minimizing the $D$-term potential \eqref{late1} defines $D$-flat, $N=1$ supersymmetry preserving vacuum states for which the moment map vanishes,
\begin{equation}
\label{umb2}
\langle {\cal{P}} \rangle = 0 \ .
\end{equation}
Therefore, in order to preserve $N=1$ supersymmetry, it follows from \eqref{burt1} that the VEVs of the dilaton, the K\"ahler modulus and the matter scalars are constrained to satisfy
\begin{equation}
\label{umb3}
-\frac{ a\epsilon_S\epsilon_R^2}{\kappa^{2}_{4}}
\left( -\frac{1}{\langle s \rangle}\pi\beta \epsilon_Sl+\frac{3l}{\langle t \rangle} \right) - e^{\kappa_{4}^{2}\langle K_{T} \rangle /3} \left(\langle \tilde Q_1 \rangle  \langle C^1\rangle \langle\bar C^1\rangle
+\langle \tilde Q_2 \rangle  \langle C^2\rangle \langle\bar C^2\rangle \right)  = 0 \ .
\end{equation}
The first term in \eqref{umb3} corresponds to the Fayet-Iliopoulos (FI) term. That is
\begin{equation}
\label{cup1}
\text{FI}=-\frac{ a\epsilon_S\epsilon_R^2}{\kappa^{2}_{4}}
\left(- \frac{1}{\langle s \rangle}\pi\beta \epsilon_Sl+\frac{3l}{\langle t \rangle} \right)\ .
\end{equation}
Therefore, the D-term flatness condition $\langle V_{D} \rangle =0$--required to preserve unbroken $N=1$ supersymmetry in the vacuum-- sets
\begin{align}
\label{cup4}
\langle \mathcal{P}\rangle =0\quad
\Rightarrow \quad \text{FI}=e^{\kappa_{4}^{2}\langle K_{T} \rangle /3} \left(\langle \tilde Q_1 \rangle  \langle C^1\rangle \langle\bar C^1\rangle
+\langle \tilde Q_2 \rangle  \langle C^2\rangle \langle\bar C^2\rangle \right) \ .
\end{align}

Further analysis of the vacuum state requires one to expand the scalar fields around a chosen solution of the D-term flatness condition. The physical results depend heavily on whether one chooses the vacuum to satisfy a) $\text{FI}=0$ or b) $\text{FI}\neq 0$. These two types of $N=1$ supersymmetric vacua correspond to very different low energy physics and, therefore, we will analyze them seperately throughout the remainder of this paper. Furthermore, the addition of non-perturbative effects--to be discussed in later sections-- produces different final mass spectra for each of these two types of vacua.  In the rest of this section, however, we analyze the field spectrum which results after only the $D$-flatness condition is satisfied. We begin with the case when the FI term vanishes.

\subsection{Vanishing FI Term}

\begin{figure}[t]
   \centering
\includegraphics[width=0.45\textwidth]{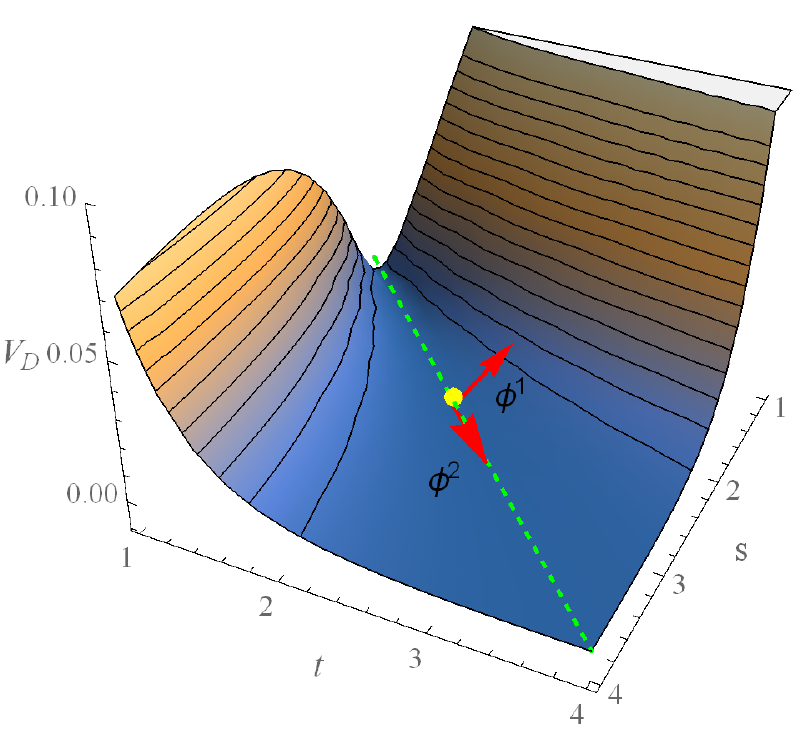}
\caption{For the $FI=0$ case, where the $C^1$ and $C^2$ matter fields decouple, we plot the D-term potential $V_{D}$ given in eq. \eqref{late1} in $s,t$ moduli space. The potential is given in units of $M_U^4$. The $V_{D}=0$ locus is displayed by the dashed green line. This D-flat direction is defined by eq. \eqref{cof1}, where we here assume, for specificity, that $\frac{\pi\epsilon_S\beta}{3}=1$. The D-flat vacuum can be chosen to be anywhere along the dashed green line. One such point, displayed in yellow, is at $\langle s\rangle=\langle t\rangle =2.3$. Scalar perturbations around this vacuum define a massless field, $\phi^2$, aligned along the green line, and a massive field $\phi^1$, aligned along the maximum variation of the potential around the chosen vacuum. }
\label{fig:D_flat11}
\end{figure}

It follows from \eqref{cup1} that a vanishing  Fayet-Iliopoulos term requires
\begin{equation}
\label{cup2}
\left( -\frac{1}{\langle s \rangle}\pi\beta \epsilon_Sl+\frac{3l}{\langle t \rangle} \right) = 0
\end{equation}
or, equivalently, that
\begin{equation}
\label{cof1}
\langle s \rangle =\frac{\pi\epsilon_{S} \beta}{3} \langle t \rangle \ .
 \end{equation}
 Of course, if the FI term vanishes, the $D$-flatness condition \eqref{cup4} implies that the matter field VEVs must vanish; that is
\begin{equation}
\label{cup3}
\langle C^1\rangle= \langle C^2\rangle= 0.
\end{equation}
This latter condition ``decouples'' the hidden sector matter fields from the $S$ and $T$ moduli. Ignoring the hidden sector matter scalars, one can compute the  D-term potential $V_{D}$ over the $s$ and $t$ component scalars using \eqref{late1} and \eqref{burt1}. This is plotted in Figure \ref{fig:D_flat11}--where, for specificity, we have chosen $\pi\epsilon_{S}\beta/3=1$. The $\langle s \rangle$, $\langle t \rangle$ VEVs satisfying the $D$-flatness condition \eqref{cof1} form the dashed green flat line in the figure. 

Let us now choose any point $\langle s \rangle$, $\langle t \rangle$  satisfying the $D$-flatness condition \eqref{cof1}; that is, any point on the dashed green line. Expanding $S=\langle s \rangle+\delta S$ and $T=\langle t \rangle+\delta T$, one finds that the Lagrangian for $\delta S$ and $\delta T$  has off-diagonal kinetic energy and mass terms.  However, it was shown in \cite{Dumitru:2021jlh} that one can define two new complex scalar fields $\xi^{1}$ and $\xi^{2}$ with canonically normalized kinetic terms which are mass eigenstates. Specifically, we define a unitary matrix $\mathbf U$ which rotates the scalar perturbations $\delta S$ and $\delta T$ into a massive scalar $\xi^1$ and a massless scalar $\xi^2$, such that
\begin{equation}
\label{light1}
\left(\begin{matrix}
\xi^1\\\xi^2
\end{matrix}
\right)={\mathbf U}
\left(\begin{matrix}
\delta S\\\delta T
\end{matrix}
\right)\ ,\quad 
\left(\begin{matrix}
\delta S\\\delta T
\end{matrix}
\right)={\mathbf U}^{-1}\left(\begin{matrix}
\xi^1\\\xi^2
\end{matrix}
\right)
\ ,
\end{equation}
with ${\mathbf U}$ and its inverse ${\mathbf U}^{-1}$ given by
\begin{align}
\label{eq:U22}
{\mathbf U}=&
\frac{1}{\langle \Sigma \rangle }\begin{pmatrix}[1.4]
\langle g_{S\bar S}\bar k_S \rangle &\quad\langle g_{T\bar T}\bar k_T \rangle\\
 \sqrt{\langle g_{S\bar S}g_{T\bar T} \rangle} \langle \bar k_T \rangle&\quad -\sqrt{\langle g_{S\bar S}g_{T\bar T} \rangle} \langle \bar k_S \rangle
\end{pmatrix}
\end{align}
and
\begin{align}
{\mathbf U}^{-1}&=
\frac{1}{\langle \Sigma \rangle}\begin{pmatrix}[1.4]
\langle k_S \rangle&\quad\sqrt{\langle \frac{g_{T\bar T}}{g_{S\bar S}}\rangle} \langle k_T \rangle\\
 \langle k_T \rangle&\quad-\sqrt{\langle \frac{g_{S\bar S}}{g_{T\bar T}}\rangle}  \langle k_S \rangle
\end{pmatrix}\ ,
\label{eq:U22A}
\end{align}
respectively, where
\begin{equation}
\label{pad3}
\Sigma^2=g_{S\bar S}k_S\bar k_S+g_{T\bar T}k_T\bar k_T\ .
\end{equation}
The field
\begin{equation}
\label{pad1}
\xi^1=\left\langle g_{S\bar S}\frac{\bar k_S}{\Sigma} \right\rangle \delta S+\left\langle g_{T\bar T}\frac{\bar k_T}{\Sigma}\right\rangle \delta T
\end{equation}
is a massive complex field with mass  $m_{\text{anom}}\sim \mathcal{O}(M_U)$, while the complex field
\begin{equation}
\label{pad2}
\xi^2=\sqrt{\langle g_{S\bar S}g_{T\bar T}\rangle }\left\langle \frac{ \bar k_T}{\Sigma} \right\rangle \delta S-\sqrt{\left\langle g_{S\bar S}g_{T\bar T}\right\rangle }\left\langle \frac{\bar k_S}{\Sigma} \right\rangle \delta T
\end{equation}
is massless. A brief summary of that analysis is the following.

First consider the complex scalar $\xi^{1}$.  Expressed in terms of its real component fields
\begin{equation}
\label{pink1}
\xi^{1}=\eta^{1}+i \phi^{1} \ ,
\end{equation}
where
\begin{equation}
\label{pink4}
\phi^{1}=\left\langle g_{S\bar S}\frac{i\bar k_S}{\Sigma} \right\rangle \delta s+\left\langle g_{T\bar T}\frac{i\bar k_T}{\Sigma}\right\rangle \delta t
\end{equation}
is a canonically normalized real scalar field with mass
\begin{equation}
m_{\phi^{1}}= \sqrt{2\langle g_{2}^{2} \Sigma^{2} \rangle} \ .
\label{pink2}
\end{equation}
On the other hand, the real scalar $\eta^{1}$ was shown to simply be the $U(1)$ Goldstone boson which can be gauged away, giving the $U(1)$ gauge field $A_{2 \mu}$ the anomalous mass
\begin{equation}
m_{\text{anom}}=m_{\phi^{1}}= \sqrt{2\langle g_{2}^{2} \Sigma^{2} \rangle} \ .
\label{pink3}
\end{equation}

The scalar field $\phi^{1}$ forms the real bosonic degree of freedom of a massive vector superfield with mass \eqref{pink3} 
\begin{equation}
\label{day1}
m_{\text{anom}}=\sqrt{2\langle g_{2}^{2} \Sigma^{2} \rangle} \sim \mathcal{O}(M_U) \ ,
\end{equation}
where $M_{U}$ is the unification scale of the non-Abelian complement of the anomalous $U(1)$ which, although model dependent, typically exceeds $10^{16}$GeV.
Hence, $\xi^{1}$  disappears from the low energy matter spectrum leaving $\xi^{2}$ as a canonically normalized massless scalar. Writing 
\begin{equation}
\label{pink5}
\xi^{2}=\eta^{2}+i \phi^{2} \ ,
\end{equation}
it follows from \eqref{pad2} that
\begin{equation}
\label{pad2A}
\phi^2=\sqrt{\langle g_{S\bar S}g_{T\bar T}\rangle }\left\langle \frac{i \bar k_T}{\Sigma} \right\rangle \delta s-\sqrt{\left\langle g_{S\bar S}g_{T\bar T}\right\rangle }\left\langle \frac{i\bar k_S}{\Sigma} \right\rangle \delta t 
\end{equation}
is a massless  real scalar field. That is, $\phi^{2}$ is a fluctuation along the dashed green line in Figure 1. Since, by construction, the scalar fields $\phi^{1}$ and $\phi^{2}$ are orthogonal to each other, it follows that field $\phi^{1}$ is always orthogonal to the dashed green line in Figure 1 for any initial values of $\langle s \rangle$ and $\langle t \rangle$. This is indicated in Figure 1.
On the other hand, the real scalar field
\begin{equation}
\label{pad2B}
\eta^2=\sqrt{\left\langle g_{S\bar S}g_{T\bar T} \right\rangle }\left\langle \frac{i \bar k_T}{\Sigma} \right\rangle \sigma-\sqrt{\langle g_{S\bar S}g_{T\bar T}\rangle }\left\langle \frac{i\bar k_S}{\Sigma} \right\rangle \chi \ ,
\end{equation}
which is orthoganal to $\eta^{1}$, remains in the effective theory as a massless axion. As discussed above, since $\eta^{2}$ is a linear combination of the $S$ and $T$ axions, $\sigma$ and $\chi$ respectively, it does not enter the potential $V_{D}$.

Using the same formalism, the fermions $\psi_{S}$ and $\psi_{T}$ associated with the chiral superfields $\tilde{S}$ and $\tilde{T}$ respectively, can be rotated to a canonically normalized basis of mass eigenstates $\psi_{\xi}^1$, $\psi_{\xi}^2$ using the same unitary matrix $\mathbf U$. That is,
\begin{equation}
\left(\begin{matrix}
\psi_\xi^1\\ \psi_\xi^2
\end{matrix}
\right)=\mathbf{U}
\left(\begin{matrix}
\psi_S\\\psi_T
\end{matrix}
\right)\ ,\quad 
\left(\begin{matrix}
\psi_S\\\psi_T
\end{matrix}
\right)=\mathbf{U}^{-1}\left(\begin{matrix}
\psi_\xi^1\\ \psi_\xi^2
\end{matrix}
\right)
\ .
\end{equation}
The fermion  $\psi_{\xi}^1$combines with the $U(1)$ gaugino field to form a Dirac fermion. Since $N=1$ supersymmetry remains unbroken, this Dirac fermion acquires the same mass $m_{\text{anom}}\sim \mathcal{O}(M_U)$ and becomes part of the massive $U(1)$ vector supermultiplet, while $\psi_{\xi}^2$ remains massless. Hence, $\psi_{\xi}^{1}$  disappears from the low energy matter spectrum leaving $\psi_{\xi}^{2}$ as a canonically normalized massless fermion. 

Therefore, ignoring interaction terms, the hidden sector low energy moduli Lagrangian when $FI=0$ is simply
\begin{equation}
\mathcal{L} \supset  -\partial^{\mu} \bar{\xi}^{2} \partial_{\mu} \xi^{2} -i \psi_{\xi}^{2} \slashed\partial \psi_{\xi}^{2\dagger}  \ .
\label{help1}
\end{equation}
At this point, we recall that when $\text{FI}=0$ the hidden sector matter chiral multiplets $\tilde{C}_{1}$ and $\tilde{C}_{2}$ have ``decoupled'' from the $\tilde{S}$ and $\tilde{T}$ superfields. Hence, they remain massless multiplets with their own kinetic energy Lagrangian. This massless matter field Lagrangian is given by
\begin{equation}
\mathcal{L} \supset - G_{L\bar M}\partial_\mu C^L \partial^\mu \bar C^{\bar M}
 -iG_{L\bar M}\psi^L \slashed{\partial} \psi^{\bar M\dag} \ ,
 \label{ny1}
\end{equation}  
where
\begin{equation}
G_{L\bar M}= 
\frac{\partial^2 K_{\text{matter}}^{\text{(hid)}}}
{\partial C^L\partial \bar C^{\bar M}} = e^{\kappa_{4}^{2}K_{T}/3} \delta_{L\bar{M}}
\label{black1}
\end{equation}
and $L,M=1,2$ .

\subsection{Non-Vanishing FI Term}

It follows from \eqref{cup4} that if $\text{FI} \neq 0$, then at least one of the scalar matter field VEVs $\langle C^{i} \rangle, i=1,2$ must  be non-vanishing so as to cancel the FI term and set $\langle V_{D} \rangle =0$ . Assuming, for simplicity, that the $U(1)$ charges of the matter supermultiplets are identical, that is, $Q_1=Q_2$--as was the case in the $B-L$ MSSM vacuum presented in \cite{Ashmore:2020ocb}--then one can always rotate $C^1$ and $C^2$ so that, for example, 
\begin{equation}
\langle C^1 \rangle \neq 0~, \quad \langle C^2 \rangle =0 \ .
\label{book1}
\end{equation}
In this case, unlike when $\text{FI}=0$, the $C^1$ matter field does {\it not} decouple from the $S$ and $T$ moduli. Rather it mixes with them in a complicated way. Note, however, that the $C^2$ matter scalar continues to completely decouple.

For each set of VEVs $\langle s \rangle$, $\langle t \rangle$ and $\langle C^1 \rangle$ satisfying the $D$-flatness condition, there again exists a unitary matrix $\mathbf{U}$ that rotates the scalar fluctuation basis $\delta S$, $\delta T$ and $\delta C^1$ to a new basis $\xi^{1}, \xi^{2}, \xi^{3}$
whose kinetic terms in the effective Lagrangian are canonically normalized and are mass eigenstates. Both the real and imaginary parts of $\xi^{1}$ are ``eaten''  by the anomalous $U(1)$ vector superfield which becomes massive. Now, however, its mass is the sum of two different terms. The first is the anomalous mass $m_{\text{anom}} \sim \mathcal{O}(M_U)$ arising as in the $\text{FI}=0$ case. The second mass, however, is due to the spontaneous breaking of the $U(1)$ symmetry by the non-zero VEV of $C^1$ and, hence, has the form $m_{\text{matter}} \sim \langle C^1 \rangle $.  Therefore, $\xi^{1}$  disappears from the low energy matter spectrum leaving $\xi^{2}$ and $\xi^{3}$ as canonically normalized massless scalars. 

Since $N=1$ supersymmetry is unbroken, it follows that the three fermions $\psi_{S}, \psi_{T}$ and $\psi_{C^1}$  are also rotated by matrix $\mathbf{U}$ to a new basis $\psi_{\xi}^{1}, \psi_{\xi}^{2}, \psi_{\xi}^{3}$ with canonical kinetic energy terms. The fermion $\psi_{\xi}^{1}$ also has mass $m_{\text{anom}} + m_{\text{matter}}$ and is ``eaten'' by the gaugino part of the anomalous $U(1)$ vector superfield. This leaves only two massless fermions, $\psi_{\xi}^{2}$ and $\psi_{\xi}^{3}$, in the low energy effective Lagrangian. 

Therefore, ignoring interaction terms, the hidden sector low energy Lagrangian when $\text{FI} \neq 0$ has two separate contributions. The first is due to the $\tilde{S,}\tilde{T,}$ and $\tilde{C^1}$ chiral superfields and is given by
\begin{equation}
\mathcal{L} \supset \sum_{i=2,3} \big(-\partial^{\mu} \bar{\xi}^{i} \partial_{\mu} \xi^{i} -i \psi_{\xi}^{i} \slashed\partial \psi_{\xi}^{i\dagger} \big) \ .
\label{help1a}
\end{equation}
The second contribution is due to the $\tilde{C}_{2}$ chiral superfield which, since it does not have a non-vanishing VEV, has ``decoupled'' from the other fields. 
Hence, it remains a massless multiplet with its own kinetic energy Lagrangian. This massless matter field Lagrangian is given by
\begin{equation}
\mathcal{L} \supset -G_{22}\partial_\mu C^2 \partial^\mu \bar C^{2}
 -iG_{22}\psi^2 \slashed{\partial} \psi^{\bar 2\dag} \ ,
 \label{ny1b}
\end{equation}  
where
\begin{equation}
G_{22}= 
\frac{\partial^2 K_{\text{matter}}^{\text{(hid)}}}
{\partial C^2\partial \bar C^{2}}  = e^{\kappa_{4}^{2}K_{T}/3} \ .
\label{black1b}
\end{equation}

\section{Mass Spectrum after SUSY breaking}\label{mass_spect_sec}

In this section, we analyze the mass spectrum of the low energy theory after turning on non-perturbative effects. That is, we introduce the non-vanishing superpotential
\begin{equation}
\hat W_{np}(S,T) \neq 0 
\label{view1}
\end{equation}
which generates non-vanishing F-terms $F_S$ and $F_T$ that break $N=1$ supersymmetry and produce a potential energy term $V_{F}$ that, as has the form 
given by
\begin{equation}
V_F=e^{\kappa^2_4K}\left[ g^{A\bar B}(D_A\hat W_{np})(D_{\bar B}\hat W_{np}^*)-3\kappa_4^2{|\hat W_{np}|^2}  \right] 
\label{fly4}
\end{equation}
where
\begin{equation}
D_A\hat W_{np}=\partial_{A}\hat W_{np}+\kappa_4^2K_A\hat W_{np}\ .
\label{fly4A}
\end{equation}
The indices $A,B$ each run over $S,T$. 
This non-perturbative potential potentially stabilizes the remaining massless scalar fields  
defined in the previous section. Recall that before turning on these non-perturbative effects, the matter content in both the observable and the hidden sectors was massless. Furthermore--as discussed in Section 2--the theory contained one massless modulus supermultiplet $(\xi^2,\psi_\xi^2)$ when $FI=0$, and several such massless supermultiplets $(\xi^i,\psi_\xi^i)$, $i \geq 2$ for $FI \neq 0$. 

After the non-perturbative effects which break supersymmetry are turned on, the mass terms of the low energy theory are the following. As discussed in \cite{Choi:1997cm,Soni:1983rm,Kaplunovsky:1993rd,Brignole:1997wnc,Martin:1997ns}:
\begin{itemize}

\item The gravitino mass 

\begin{equation}
m_{3/2}=\kappa_{4}^{2}e^{(K_{S}+K_{T})/2}|\hat W_{np}|\ .
\end{equation}
The gravitino mass is a good indicator of the scale of the masses aquired by the low-energy spectrum after supersymmetry is broken. Therefore, we define
\begin{equation}
m_{\text{SUSY}}=m_{3/2}=\kappa_{4}^{2}e^{(K_{S}+K_{T})/2}|\hat W_{np}|\ .
\end{equation}

\item Soft SUSY breaking mass terms on the observable sector. These include the  universal gaugino mass term
\begin{equation}
M_{1/2}=\frac{1}{2{\rm Re} f_1}F^{A}\partial_{A}{\rm Re}f_{1}\ ,
\end{equation}
as well as the quadratic scalar masses
\begin{equation}
m_{\cal{I} \bar{\cal{J}}}^{2}=m^{2}_{3/2}Z_{\cal{I} \bar{\cal{J}}}-F^{A}\bar{F}^{\bar{B}}R_{A \bar{B} \cal{I}\bar{\cal{J}}}\ ,
\end{equation}
where
\begin{equation}
R_{A \bar{B} \cal{I}\bar{\cal{J}}}=\partial_{A}\partial_{\bar{B}}Z_{\cal{I}\bar{\cal{J}}}-\Gamma^{\cal{N}}_{A\cal{I}}Z_{\cal{N}\bar{\cal{L}}}{\bar{\Gamma}}^{\bar{\cal{L}}}_{\bar{B}\bar{\cal{J}}}
\label{r1}
\end{equation}
and
\begin{equation}
\Gamma^{\cal{N}}_{A\cal{I}}=Z^{\cal{N}\bar{\cal{J}}}\partial_{A}Z_{\bar{\cal{J}}\cal{I}} \ .
\label{r2}
\end{equation}
In our model
\begin{equation}
Z_{\cal{I} \bar{\cal{J}}}=e^{\kappa_4^2K_T/3}\mathcal{G}_{\cal{I}\bar{\cal{J}}}\ .
\end{equation}

\item Hidden scalar mass terms. The hidden sector scalar fields $C^1$ and $C^2$ also obtain soft SUSY breaking mass contributions. The formulas are identical to the scalars on the observable sector shown above with, however, the $\cal{I}, \cal{J}, \dots$ indices replaced by $L,M, \dots$. In addition, on the hidden sector, we have
\begin{equation}
Z_{L\bar M}\equiv G_{L\bar M}=e^{\kappa_4^2K_T/3}\delta_{L\bar M}\ , \quad \text{for}\quad L,M=1,2\ .
\end{equation}

\item Moduli and hidden scalar mass terms. All these scalar masses are obtained by studying the second derivatives of the potential $V_D+V_F$ with respect to moduli fields $S$, $T$  and the matter fields $C^{1}$, $C^{2}$ evaluated at the vacuum state. We get
\begin{multline}
V=\Lambda+\frac{1}{2}\left\langle\frac{\partial^2 V}{\partial z^A\partial \bar z^{\bar B}}\right\rangle\delta z^A\delta \bar z^{\bar B}+\frac{1}{2}\left\langle\frac{\partial^2 V}{\partial z^A\partial z^{B}}\right\rangle\delta z^A\delta  z^{B}+\frac{1}{2}\left\langle\frac{\partial^2 V}{\partial \bar z^{\bar A}\partial \bar z^{\bar B}}\right\rangle\delta \bar z^{\bar{A}}\delta  \bar{z}^{\bar B}+\dots\ ,
\end{multline}
where $z^A, z^B=S,T, C^{1}, C^{2}$ are the scalar perturbations and 
\begin{equation}
\left\langle V \right\rangle=\Lambda\ .
\end{equation}

The total potential has to satisfy the following conditions in order to define a stable vacuum state:
\begin{equation}
 \left \langle \frac{\partial V}{\partial z^A}\right\rangle=0\ ,\quad  \left \langle \frac{\partial V}{\partial \bar z^A}\right\rangle=0 \ . 
\end{equation}
In this section, we will assume that it is always possible to find a solution which satisfy the above conditions in both classes of vacua that we study, that is, when $\text{FI}=0$ and when $\text{FI}\neq 0$. In Section \ref{Sec:Mod_Stab}, we will discuss a few simple examples of scalar potentials which do satisfy these conditions.

We can separate the mass contributions from the D-term and F-term potentials and write
\begin{multline}
V=\Lambda+m_{\text{anom}}^2\bar\xi^1\xi^1+\\
\frac{1}{2}\left\langle\frac{\partial^2 V_{F}}{\partial z^A\partial \bar z^{\bar B}}\right\rangle\delta z^A\delta \bar z^{\bar B}+\frac{1}{2}\left\langle\frac{\partial^2 V_{F}}{\partial z^A\partial z^{B}}\right\rangle\delta z^A\delta  z^{B}+\frac{1}{2}\left\langle\frac{\partial^2 V_{F}}{\partial \bar z^{\bar A}\partial \bar z^{\bar B}}\right\rangle\delta \bar z^{\bar{A}}\delta  \bar{z}^{\bar B}+\dots\ ,
\end{multline}
After computing the mass matrices in the scalar pertrurbations, it is necessary to project these perturbations into a mass eigenstate basis, such that the mass matrix becomes diagonal. One of these mass eigenstates must be the state $\xi^1$, given in eq. \eqref{pad1}, which is fixed at compactification by the D-term potential. The rest of the moduli scalar states must be orthogonal to it. We will outline this process in the rest of this section, for both the vanishing FI and non-vanishing FI cases.

\item Observable sector fermion mass terms. Fermion masses arise in the observable sector from the Higgs mechanism--and are are not directly generated by soft SUSY breaking fermion mass terms due to gaugino condensation. 

\item Hidden sector matter fermion masses. As in the observable sector, for the hidden sector matter chiral superfields with $\langle C_{L} \rangle =0$, there are no soft SUSY breaking fermion mass terms generated by gaugino condensation. Furthermore, within the context we study, there is no hidden sector Higgs mechanism. It follows that the hidden sector matter fermions whose scalar partners satisfy $\langle C_{L} \rangle = 0$ are all massless--that is, 
\begin{equation}
m_{\psi_{L}}=0 \ .
\label{blue1}
\end{equation}

\item Moduli fermions mass terms. Adding a non-perturbative superpotential generates new moduli fermion masses in the low-energy effective theory. These originate from the fermion bilinear terms
 \begin{equation}
 \mathcal{L}\supset
-\frac{1}{2}e^{\kappa_4^2K/2}\mathcal{D}_A{D}_B\hat W_{np} \psi^{A}\psi^{B}+h.c.\ 
\end{equation}
 in the supergravity Lagrangian, where $A,B$ each run over $S,T$ and
\begin{multline}
\mathcal{D}_AD_B\hat W_{np}=\partial_{A}\partial_{B}\hat W_{np}+\kappa_4^2(\partial_{A}\partial_{B}K\hat W_{np}+\partial_{A}KD_B\hat W_{np}+\partial_{B}KD_A\hat W_{np})\\-\Gamma^C_{AB}D_C\hat W_{np}+\mathcal{O}(M_P^{-3})\ .
\end{multline}
These moduli fermion mass terms are non-vanishing in vacua in which the F-terms $F_S=D_S\hat W_{np}$ and $F_T=D_T\hat W_{np}$ are non-vanishing.

\end{itemize}

Since the form of the soft SUSY breaking terms in the both the observable and $\langle C^{L} \rangle =0$ hidden sector which result from gaugino condensation are well-known, in the rest of this section we will study the mass spectrum of the moduli fields as well as matter fields with $\langle C^{L} \rangle \neq 0$ . The results, once again, depend on whether or not the Fayet-Iliopoulos term vanishes. As before, we will begin our analysis for the $\text{FI}=0$ case.

\subsection {Vanishing $\text{FI}$ Term}

\subsubsection{Scalar Moduli Eigenstates}\label{sec:scalar_eigen1}

Prior to turning on the non-perturbative gaugino condensate superpotential, the canonically normalized complex scalar mass eigenstates $\xi^{1}$ and $\xi^{2}$ were presented in Subsection 2.2. Eigenstate $\xi^{1}$ was shown to have the non-vanishing mass $m_{\text{anom}} \sim\mathcal{O}(M_U)$, while $\xi^{2}$ was shown to be massless. In Section 2--following the analysis in \cite{Dumitru:2021jlh}--we absorbed $\xi^{1}$ into a massive $U(1)$ vector superfield and considered only the non-interacting effective Lagrangian for the zero mass $\xi^2$ scalar. This was given by the first term in \eqref{help1}. However, it will be useful in the following analysis to consider the effective Lagrangian for both scalars $\xi^1$ and $\xi^2$--prior to $\xi^1$ being absorbed. 
It is straightforward to show that the scalar part of \eqref{help1} then becomes 
\begin{equation}
\mathcal{L} \supset  -\partial^{\mu} \bar{\xi}^{1} \partial_{\mu} \xi^{1} -\partial^{\mu} \bar{\xi}^{2} \partial_{\mu} \xi^{2}-m_{\text{anom}}^2{\xi^1}\bar \xi^1\   \ .
\label{help1A}
\end{equation}

After turning on the non-perturbative potential term $V_{F}$, however, it is necessary to completely reanalyze the mass eigenstates. In this case, ignoring the kinetic terms, which will remain canonically normalized, the non-interaction part of the scalar Lagrangian in the $\text{FI}=0$ case is given by

\begin{equation}
\begin{split}
\label{light3}
\mathcal{L}\supset -m_{\text{anom}}^2{\xi^1}\bar \xi^1
-&2(\delta \bar S,\>\delta \bar T)
{\mathbf{M_s}}
\left( \begin{matrix}\delta S\\\delta  T \end{matrix}\right)\\
&-(\delta  S,\>\delta T)
{\mathbf{M_s^\prime}}
\left( \begin{matrix}\delta S\\\delta  T \end{matrix}\right)-( \delta \bar S,\>\delta \bar T)
{\mathbf{M_s^{\prime \prime}}}
\left( \begin{matrix}\delta \bar S\\\delta \bar T \end{matrix}\right)\ ,
\end{split}
\end{equation}
where the $2\times 2$ scalar moduli squared mass matrices are given by
\begin{equation}
\label{eq:M_s}
{\mathbf{M_s}}=\left(  
\begin{matrix}
m^2_{S\bar S}&m^2_{S\bar T}\\
m^2_{T\bar S}&m^2_{T\bar T}\\
\end{matrix}
\right)\ , \quad {\mathbf{M_s^\prime}}=\left(  
\begin{matrix}
m^2_{S S}&m^2_{S T}\\
m^2_{T S}&m^2_{T T}\\
\end{matrix}
\right)\ ,\quad {\mathbf{M_s^{\prime \prime}}}=\left(  
\begin{matrix}
m^2_{\bar S\bar S}&m^2_{\bar S\bar T}\\
m^2_{\bar T\bar S}&m^2_{\bar T\bar T}\\
\end{matrix}
\right)\ .
\end{equation}
The matrix elements have the expressions
\begin{equation}
\begin{split}
\label{eq:s_mass_1}
\left[{\mathbf{M_s}}\right]_{A\bar B}&\equiv m_{z^A\bar z^{\bar B}}^2=\frac{1}{2}\left\langle\frac{\partial^2 V_F}{\partial z^A\partial \bar z^{\bar B}}\right\rangle\ ,\\
 \left[{\mathbf{M_s}^\prime}\right]_{A B}&\equiv m_{z^A z^{ B}}^2=\frac{1}{2}\left\langle\frac{\partial^2 V_F}{\partial z^A\partial  z^{ B}}\right\rangle\ ,
 \\
\left[{\mathbf{M_s}^{\prime \prime}}\right]_{\bar A\bar B}&\equiv m_{\bar z^{\bar A}\bar z^{\bar B}}^2=\frac{1}{2}\left\langle\frac{\partial^2 V_F}{\partial \bar  z^{\bar A}\partial \bar z^{\bar B}}\right\rangle\ ,
\end{split}
\end{equation}
where $z^A\in [S, T]$ and $A,B=1,2$. 
These mass matrix elements are the order of the SUSY breaking scale, $m_{\text{SUSY}}$, discussed in detail in \ref{AppendixB1}. For example, for the matrix elements of $\mathbf{M_s}$ we have
\begin{equation}
m_{S\bar S},\>m_{S\bar T},\>m_{T\bar T}\sim \mathcal{O}(m_{\text{SUSY}})\ .
\label{plane1}
\end{equation}
It is generically true that
\begin{equation}
m_{S\bar S},\>m_{S\bar T},\>m_{T\bar T}\ll m_{\text{anom}}\sim \mathcal{O}(M_U) \ .
\label{plane2}
\end{equation}
Rewriting $\delta S$ and $\delta T$ in terms of $\xi^1$ and $\xi^2$  using \eqref{light1} and \eqref{eq:U22}, \eqref{eq:U22A}, expression \eqref{light3} becomes
{\small
\begin{equation}
\begin{split}
&\mathcal{L}\supset=-m_{\text{anom}}^2{\xi^1}{\bar \xi}^1+
\\
&-2({\bar \xi}^1,\>{\bar \xi}^2)
{\mathbf{U}^{-1}}^\dag \mathbf{M_s}\mathbf{U}^{-1}
\left( \begin{matrix} \xi^1\\   \xi^2 \end{matrix}\right)
-({ \xi}^1,\>{ \xi}^2)
{\mathbf{U}^{-1 {\text{T}}}} \mathbf{M_s^\prime}\mathbf{U}^{-1}
\left( \begin{matrix} \xi^1\\   \xi^2 \end{matrix}\right)
-({\bar \xi}^1,\>{\bar \xi}^2)
{\mathbf{U}^{-1}}^{*} \mathbf{M_s^{\prime \prime}}\mathbf{U}^{-1\dag}
\left( \begin{matrix}\bar \xi^1\\   \bar \xi^2 \end{matrix}\right)
\\
&
=-m_{\text{anom}}^2{\xi^1}{\bar \xi}^1-({\bar \xi}^1,\>{\bar \xi}^2)
\mathbf{M}_{\mathbf{s}}^{\boldmath{\xi}}
\left( \begin{matrix} \xi^1\\   \xi^2 \end{matrix}\right)
-({ \xi}^1,\>{ \xi}^2)
\mathbf{M}_{\mathbf{s}}^{\boldmath{\xi}\prime}
\left( \begin{matrix} \xi^1\\   \xi^2 \end{matrix}\right)
-({\bar \xi}^1,\>{\bar \xi}^2)
\mathbf{M}_{\mathbf{s}}^{\boldmath{\xi}\prime\prime}
\left( \begin{matrix}\bar \xi^1\\   \bar \xi^2 \end{matrix}\right)\ .
\end{split}
\end{equation}
}
The mass matrices $\mathbf{M}_{\mathbf{s}}^{\boldmath{\xi}}$, $\mathbf{M}_{\mathbf{s}}^{\boldmath{\xi}\prime}$ and $\mathbf{M}_{\mathbf{s}}^{\boldmath{\xi}\prime\prime}$
are all non-diagonal. Therefore, to get the new mass eigenstates after adding the non-perturbative effects, we need to diagonalize these matrices. However, since $m_{SUSY} \ll M_{U}$, it is straightforward to give a highly accurate approximation to the result. This is the following. Since $m_{\text{anom}}$ is so large, 
the $\xi^{1}$ scalar and its mass remain essentially unchanged after turning on the non-perturbative effects. Hence, we will assume that these quantities remain strictly unchanged and, since $\xi^1$ is so heavy, that it effectively decouples from the low energy theory. It follows that the direction of the eigenstate $\xi^2$ is fixed. The mass $m_{22}^2$ is then nothing more than the variation of the non-peturbative potential $V_F$ along this fixed direction. In conclusion, at the scale of SUSY breaking, we are left with a single scalar field $\xi^{2}$ and its conjugate, which form mass terms
\begin{align}
\mathcal{L}\supset - \left[ \mathbf{M}_{\mathbf{s}}^{\boldmath{\xi}}\right]_{22}\bar {\xi^2} \xi^2- \left[ \mathbf{M}_{\mathbf{s}}^{\boldmath{\xi}\prime}\right]_{22} {\xi^2}\xi^2-  \left[ \mathbf{M}_{\mathbf{s}}^{\boldmath{\xi}\prime\prime}\right]_{22}\bar {\xi^2}\bar \xi^2\ .
\end{align}
\begin{equation}
\begin{split}
\label{eq:first_sc_masses}
\left[ \mathbf{M}_{\mathbf{s}}^{\boldmath{\xi}}\right]_{22}&= 2m_{S\bar S}^2\langle\frac{g_{T\bar T}}{g_{S\bar S}}\frac{k_T\bar k_T}{\Sigma^2}\rangle-2m_{S\bar T}^2\langle\frac{k_S\bar k_T}{\Sigma^2}\rangle-2m_{T\bar S}^2\langle\frac{k_T\bar k_S}{\Sigma^2}\rangle
+2m_{T\bar T}^2\langle\frac{g_{S\bar S}}{g_{T\bar T}}\frac{k_S\bar k_S}{\Sigma^2} \rangle \ ,\\
\left[ \mathbf{M}_{\mathbf{s}}^{\boldmath{\xi}^\prime}\right]_{22}&=m_{SS}^2 \langle \frac{g_{T\bar T}}{g_{S\bar S}}\frac{k_T k_T}{\Sigma^2}\rangle-m_{S T}^2 \langle \frac{k_S k_T}{\Sigma^2}\rangle-m_{T S}^2 \langle \frac{k_S k_T}{\Sigma^2}\rangle
+m_{T T}^2 \langle \frac{g_{S\bar S}}{g_{T\bar T}}\frac{k_S k_S}{\Sigma^2} \rangle\ ,\\
\left[ \mathbf{M}_{\mathbf{s}}^{\boldmath{\xi}\prime \prime}\right]_{22}&=  m_{\bar S\bar S}^2 \langle \frac{g_{T\bar T}}{g_{S\bar S}}\frac{\bar k_T\bar k_T}{\Sigma^2}\rangle-m_{\bar S\bar T}^2 \langle \frac{ \bar k_S\bar k_T}{\Sigma^2}\rangle-m_{\bar T\bar S}^2 \langle \frac{\bar k_S\bar k_T}{\Sigma^2}\rangle
+m_{\bar T\bar T}^2\ \langle \frac{g_{S\bar S}}{g_{T\bar T}}\frac{\bar k_S\bar k_S}{\Sigma^2} \rangle\ .\\
\end{split}
\end{equation}
Using that $k_Ak_B=\bar k_A\bar k_B=-k_A\bar k_B$, where the indices $A, B$ run over $S$ and $T$, we find the Lagrangian mass terms
\begin{equation}
\begin{split}
\label{eq:first_sc_masses}
\mathcal{L}\supset &
-2 \left(   m_{S\bar S}^2\langle\frac{g_{T\bar T}}{g_{S\bar S}}\frac{k_T\bar k_T}{\Sigma^2}\rangle-m_{S\bar T}^2\langle\frac{k_S\bar k_T}{\Sigma^2}\rangle-m_{T\bar S}^2\langle\frac{k_T\bar k_S}{\Sigma^2}\rangle
+m_{T\bar T}^2\langle\frac{g_{S\bar S}}{g_{T\bar T}}\frac{k_S\bar k_S}{\Sigma^2} \rangle \right)  \bar {\xi^2} \xi^2\\
&+ \left(   m_{SS}^2\langle\frac{g_{T\bar T}}{g_{S\bar S}}\frac{k_T \bar k_T}{\Sigma^2}\rangle-m_{S T}^2\langle\frac{k_S \bar k_T}{\Sigma^2}\rangle-m_{T S}^2\langle\frac{k_S \bar  k_T}{\Sigma^2}\rangle
+m_{T T}^2\langle\frac{g_{S\bar S}}{g_{T\bar T}}\frac{k_S\bar  k_S}{\Sigma^2}  \rangle \right) {\xi^2} \xi^2\\
&+ \left( m_{\bar S\bar S}^2\langle\frac{g_{T\bar T}}{g_{S\bar S}}\frac{k_T\bar k_T}{\Sigma^2}\rangle-m_{\bar S\bar T}^2\langle\frac{ k_S\bar k_T}{\Sigma^2}\rangle-m_{\bar T\bar S}^2\langle\frac{ k_S\bar k_T}{\Sigma^2}\rangle
+m_{\bar T\bar T}^2\langle\frac{g_{S\bar S}}{g_{T\bar T}}\frac{ k_S\bar k_S}{\Sigma^2}  \rangle \right)\bar {\xi^2}\bar \xi^2\ .
\end{split}
\end{equation}
The mass terms above can also be written in terms of the real and imaginary components of $\xi^2$, that is
\begin{equation}
\xi^2=\text{Re}(\xi^2)+i\text{Im}(\xi^2)\equiv {\eta^2}+i{\phi^2}\ .
\end{equation}
We obtain
\begin{equation}
\begin{split}
\label{eq:first_sc_masses2}
\mathcal{L}\supset &
-\bigg[ (2m_{S\bar S}^2-m_{S S}^2-m_{\bar S\bar S}^2)\langle\frac{g_{T\bar T}}{g_{S\bar S}}\frac{k_T\bar k_T}{\Sigma^2}\rangle-
(2m_{S\bar T}^2-m_{S T}^2-m_{\bar S\bar T}^2)\langle\frac{k_S\bar k_T}{\Sigma^2}\rangle\\
&-(2m_{T\bar S}^2-m_{T S}^2-m_{\bar T\bar S}^2)\langle\frac{k_T\bar k_S}{\Sigma^2}\rangle
+(2m_{T\bar T}^2-m_{T T}^2-m_{\bar T\bar T}^2)\langle\frac{g_{S\bar S}}{g_{T\bar T}}\frac{k_S\bar k_S}{\Sigma^2} \rangle \bigg]  {\eta^2}^2\\
&-\bigg[ (2m_{S\bar S}^2+m_{S S}^2+m_{\bar S\bar S}^2)\langle\frac{g_{T\bar T}}{g_{S\bar S}}\frac{k_T\bar k_T}{\Sigma^2}\rangle-
(2m_{S\bar T}^2+m_{S T}^2+m_{\bar S\bar T}^2)\langle\frac{k_S\bar k_T}{\Sigma^2}\rangle\\
&-(2m_{T\bar S}^2+m_{T S}^2+m_{\bar T\bar S}^2)\langle\frac{k_T\bar k_S}{\Sigma^2}\rangle
+(2m_{T\bar T}^2+m_{T T}^2+m_{\bar T\bar T}^2)\langle\frac{g_{S\bar S}}{g_{T\bar T}}\frac{k_S\bar k_S}{\Sigma^2} \rangle \bigg] {\phi^2}^2\\
&-2i\bigg[ (m_{S S}^2-m_{\bar S\bar S}^2)\langle\frac{g_{T\bar T}}{g_{S\bar S}}\frac{k_T\bar k_T}{\Sigma^2}\rangle-
(m_{S T}^2-m_{\bar S\bar T}^2)\langle\frac{k_S\bar k_T}{\Sigma^2}\rangle\\
&-(m_{T S}^2-m_{\bar T\bar S}^2)\langle\frac{k_T\bar k_S}{\Sigma^2}\rangle
+(m_{T T}^2-m_{\bar T\bar T}^2)\langle\frac{g_{S\bar S}}{g_{T\bar T}}\frac{k_S\bar k_S}{\Sigma^2} \rangle \bigg] {{\phi^2}} {\eta^2}\ .\\
\end{split}
\end{equation}
Recalling from \eqref{pad2B} that $\eta^{2}$ is a linear combination of the $S$ and $T$ axions, $\sigma$ and $\chi$ respectively, it follows that
if the $F$-term potential generated by non-perturbative effects depends on the real parts of the moduli fields $s=\text{Re}(S)$ and $t=\text{Re}(T)$ only, 
the mass term coefficient in front of ${\eta^2}^2$, as well as the coefficient of the mixing term ${\phi^2}{\eta^2}$ in the above expression vanish. This is, of course, expected, since such a potential cannot generate a non-flat direction along the axion components of the moduli fields. 

In general, however, the non-perturbative potential can depend on the axionic components $\sigma$ and $\chi$ as well. In this case, the one must compute the mass coefficients in eq. \eqref{eq:first_sc_masses2} in front of ${\eta^{2}}^{2}$ and ${\phi^2}{\eta^2}$. Note that the potentially problematic mixing term ${\phi^2}{\eta^2}$ is possibly non-zero. However, we find that the dependence of the allowed scalar potentials on the axion fields is proportional to
$\cos \sigma, \>\cos \chi$. In general, these potentials are minimized when the cosine functions equal $-1$. Therefore, in vacuum states defined as having minimal energy, any cross terms such as ${\phi^2}{\eta^2}$, which couple axion fields and real scalar components, will vanish. In Section \ref{Sec:Mod_Stab}, we will show that this indeed the case in a set of explicit examples. From here on, we will always assume that the axionic degrees of freedom are separated from the real scalar ones in the mass matrices.

The complexity of these mass computations increases quickly 
for $FI \neq 0 $ (as well as when manifolds with $h^{1,1}>1$ are considered). In this case, the matter fields from the hidden sector are mixed with the moduli--as we will show in the next section. It is, therefore, easier to separate the real scalar and axion components in the potential from the start. That is, instead of expressing the mass mixing matrices as in eq \eqref{light3}, we write
\begin{equation}
\begin{split}
\label{light23}
\mathcal{L}\supset -m_{\text{anom}}^2{\phi^1}^2-m_{\text{anom}}^2{\eta^1}^2
-&(\delta s,\>\delta t)
{\mathbf{M^{(r)}_s}}
\left( \begin{matrix}\delta s\\\delta  t \end{matrix}\right)-(\delta \sigma,\>\delta \chi)
{\mathbf{M^{(i)}_s}}
\left( \begin{matrix}\delta \sigma \\ \delta  \chi \end{matrix}\right)\ .\\
\end{split}
\end{equation}
The symbols $(r)$ and $(i)$ label the real and the imaginary components of the moduli fields, respectively. 
In this case, the $2\times 2$ scalar moduli squared mass matrices are given by
\begin{equation}
\label{eq:M_s2}
{\mathbf{M_s^{(r)}}}=\left(  
\begin{matrix}
m^2_{s s}&m^2_{st}\\
m^2_{t s}&m^2_{tt}\\
\end{matrix}
\right)\ , \quad {\mathbf{M_s^{(i)}}}=\left(  
\begin{matrix}
m^2_{\sigma \sigma}&m^2_{\sigma \chi}\\
m^2_{\chi \sigma}&m^2_{\chi \chi}\\
\end{matrix}
\right)\ .
\end{equation}
Note that, from the above discussion, no mixing between the real scalars and axion components exists. The expressions for the matrix elements of ${\mathbf{M_s^{(r)}}}$ are obtained by doubly differentiating the $F$-term potential with respect to $s$ and $t$, while the matrix elements of ${\mathbf{M_s^{(i)}}}$ are obtained by doubly differentiating the $F$-term potential with respect to $\sigma$ and $\chi$. For example
\begin{equation}
\begin{split}
\label{eq:s_mass_potential}
\left[{\mathbf{M_s^{(r)}}}   \right]_{11}&\equiv m_{\sigma \sigma}^2=\frac{1}{2}\left\langle\frac{\partial^2 V_F}{\partial \sigma\partial \sigma}\right\rangle\ .
\end{split}
\end{equation}
The next step in our analysis is to rotate these scalar perturbations into a base for the real and imaginary parts of the eigenstates $\xi^1$ and $\xi^2$. From equation \eqref{light1}, we derive the the relations
\begin{equation}
\label{bird1}
\left(\begin{matrix}
\delta s\\\delta t
\end{matrix}\right)
=i\mathbf{U}^{-1}
\left(\begin{matrix}
{\phi^1}\\ {\phi^2}
\end{matrix}\right)\ ,\quad 
\left(\begin{matrix}
\delta \sigma\\\delta \chi
\end{matrix}\right)
=-i\mathbf{U}^{-1}
\left(\begin{matrix}
{\eta^1}\\ {\eta^2}
\end{matrix}\right)
\end{equation}
where matrix  $\mathbf{U}^{-1}$ is given in \eqref{eq:U22A}.
After decoupling the heavier states ${\phi^1}$ and ${\eta^1}$, which have mass close to the unification scale, we are left with the low energy mass terms
\begin{multline}
\mathcal{L}\supset
-\left(  m_{ss}^2\langle\frac{g_{T\bar T}}{g_{S\bar S}}\frac{k_T\bar k_T}{\Sigma^2}\rangle-2m_{st}^2\langle\frac{k_S\bar k_T}{\Sigma^2}\rangle
+2m_{tt}^2\langle\frac{g_{S\bar S}}{g_{T\bar T}}\frac{k_S\bar k_S}{\Sigma^2} \rangle \right){\phi^2}^2\\
-
\left(  m_{\sigma\sigma}^2\langle\frac{g_{T\bar T}}{g_{S\bar S}}\frac{k_T\bar k_T}{\Sigma^2}\rangle-2m_{\sigma \chi}^2\langle\frac{k_S\bar k_T}{\Sigma^2}\rangle
+m_{\chi \chi}^2\langle\frac{g_{S\bar S}}{g_{T\bar T}}\frac{k_S\bar k_S}{\Sigma^2} \rangle \right){\eta^2}^2\ ,
\end{multline}
which have the same form as in eq. \eqref{eq:first_sc_masses2} with vanishing mixing term ${\phi^2}{\eta^2}$. 

\subsubsection{Fermion Moduli Eigenstates}

The analysis for the fermions is similar. The matrix $\mathbf U$ which rotates the moduli fermions $\psi_S$ and $\psi_T$ into the massive state $\psi_\xi^1$ and the massless state $\psi_\xi^2$, such that
\begin{equation}
\label{bird2}
\left(\begin{matrix}
\psi_\xi^1\\ \psi_\xi^2
\end{matrix}
\right)=\mathbf{U}
\left(\begin{matrix}
\psi_S\\\psi_T
\end{matrix}
\right)\ ,\quad 
\left(\begin{matrix}
\psi_S\\\psi_T
\end{matrix}
\right)=\mathbf{U}^{-1}\left(\begin{matrix}
\psi_\xi^1\\ \psi_\xi^2
\end{matrix}
\right)
\ ,
\end{equation}
 is the same as for the scalars. 
During the superHiggs mechanism described in Section 2, $\psi_{\xi}^{1}$ formed a Dirac fermion with the $U(1)$ gaugino $\lambda_{2}$ and became part of a massive $U(1)$ vector superfield. We considered only the non-interacting effective Lagrangian for the zero mass $\psi_{\xi}^2$ fermion. This was given by the second term in \eqref{help1}. However, it will be useful in the following analysis to consider the effective Lagrangian for both fermions $\psi_\xi^1$ and $\psi_{\xi}^2$--prior to $\psi_{\xi}^1$ being absorbed. 
It is straightforward to show that the fermion part of \eqref{help1} then becomes 
\begin{equation}
\mathcal{L} \supset  -i \psi_{\xi}^{1} \slashed\partial \psi_{\xi}^{1\dagger} -i \psi_{\xi}^{2} \slashed\partial \psi_{\xi}^{2\dagger} -\frac{i}{\langle g_2^2\rangle}\lambda_{2} \slashed{\partial} \lambda_{2}^\dag - m_{\text{anom}}  \left(\psi_{\xi}^{1\dagger}\frac{\lambda_{2}^\dag}{\langle g_2\rangle} + \psi_{\xi}^{1} \frac{\lambda_{2}}{\langle g_2\rangle}  \right)
\ .
\label{help1A}
\end{equation}

After turning on the non-perturbative potential term $V_{F}$, however, it is necessary to completely reanalyze the mass eigenstates.  As explained in the introduction of this section, adding a non-perturbative superpotential generates new fermion masses in the low-energy effective theory, originating from the fermion bilinear terms
 \begin{equation}
 \mathcal{L}\supset
-\frac{1}{2}e^{\kappa_4^2K/2}\mathcal{D}_A{D}_B\hat W_{np} \psi^{A}\psi^{B}+h.c.\ 
\end{equation}
 in the supergravity Lagrangian, where $A,B$ each run over $S,T$.
In this case, ignoring the kinetic terms which will remain canonically normalized, the non-interaction part of the fermion Lagrangian in the $\text{FI}=0$ case is given by
\begin{equation}
\mathcal{L} \supset -m_{\text{anom}}  \left(\psi_{\xi}^{1\dagger}\frac{\lambda_{2}^\dag}{\langle g_2\rangle} + \psi_{\xi}^{1} \frac{\lambda_{2}}{\langle g_2\rangle}  \right)-\left[
\left(\begin{matrix}
\psi_S&\psi_T
\end{matrix}
\right)
{\mathbf{M_f}}
\left(\begin{matrix}
\psi_S\\\psi_T
\end{matrix}
\right)+h.c.\right]\ ,
\end{equation}
where we have defined the fermion mass matrix
\begin{equation}
\label{mfm}
{\mathbf{M_f}}=\left(
\begin{matrix}
M_{S S}& M_{ST}\\
M_{T S}& M_{TT}
\end{matrix}
\right) \ .
\end{equation}
These matrix elements are defined in the Appendix \ref{AppendixB2}. For the $FI=0$ case, where $\langle C^1\rangle=\langle C^2\rangle=0$, these are given by
\begin{equation}
\begin{split}
\label{eq:f_mass_1}
&M_{SS}=\tfrac{1}{2} e^{\kappa_4^2 \langle K_{\text{mod}}\rangle /2} \langle \partial^2_S\hat {W}_{np}+\kappa_4^2(\partial^2_SK_S {\hat W}_{np}+2\partial_SK_S{ \partial_S\hat W}_{np}\\
&\hspace{6cm}
+\kappa_4^2(\partial_SK_S)^2\hat W_{np})+\Gamma_{SS}^AD_A\hat W_{np} \rangle \ ,\\
&M_{S T},\>M_{T S}=\tfrac{1}{2} e^{\kappa_4^2 \langle K_{\text{mod}}\rangle/2}\langle \partial_S\partial_{T}{\hat W_{np}}+\kappa_4^2(\partial_SK_S   \partial_{T}{\hat W_{np}}+\partial_{T}K_T \partial_S{\hat W_{np}} \\
&\hspace{6cm}+
\kappa_4^2\partial_SK_S\partial_TK_T\hat W_{np})+\Gamma^A_{ST}D_A\hat W_{np}\rangle\ , \\
&M_{T T}=\tfrac{1}{2} e^{\kappa_4^2 \langle K_{\text{mod}}\rangle/2}\langle \partial^2_{T}{\hat W_{np}}
+\kappa_4^2(\partial^2_{T}K_T \hat W_{np}+2\partial_{T}K_T \partial_{T}{\hat W_{np}}\\
&\hspace{6cm}+\kappa_4^2
(\partial_TK_T)^2\hat W_{np})+\Gamma^A_{TT}D_A\hat W_{np}\rangle \ .\\
\end{split}
\end{equation}
Written in terms of the states $\psi_\xi^1,\>\psi_\xi^2$ only,  using \eqref{light1} and \eqref{eq:U22} , \eqref{eq:U22A}, the fermion mass terms in the Lagrangian become

\begin{equation}
\begin{split}
\mathcal{L}&\supset-m_{\text{anom}}\left(\psi_{\xi}^{1\dagger}\frac{\lambda_{2}^\dag}{\langle g_2\rangle} + \psi_{\xi}^{1} \frac{\lambda_{2}}{\langle g_2\rangle}  \right)-\left(\begin{matrix}[1.5]
\psi_\xi^{1}&\psi_\xi^{2}
\end{matrix}
\right)  \left(\begin{matrix}[1.5]
\begin{matrix}
\langle\frac{ \bar k_S}{\Sigma}\rangle &\quad \langle\frac{\bar k_T}{\Sigma}\rangle \\
 \quad\ \sqrt{\langle\frac{g_{T\bar T}}{g_{S\bar S}}\rangle}\langle\frac{\bar k_T}{\Sigma}\rangle&\quad- \langle\frac{\bar k_S}{\Sigma}\rangle\sqrt{\langle\frac{g_{S\bar S}}{g_{T\bar T}}\rangle}\\
\end{matrix}
\end{matrix}
\right)\times \\&\qquad\qquad\qquad\times
\left(
\begin{matrix}[1.2]
M_{S S}& M_{S T}\\
M_{T S}& M_{T T}
\end{matrix}
\right)
 \left(\begin{matrix}[1.5]
\begin{matrix}
\frac{  \langle k_S}{\Sigma}\rangle &\quad \langle\frac{ k_T}{\Sigma}\rangle \\
 \quad\ \sqrt{\langle\frac{g_{T\bar T}}{g_{S\bar S}}\rangle}\langle\frac{ k_T}{\Sigma}\rangle&\quad- \langle\frac{ k_S}{\Sigma}\rangle\sqrt{\langle\frac{g_{S\bar S}}{g_{T\bar T}}\rangle}\\
\end{matrix}
\end{matrix}
\right) \left(\begin{matrix}
\psi_\xi^1\\\psi_\xi^2
\end{matrix}
\right)+h.c.
\\
&
=-m_{\text{anom}}\left(\psi_{\xi}^{1\dagger}\frac{\lambda_{2}^\dag}{\langle g_2\rangle} + \psi_{\xi}^{1} \frac{\lambda_{2}}{\langle g_2\rangle}  \right)-\left[( \psi_\xi^{1},\> \psi_\xi^{2 })
\left(  
\begin{matrix}
M_{11}&M_{1 2}\\
M_{2 1}&M_{2 2}\\
\end{matrix}
\right)\left( \begin{matrix} \psi_\xi^1\\   \psi_\xi^2 \end{matrix}\right)+h.c.\right]\\
&=-m_{\text{anom}}\left(\psi_{\xi}^{1\dagger}\frac{\lambda_{2}^\dag}{\langle g_2\rangle} + \psi_{\xi}^{1} \frac{\lambda_{2}}{\langle g_2\rangle}  \right)
-\left[\mathbf{M}_{\mathbf{f}}^{\boldmath{\xi}}\right]_{AB}\psi_\xi^A\psi_\xi^{B }
-\left[\mathbf{M}_{\mathbf{f}}^{\boldmath{\xi}}\right]_{\bar A\bar B}\psi_\xi^{\bar A\dag}\psi_\xi^{\bar B\dag }\ .
\end{split}
\end{equation}
The elements of the fermion mass matrix $\mathbf{M}_{\mathbf{f}}^{\boldmath{\xi}}={\mathbf{U}^{-1}}^\dag \mathbf{M_f}\mathbf{U}^{-1}$ are
\begingroup
\allowdisplaybreaks
\begin{equation}
\begin{split}
M_{11}&=  M_{S S}\langle\frac{k_S\bar k_S}{\Sigma^2}\rangle+2M_{S T}\langle\frac{k_S\bar k_T}{\Sigma^2}\rangle
+M_{TT}\frac{k_T\bar k_T}{\Sigma^2} \rangle \ ,\\
M_{12}&= M_{SS}\langle\sqrt{\frac{g_{T\bar T}}{g_{S\bar S}}}\frac{k_S\bar k_T}{\Sigma^2}\rangle-M_{S T}\langle\sqrt{\frac{g_{S\bar S}}{g_{T\bar T}}}\frac{k_S\bar k_S}{\Sigma^2}\rangle
+M_{T S}\langle\sqrt{\frac{g_{T\bar T}}{g_{S\bar S}}}\frac{k_T\bar k_T}{\Sigma^2}\rangle-M_{T T}\langle\sqrt{\frac{g_{S\bar S}}{g_{T\bar T}}}\frac{k_T\bar k_S}{\Sigma^2} \rangle \ ,\\\
\label{eq:first_f_masses}
M_{22}&= M_{SS}\langle\frac{g_{T\bar T}}{g_{S\bar S}}\frac{k_T\bar k_T}{\Sigma^2}\rangle-2M_{S T}\langle\frac{k_S\bar k_T}{\Sigma^2}\rangle
+M_{T T}\langle\frac{g_{S\bar S}}{g_{T\bar T}}\frac{k_S\bar k_S}{\Sigma^2} \rangle \ .
\end{split}
\end{equation}
\endgroup
The mass $m_{\text{anom}}$ of the Dirac fermion $\Psi= \dbinom{\lambda_{2}^{\dagger}/\langle g_2\rangle}{\psi_{\xi}^{1}} $ is much larger than $M_{12}(=M_{21})$ and $\>M_{22}$. Therefore, the state $\psi_\xi^1$, together with the gaugino $\lambda_2$, are decoupled at the SUSY breaking scale, leaving only $\psi_{\xi}^{2}$ in the low energy Lagrangian. Hence, the only fermion mass terms present in the effective theory are
\begin{equation}
\mathcal{L}\supset -M_{22}\left(\psi_\xi^2\psi_\xi^{2}+\psi_\xi^{2\dag}\psi_\xi^{2\dag}\right)\ .
\end{equation}
This is the mass of a Majorana fermion 
\begin{equation}
\mathcal{L}\supset -M_{\Psi^2_\xi}\Psi_\xi^2\Psi_\xi^{2\dag}\ ,
\end{equation}
where
\begin{equation}
\Psi_\xi^2= \dbinom{\psi_\xi^2}{\psi_\xi^{2\dag}} \ , \quad M_{\Psi^2_\xi}=2M_{22}\ .
\end{equation}
It follows from the above expression for $M_{22}$ that
\begin{equation}
\label{books2}
M_{\Psi_\xi^2} \sim {\mathcal{O}}(m_{\text{SUSY}}) \ll {\cal{O}}(M_{U}) \ .
\end{equation}

Finally, we note that turning on non-perturbative gaugino condensation leads to $N=1$ SUSY breaking. As a consequence, the masses of the scalars $\phi^2$, $\eta^2$ and the fermion $\psi^2_\xi$, which used to be identical--that is, were all vanishing-- prior to supersymmetry breaking, now differ. That is,
\begin{equation}
m_{\phi^2}\neq m_{\eta^2}\neq M_{\Psi_\xi^2}\ .
\end{equation}

\subsubsection{Final Low-Energy States}\label{sec:final_states1}

We conclude this subsection by displaying, in the case that $\text{FI}=0$, the Lagrangian for the low-energy spectrum of the moduli and hidden sector after supersymmetry breaking. Ignoring all interaction terms, this Lagrangian is given by
\begin{equation}
\begin{split}
\mathcal{L}&= -\partial^{\mu} \phi^{2} \partial_{\mu} \phi^{2}-\partial^{\mu} \eta^{2} \partial_{\mu} \eta^{2}- G_{L\bar M}\partial_\mu C^L \partial^\mu \bar C^{\bar M} -i \Psi_{\xi}^{2} \slashed\partial \Psi_{\xi}^{2\dagger}-iG_{L\bar M}\psi^L \slashed{\partial} \psi^{\bar M\dag}\\
&- m_{\phi^2}^2{\phi^2}^2- m_{\eta^2}^2{\eta^2}^2-\tfrac{1}{2}M_{\Psi_\xi^2}\Psi_\xi^2\Psi_\xi^{2\dag}-m_{L\bar M}^2C^{L}\bar C^{\bar M}\ .
 \end{split}
\end{equation}  
where 
\begin{align}
\label{eq:mphi2}
m_{\phi^2}^2&= m_{s s}^2\langle\frac{g_{T\bar T}}{g_{S\bar S}}\frac{k_T\bar k_T}{\Sigma^2}\rangle-2m_{s t}^2\langle\frac{k_S\bar k_T}{\Sigma^2}\rangle
+m_{tt}^2\langle\frac{g_{S\bar S}}{g_{T\bar T}}\frac{k_S\bar k_S}{\Sigma^2} \rangle \  , \\
\label{eq:meta2}
m_{\eta^2}^2&= m_{\sigma \sigma}^2\langle\frac{g_{T\bar T}}{g_{S\bar S}}\frac{k_T\bar k_T}{\Sigma^2}\rangle-2m_{\sigma\chi}^2\langle\frac{k_S\bar k_T}{\Sigma^2}\rangle
+m_{\chi \chi}^2\langle\frac{g_{S\bar S}}{g_{T\bar T}}\frac{k_S\bar k_S}{\Sigma^2} \rangle \  , \\
M_{\Psi_\xi^2}&= 2M_{S S}\langle\frac{g_{T\bar T}}{g_{S\bar S}}\frac{k_T\bar k_T}{\Sigma^2}\rangle-4M_{ST}\langle\frac{k_S\bar k_T}{\Sigma^2}\rangle
+2M_{T T}\langle\frac{g_{S\bar S}}{g_{T\bar T}}\frac{k_S\bar k_S}{\Sigma^2} \rangle\ 
\label{eq:mPsi2}
\end{align}
are the moduli masses computed above. The hidden matter scalars $C^1$ and $C^2$ obtain the masses
\begin{equation}
m^2_{L\bar M}=m^{2}_{3/2}G_{L\bar M}-F^{A}\bar F^{\bar B}R_{A\bar BL\bar M}\ ,
\label{pro1}
\end{equation}
while the hidden matter fermions $\psi_1$ and $\psi_2$ remain massless. An equivalent expression for the hidden scalar masses has been derived in Appendix \ref{AppendixB1}.

We now continue to the case in which $\text{FI} \neq 0$.

\subsection{Non-vanishing $\text{FI}$ Term}\label{non-vanish_mass}

Let us now allow consider the case when $\text{FI}\neq 0$. That is,
\begin{equation}
- \frac{a\epsilon_S\epsilon_R^2}{\kappa_{4}^{2}}
\left( -\frac{1}{\langle s \rangle}\pi \beta \epsilon_S l+\frac{3l}{\langle t \rangle} \right) \neq 0 \ .
\label{clip1}
\end{equation}
Then, as discussed in Subsection 2.3, in order to satisfy the $D$-flatness condition $\langle V_{D} \rangle=0$ it is necessary for at least one of the hidden sector matter field VEVs to be non-zero. Following the discussion in that subsection, we will henceforth assume that 
\begin{equation}
\langle C^1 \rangle \neq 0 , \quad \langle C^2 \rangle  =0 \ .
\label{clip2}
\end{equation}
As discussed previously, it follows that the matter field $C^1$ mixes with the $S$ and $T$ moduli, while only the $C^2$ hidden matter field completely decouples. The condition \eqref{umb3} to preserve $N=1$ supersymmetry is then given by
\begin{equation}
\langle \mathcal{P}\rangle =0\quad \Rightarrow \quad - \frac{a\epsilon_S\epsilon_R^2}{\kappa_{4}^{2}}
\left( -\frac{1}{\langle s \rangle}\pi \beta \epsilon_S l+\frac{3l}{\langle t \rangle} \right) -e^{\kappa_{4}^{2}\langle K_{T} \rangle /3} \langle \tilde Q_1 \rangle \langle C^1\rangle \langle\bar C^1\rangle=0\ ,
\end{equation}
This vacuum is defined by the expectation values of three scalar fields, $T$, $S$ and $C^1$. The scalar perturbations around the vacuum are $\delta S$, $\delta T$ and $\delta C^1$. Based on the results of our work in \cite{Dumitru:2021jlh}, we find prior to turning on any non-perturbative effects,  a linear combination of these scalar perturbations
given by
\begin{equation}
\xi^1=\langle g_{S\bar S}\frac{\bar k_S}{\Sigma^\prime} \rangle \delta S+\left\langle g_{T\bar T}\frac{\bar k_T}{\Sigma^\prime}
+g_{T\bar C^1}\frac{\bar k_1}{\Sigma^\prime}\right\rangle \delta T+\left\langle g_{C^1\bar T}\frac{\bar k_T}{\Sigma^\prime}+g_{C^1\bar C^1}\frac{\bar k_1}{\Sigma^\prime}\right\rangle \delta C^1
\end{equation}
acquires the mass $m_{\text{anom}} \sim {\cal{O}(}M_{U})$. One can then form two other states, $\xi^{2}$ and $\xi^{3}$, as linear combinations of these perturbations which remain massless. As a consequence, one must extend the $2 \times 2$ rotation matrix $\mathbf{U}$ defined in the previous sections to a $3 \times 3$ matrix
\begin{equation} 
\label{fdr1}
\left(\begin{matrix}
\xi^1\\ \xi^2\\ \xi^3
\end{matrix}
\right)=\mathbf{U}
\left(\begin{matrix}
 \delta S \\ \delta T\\ \delta C^1
\end{matrix}
\right) \ .
\end{equation}
As above, it is useful to write $\xi^{i}=\eta^{i}+i\phi^{i}$, $i=1,2,3$ and let
\begin{equation}
\label{fdr2}
\delta S= \delta s +i \delta \sigma , \quad \delta T= \delta t +i \delta \chi , \quad \delta C^1 = {\text{Re}} (\delta C^1) +i {\text{Im}} (\delta C^1) \ .
\end{equation}
Rotation \eqref{fdr1} can then be expressed as
\begin{equation}
\left(\begin{matrix}
{\phi^1}\\{\phi^2}\\{\phi^3}
\end{matrix}
\right)=-i\mathbf{U^{(r)}}
\left(\begin{matrix}
\delta s\\\delta t\\ \text{Re}(\delta C^1)
\end{matrix}
\right)\ ,\quad 
\left(\begin{matrix}
{\eta^1}\\{\eta^2}\\{\eta^3}
\end{matrix}
\right)=i\mathbf{U^{(i)}}
\left(\begin{matrix}
\delta \sigma\\\delta \chi\\ \text{Im}(\delta C^1)
\end{matrix}
\right)\ .
\label{fdr3}
\end{equation}
Similarly, prior to turning on any non-perturbative effects, it follows from $N=1$ supersymmetry that the associated fermions $\psi_{S}, \psi_{T}, \psi_{C^1}$ also transform as
\begin{equation}
\label{bird3}
\left(\begin{matrix}
\psi_\xi^1\\\psi_\xi^2\\\psi_\xi^3
\end{matrix}
\right)=\mathbf{U}
\left(\begin{matrix}
\psi_S\\ \psi_T\\ \psi_1
\end{matrix}
\right)\ ,
\end{equation}
with the same unitary matrix $\mathbf{U}$.

When $\langle C^1\rangle=0$, we are in the vanishing $\text{FI}$ case we studied earlier. The matter scalar perturbations are not coupled to the moduli perturbations and, hence. the rotation matrices have the form
\begin{equation}
\begin{split}
&\mathbf{U^{(r)}}=\mathbf{U^{(i)}}=\mathbf{U}=
 \left(\begin{matrix}
\frac{1}{\langle \Sigma \rangle }\begin{pmatrix}[1.3]
\langle g_{S\bar S}\bar k_S \rangle &\quad\langle g_{T\bar T}\bar k_T \rangle\\
 \sqrt{\langle g_{S\bar S}g_{T\bar T} \rangle} \langle \bar k_T \rangle&\quad -\sqrt{\langle g_{S\bar S}g_{T\bar T} \rangle} \langle \bar k_S \rangle
\end{pmatrix}&0\\
0&i
\end{matrix}
\right) \ ,\\
&\mathbf{U^{(r)}}^{-1}=\mathbf{U^{(i)}}^{-1}=\mathbf{U}^{-1}=
 \left(\begin{matrix}
\frac{1}{\langle \Sigma \rangle }\begin{pmatrix}[1.4]
\langle k_S \rangle&\quad\sqrt{\langle \frac{g_{T\bar T}}{g_{S\bar S}}\rangle} \langle k_T \rangle\\
 \langle k_T \rangle&\quad-\sqrt{\langle \frac{g_{S\bar S}}{g_{T\bar T}}\rangle}  \langle k_S \rangle
\end{pmatrix}&0\\
0&-i
\end{matrix}
\right)\ .
\end{split}
\label{river1}
\end{equation}
However,  when we turn on a scalar matter field VEV $\langle C^1 \rangle \neq 0$, 
we expect the rotation matrices for the real scalar components, the axions and the fermions--henceforth denoted by $\mathbf{U_s^{(r)}}$, $\mathbf{U_s^{(i)}}$ and $\mathbf{U_f}$ respectively for clarity--to differ; that is $\mathbf{U_s^{(r)}} \neq\mathbf{U_s^{(i)}} \neq\mathbf{U_f}$. This was not the case in the previous example when $\text{FI}=0$, because one of the eigesntates, $\xi^1={\eta^1}+i{\phi^1}$, was already fixed at the compactification scale, while the remaining one, $\xi^2={\eta^2}+i{\phi^2}$ was unique and orthogonal to it. 

Let us assume that when we turn on the $\langle C^1 \rangle \neq 0$ VEV, the rotation matrices have the form
\begin{align}
&\mathbf{U}\mapsto \mathbf{U_s^{(r)}}=\mathbf{R_s^{(r)}}\mathbf{U}\ , \label{br1}\\
&\mathbf{U}\mapsto \mathbf{U_s^{(i)}}=\mathbf{R_s^{(i)}}\mathbf{U}\ ,\label{br2}\\
&\mathbf{U}\mapsto \mathbf{U_f}=\mathbf{R_f}\mathbf{U}\ ,
\label{eq:fermion_matrix}
\end{align}
where $\mathbf{R_s^{(r)}}$, $\mathbf{R_s^{(i)}}$ and $\mathbf{R_f}$ are $3\times 3$ matrices and $\mathbf{U}$ is given in \eqref{river1}.
The form of these $\mathbf{R}$ matrices must be such that $\mathbf{U_s^{(r)}}$, $\mathbf{U_s^{(i)}}$ and $\mathbf{U_f}$ normalize the kinetic terms. For example, the matrix $\mathbf{U_s^{(r)}}$ must rotate the real scalar perturbations
$(\delta s,\>\delta t\ ,\delta \text{Re}(C^1)$ into the eigenstates $({\phi^1},\>{\phi^2},\>{\phi^3})$, such that
\begin{equation}
g_{A\bar B} \partial^{\mu} \text{Re}(\delta z^A)\partial _{\mu} \text{Re}(\delta \bar z^{\bar B})=\delta_{AB}\partial^{\mu}\phi^A \partial_{\mu} \phi^{B}\ .
\end{equation}
The kinetic energy normalization condition shown above is satisfied for
\begin{equation}
g_{A\bar B}\left[\mathbf{U_s^{(r)}}^{-1}\right]^{ A}_{ C}\left[{\mathbf{U_s^{(r)}}^{-1}}^{\dag}\right]^{\bar B}_{\bar D}=\delta_{C\bar D}\ ,
\end{equation}
from which we recover an orthogonality condition for the rotation matrix $\mathbf{R_s^{(r)}}$; that is
\begin{equation}
\begin{split}
&g_{A\bar B}\left[\mathbf{U_s^{(r)}}^{-1}\right]^{ A}_{ C}\left[{\mathbf{U_s^{(r)}}^{-1}}^{\dag}\right]^{\bar B}_{\bar D}=\delta_{C\bar D}\ ,\\
&\Rightarrow g_{A\bar B}\left[\mathbf{U_s^{(r)}}^{-1}\right]^{ A}_{ E}\left[{\mathbf{U_s^{(r)}}^{-1}}^{\dag}\right]^{\bar B}_{\bar F}[{\mathbf{R_s^{(r)}}}^{-1}]^E_C[{\mathbf{R_s^{(r)}}^{-1}}^\dag]^{\bar F}_{\bar D}=\delta_{C\bar D}\\&\Rightarrow
\delta_{E\bar F}[{\mathbf{R_s^{(r)}}}^{-1}]^E_C[{\mathbf{R_s^{(r)}}^{-1}}^\dag]^{\bar F}_{\bar D}=\delta_{C\bar D}\\
&\Rightarrow{\mathbf{R_s^{(r)}}}^\dag\mathbf{R_s^{(r)}}=\mathcal{I}\ .
\end{split}
\end{equation}
To get the third equality we have used the fact that the matrix $\mathbf{U}^{-1}$ diagonalizes the metric $g_{A\bar B}$ as well.
Hence, we learned that $\mathbf{R_s^{(r)}}$ is a $3\times 3$ unitary matrix. Choosing the VEV of $C^1$ to be real, such that $\langle C^1\rangle =v\in \mathbb{R}$, one can show that 
{\small
\begin{multline}
\mathbf{R_s^{(r)}}=\\=\left(\begin{matrix}
\cos \alpha_r \cos \beta_r \quad &\quad\cos \alpha_r \sin \beta_r \sin \gamma_r-\sin \alpha_r \cos \gamma_r&\quad\cos\alpha_r\sin \beta_r\cos \gamma_r+\sin \alpha_r \sin \gamma_r\\
\sin \alpha_r \cos \beta_r &\quad \sin \alpha_r \sin \beta_r \sin \gamma_r+\cos \alpha_r \cos \gamma_r&\quad\sin \alpha_r \sin \beta_r \cos\gamma_r-
\cos \alpha_r \sin \gamma_r\\
-\sin \beta_r \quad &\quad\cos \beta_r \sin \gamma_r&\quad\cos \beta_r \cos \gamma_r
\end{matrix}
\right)\ ,
\end{multline}
}
where $\alpha_r,\>\beta_r\ ,\gamma_r$ are arbitrary \emph{real} rotation angles in 3D. 

Continuing our analysis with this generic expression is possible, but very complicated. Therefore, for simplicity, we henceforth assume that the matter field VEV $v$, while non-vanishing, is infinitesimally small compared to the unification scale. That is, take
\begin{equation}
\label{sky1}
v=qM_U,\quad q\ll 1\ .
\end{equation}
This is equivalent to the relation
\begin{equation}
\langle k_C \rangle \ll \langle k_S \rangle, \langle k_T \rangle \ .
\end{equation}
When this is the case, the rotation angles $\alpha_r\>,\beta_r,\>\gamma_r$ are infinitesimally small. To linear order in $\alpha_r,\>\beta_r,\>\gamma_r$, the matrix $\mathbf{R_s^{(r)}}$ is given by
\begin{equation}
\mathbf{R_s^{(r)}}\approx \left(\begin{matrix}
1 \quad &\quad-\alpha_r&\quad\beta_r\\
\alpha_r &\quad 1&\quad-\gamma_r\\
-\beta_r \quad &\quad\gamma_r&\quad1
\end{matrix}
\right)=\bm{\mathcal{I}}_3+\left(\begin{matrix}
0 \quad &\quad-\alpha_r&\quad\beta_r\\
\alpha_r &\quad 0&\quad-\gamma_r\\
-\beta_r \quad &\quad\gamma_r&\quad0
\end{matrix}
\right)\ .
\end{equation}
Hence,  using \eqref{river1} and \eqref{br1} we find that
\begin{equation}
\mathbf{U_s^{(r)}}=\left(\begin{matrix}[1.4]
\langle \frac{1}{\Sigma}g_{S\bar S} \bar k_S- \frac{\alpha_r}{\Sigma}\sqrt{{g_{T\bar T}}{g_{S\bar S}}} \bar k_T \rangle&\quad\langle \frac{1}{\Sigma}g_{T\bar T} \bar k_T+\frac{\alpha_r}{\Sigma}\sqrt{{g_{T\bar T}}{g_{S\bar S}}}\bar k_S \rangle &\quad i\beta_r\\
\langle  \frac{1}{\Sigma}\sqrt{{g_{T\bar T}}{g_{S\bar S}}}\bar k_T+\frac{\alpha_r}{\Sigma}g_{S\bar S} \bar k_S \rangle &\quad \langle -\frac{1}{\Sigma}\sqrt{{g_{S\bar S}}{g_{T\bar T}}}\bar k_S+ \frac{\alpha_r}{\Sigma}\bar k_T \rangle &\quad-i\gamma_r\\
\langle -\beta_r g_{S\bar S}\bar k_S+\gamma_r \sqrt{{g_{T\bar T}}{g_{S\bar S}}}\bar k_T \rangle & \langle -\beta_r \bar k_T-\gamma_r \sqrt{{g_{T\bar T}}{g_{S\bar S}}}\bar k_S \rangle &\quad i
\end{matrix}
\right)
\end{equation}
and 
\begin{equation}
\label{eq:u_m1_matrix}
\mathbf{U_s^{(r)}}^{-1}=\left(\begin{matrix}[1.4]
\langle \frac{1}{\Sigma} k_S+ \frac{\alpha_r}{\Sigma}\sqrt{\frac{g_{T\bar T}}{g_{S\bar S}}} k_T\rangle &\quad \langle \frac{1}{\Sigma}\sqrt{\frac{g_{T\bar T}}{g_{S\bar S}}} k_T-\frac{\alpha_r}{\Sigma}k_S \rangle & \quad \langle \frac{\beta_r}{\Sigma} k_S- \frac{\gamma_r}{\Sigma}\sqrt{\frac{g_{T\bar T}}{g_{S\bar S}}}k_T \rangle\\
\langle  \frac{1}{\Sigma}k_T-\frac{\alpha_r}{\Sigma}\sqrt{\frac{g_{S\bar S}}{g_{T\bar T}}} k_S \rangle &\quad \langle - \frac{1}{\Sigma}\sqrt{\frac{g_{S\bar S}}{g_{T\bar T}}} k_S- \frac{\alpha_r}{\Sigma}k_T \rangle &\quad \langle  \frac{\beta_r}{\Sigma} k_T + \frac{\gamma_r}{\Sigma}\sqrt{\frac{g_{S\bar S}}{g_{T\bar T}}}k_S\ \rangle \\
-i\beta_r&i\gamma_r&\quad -i
\end{matrix}
\right)\ .
\end{equation}

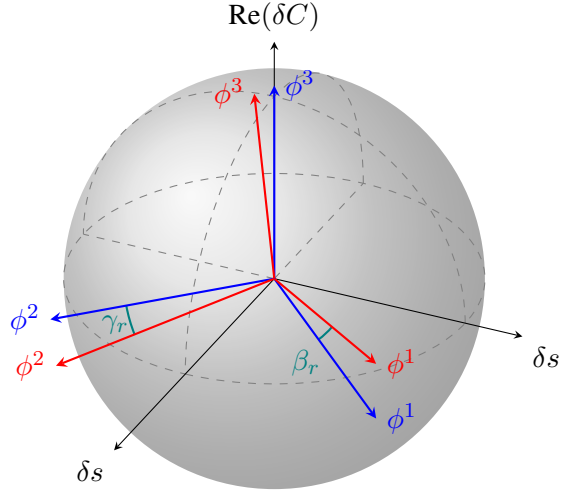
\begin{figure}[t]
\centering
\begin{tikzpicture}[tdplot_main_coords, scale = 2.8]

\coordinate (O) at (0,0,0);
\coordinate (P) at ({1/sqrt(3)},{1/sqrt(3)},{1/sqrt(3)});
\coordinate (P1) at (1,1,0);
\coordinate (P11) at (1,1,0.3);

\coordinate (P2) at (0.8,-0.8);
\coordinate (P22) at (0.78,-0.78,-0.26);

\coordinate (P3) at (0,0,1.06);
\coordinate (P33) at (-0.14,-0.17,0.9);
 
\shade[ball color = lightgray,
    opacity = 0.5
] (0,0,0) circle (1cm);
 
\tdplotsetrotatedcoords{0}{0}{0};
\draw[dashed,
    tdplot_rotated_coords,
    gray
] (0,0,0) circle (1);
 
\tdplotsetrotatedcoords{90}{90}{90};
\draw[dashed,
    tdplot_rotated_coords,
    gray
] (1,0,0) arc (0:180:1);
 
\tdplotsetrotatedcoords{0}{90}{90};
\draw[dashed,
    tdplot_rotated_coords,
    gray
] (1,0,0) arc (0:180:1);
 
\draw[-stealth] (0,0,0) -- (1.80,0,0) 
    node[below left] {$\delta s$};
 
\draw[-stealth] (0,0,0) -- (0,1.30,0)
    node[below right] {$\delta s$};
 
\draw[-stealth] (0,0,0) -- (0,0,1.30)
    node[above] {$\text{Re}(\delta C)$};
 
\draw[dashed, gray] (0,0,0) -- (-1,0,0);
\draw[dashed, gray] (0,0,0) -- (0,-1,0);

 \pic [ draw,teal, thick, "$\beta_r$", angle radius=10mm,anchor=north east,angle eccentricity=1.1] {angle = P1--O--P11 };
 \pic [ draw,teal, thick, "$\gamma_r$", angle radius=20mm ,angle eccentricity=1.1] {angle = P2--O--P22 };
 
\draw[thick,blue, -stealth] (0,0,0) -- (P1) node[right] {${\phi^1}$};
\draw[thick,red, -stealth] (0,0,0) -- (P11) node[right] {${\phi^1}$};
\draw[thick,blue, -stealth] (0,0,0) -- (P2) node[left] {${\phi^2}$};
\draw[thick,red, -stealth] (0,0,0) -- (P22) node[left] {${\phi^2}$};
\draw[thick, blue,-stealth] (0,0,0) -- (P3) node[right] {${\phi^3}$};
\draw[thick, red,-stealth] (0,0,0) -- (P33) node[left] {${\phi^3}$};

\end{tikzpicture}
\caption{With blue we show the scalar eigenstates $\phi^1,\phi^2,\phi^3$, defined in eq. \eqref{fdr3}, in the vanishing $\text{FI}$ case. With red we show the same eigenstates, after a matter field VEV $\langle C^1\rangle$ is turned on infinitesimally. The angle $\beta_r$ is fixed by the D-term stabilization condition and has the expression given in eq. \eqref{tf1}. The 
$\gamma_r$ angle is fixed after non-perturbative effects are turned on, such that the scalar perturbations $\phi^2$ and $\phi^3$ are aligned along the new mass eigenstates of the system (see eq. \eqref{align123}). The rotations of the scalars $\eta^1, \eta^2, \eta^3$ and the fermions $\psi_\xi^1,\psi_\xi^2,\psi^3_\xi$ into the new eigenstates, after turning on the $ C^1$ field VEV infinitesimally, have similar forms. In their cases, the $\beta_i$ angle, corresponding to the $\eta$ states, and the $\beta_f$ angle, corresponding to the fermions, are equal to the $\beta_r$ angle. They are all fixed by the D-flatness condition. On the other hand, the $\gamma_r$,$\gamma_i$ and $\gamma_f$ angles, defined so as to diagonalize the mass mixing matrices after SUSY is broken, differ from each other.}
\end{figure}

The $\alpha_r$, $\beta_r$ and $\gamma_r$ rotation parameters are determined after aligning the linear combinations $\phi^1$, $\phi^2$ and $\phi^3$ along the true mass eigenstates of the system. We learned that the D-term stabilization condition yields a massive scalar eigenstate $\xi^1$ in the direction
\begin{equation}
\xi^1=\langle g_{S\bar S}\frac{\bar k_S}{\Sigma^\prime} \rangle \delta S+ \left \langle g_{T\bar T}\frac{\bar k_T}{\Sigma^\prime}
+g_{T\bar C^1}\frac{\bar k_1}{\Sigma^\prime}\right\rangle \delta T+  \left \langle g_{C^1\bar T}\frac{\bar k_T}{\Sigma^\prime}+g_{C^1\bar C^1}\frac{\bar k_1}{\Sigma^\prime}\right\rangle \delta C^1\ ,
\end{equation}
The imaginary component of this scalar field is given by
\begin{equation}
\label{eq:phi_1}
\phi^1= \langle g_{S\bar S}\frac{\bar k_S}{\Sigma^\prime} \rangle \delta s+  \left \langle g_{T\bar T}\frac{\bar k_T}{\Sigma^\prime}
+g_{T\bar C^1}\frac{\bar k_1}{\Sigma^\prime}\right \rangle \delta t+  \left \langle g_{C^1\bar T}\frac{\bar k_T}{\Sigma^\prime}+g_{C^1\bar C^1}\frac{\bar k_1}{\Sigma^\prime}\right\rangle\text{Re}(\delta C^1)\ .
\end{equation}
where 
\begin{equation}
{\Sigma^\prime}^2=g_{S\bar S}k_S\bar k_S+g_{T\bar T}k_T\bar k_T+ k_1 \bar k_1\ .
\end{equation}
Therefore,
\begin{equation}
 \frac{1}{\langle \Sigma^\prime \rangle} \approx  \frac{1}{\langle \Sigma \rangle} \left(1-\langle \frac{1}{2}\frac{k_1 \bar k_1}{\Sigma^2} \rangle \right) ,
\end{equation} 
where $\Sigma$ is defined in \eqref{pad3}.
Since we compute the rotation matrix to linear order only, we can consider $\langle \Sigma^\prime \rangle=\langle \Sigma \rangle$. Comparing the first row of the matrix $\mathbf{U_s^{(r)}}$ to the linear relation in \eqref{eq:phi_1}, we learn that the $D$-term stabilization condition fixes the rotation angles $\alpha_r$ and $\beta_r$ to be
\begin{equation}
\label{tf1}
\alpha_r=0\ ,\quad i\beta_r= \langle g_{C^1\bar T}\frac{\bar k_T}{\Sigma} \rangle+\langle g_{C^1\bar C^1}\frac{\bar k_1}{\Sigma} \rangle \ .
\end{equation}
For $v=\langle C^1\rangle$ real, both $ \langle \bar k_T \rangle$ and $ \langle \bar k_1 \rangle$ are purely imaginary. Therefore, choosing $\beta_r$ as a real parameter was well motivated.
Hence, we have determined the rotation matrices $\mathbf{U_s^{(r)}}$ and $\mathbf{U_s^{(r)}}^{-1}$ up to one real parameter $\gamma_r$--which remains undetermined. The reason this parameter is still unfixed is because, so far, we have used what we learned from the D-term term stabilization condition only. At the unification scale, the D-flatness condition determines the $\phi^1$ direction, but leaves two flat directions $(\phi^2,\phi^3)$ undetermined. The orthogonality relations between these directions reduce the number of degrees of freedom in choosing these flat directions from two to one only, namely the $\gamma_r$ parameter.

Similar arguments apply for the axion (or imaginary) components of the scalar fields, as well as for the fermion components of the supermultiplets.  In those cases, one would find that the matrices $\mathbf{U_s^{(i)}}$ and $\mathbf{U_f}$ have the same form as $\mathbf{U_s^{(r)}}$, and contain the undermined parameters $\gamma_i$ and $\gamma_f$ respectively.

\subsubsection{Scalar Eigenstates}

After the D-term stabilization process alone, the scalar mass terms present in the effective theory are  
\begin{equation}
\mathcal{L}\supset  -m_{\text{anom}}^2\xi^1\bar \xi^1=-m_{\text{anom}}^2{\phi^1}^2-m_{\text{anom}}^2{\eta^1}^2\ .
\end{equation}
When non-perturbative effects are turned on, the Lagrangian gets additional scalar mass terms. Following our conclusions at the end of Subsection \ref{sec:scalar_eigen1}, we will express the mass matrices in a basis composed of the real scalar components and the corresponding imaginary component fields. That is,
\begin{equation}
\begin{split}
\label{eq:mass_terms_before}
\mathcal{L}&\supset
= -m_{\text{anom}}^2{\phi^1}^2-m_{\text{anom}}^2{\eta^1}^2\\
&-\left( \delta s,\>\delta t, \>\text{Re}(\delta  C^1)\right)
{\mathcal{M}_s^{(r)}}
\left( \begin{matrix}\delta s\\\delta  t\\ \text{Re}(\delta C^1) \end{matrix}\right)
-\left( \delta \sigma,\>\delta \chi, \>\text{Im}(\delta  C^1)\right)
{\mathcal{M}_s^{(i)}}
\left( \begin{matrix}\delta \sigma \\\delta  \chi\\\text{Im}(\delta C^1) \end{matrix}\right) \ .
\end{split}
\end{equation}
In the above equation, the scalar squared mass matrices are given by
\begin{equation}
\label{eq:calM_s}
{\mathcal{M}^{(r)}_s}=\left(  
\begin{matrix}
m^2_{ss}&m^2_{st} &0\\
m^2_{ts}&m^2_{tt}&0\\
0&0&m^2_{1}
\end{matrix}
\right)\ ,\quad 
{\mathcal{M}^{(i)}_s}=\left(  
\begin{matrix}
m^2_{\sigma \sigma}&m^2_{\sigma \chi} &0\\
m^2_{\chi \sigma}&m^2_{\chi \chi}&0\\
0&0&m^2_{1}
\end{matrix}
\right)\ ,
\end{equation}
These mass matrices have been defined in Appendix \ref{AppendixB1}. We left out the mass term of the $C^2$ field for the moment, which will be added back in the end results.

 Next, we rotate the $\left(\delta s, \delta t, \text{Re}(\delta C^1)\right)$ and $\left(\delta \sigma, \delta \chi, \text{Im}(\delta C^1)\right)$ fields into the eigenstates  $(\phi^1,\phi^2,\phi^3)$ and  $(\eta^1,\eta^2,\eta^3)$, respectively.
This set of transformations are achieved using the $\mathbf{U_s^{(r)}}^{-1}$ matrix, given--up to one undetermined angle $\gamma_{r}$--in eq. \eqref{eq:u_m1_matrix}, and $\mathbf{U_s^{(i)}}^{-1}$ which, as discussed above, is identical with the exception of one undetermined angle $\gamma_i$. Assuming that $\phi^1,\>\eta^1$ are fixed at the unification scale as presented above, the angles $\gamma_r,\gamma_i$ are fixed by aligning the linear combinations $\phi^2,\>\phi^3$ and, respectively, $\eta^2,\>\eta^3$ along the new scalar mass eigenstates of the system, such that the mass matrices become diagonal. We will outline this process in detail below. Before we begin, however, it is important to point out that because the mass matrices $\mathcal{M}^{(r)}_s$ and ${\mathcal{M}^{(i)}_s}$ differ in general, we expect the $\gamma_{r}$ and $\gamma_{i}$ angles to be different in order to  achieve the eigenstate alignment for the real and imaginary component scalar fields. Consequently,
\begin{equation}
\mathbf{U_s^{(r)}}^{-1}\neq \mathbf{U_s^{(i)}}^{-1}\ .
\end{equation}

Our next step is to rewrite the scalar mass terms in eq. \eqref{eq:mass_terms_before} in the new eigenstate basis. It follows from the above discussion that
\begin{equation}
\begin{split}
\label{eq:mass_terms_after}
\mathcal{L}\supset
& m_{\text{anom}}^2{\phi^1}^2+m_{\text{anom}}^2{\eta^1}^2+\left( {\phi^1},\>{\phi^2} , \>{\phi^3}\right)i{\mathbf{U_s^{(r)}}^{-1 {\text{T}}}}
{\mathcal{M}_s^{(r)}}i\mathbf{U_s^{(r)}}^{-1}
\left( \begin{matrix}{\phi^1}\\ {\phi^2} \\ {\phi^3} \end{matrix}\right)\\
&\hspace{3cm}+\left( {\eta^1},\>{\eta^2}, \>{\eta^3} \right)(-i){\mathbf{U_s^{(i)}}^{-1 {\text{T}}}}
{\mathcal{M}_s^{(i)}}(-i)\mathbf{U_s^{(i)}}^{-1}
\left( \begin{matrix}{\eta^1} \\ {\eta^2} \\ {\eta^3} \end{matrix}\right)\\
=&  \left( {\phi^1},\>{\phi^2} , \>{\phi^3}\right)
{\mathcal{M}_s^{\xi(r)}}
\left( \begin{matrix}{\phi^1}\\ {\phi^2} \\ {\phi^3} \end{matrix}\right)
+\left( {\eta^1},\>{\eta^2}, \>{\eta^3} \right)
{\mathcal{M}_s^{\xi(i)}}
\left( \begin{matrix}{\eta^1} \\ {\eta^2} \\ {\eta^3} \end{matrix}\right)\ .\\
\end{split}
\end{equation}
\noindent We begin by considering the first term--that is, the $\phi^{i}$, $i=1,2,3$ contribution to the Lagrangian. The elements of the (symmetric) $3\times 3$ matrix ${{\mathcal{M}_s^{\xi(r)}}}$ are 
\begingroup
\allowdisplaybreaks
\begin{align}
\left[ {{\mathcal{M}_s^{\xi(r)}}} \right]_{11}&=m^2_{\text{anom}}+m_{ss}^2\langle \frac{k_S\bar k_S}{\Sigma^2} \rangle+2m_{st}^2 \langle \frac{k_S\bar k_T}{\Sigma^2} \rangle
+m_{tt}^2 \langle \frac{k_T\bar k_T}{\Sigma^2} \rangle\\
&=m_{\text{anom}}^2+
\left[ {{\mathcal{M}_s^{\xi(r)}}} \right]^\prime_{11}\ ,\\
\left[ {{\mathcal{M}_s^{\xi(r)}}} \right]_{12}&=m_{ss}^2 \langle \sqrt{\frac{g_{T\bar T}}{g_{S\bar S}}}\frac{k_S\bar k_T}{\Sigma^2} \rangle-m_{st}^2 \langle \sqrt{\frac{g_{S\bar S}}{g_{T\bar T}}}\frac{k_S\bar k_S}{\Sigma^2} \rangle\\
&\qquad\qquad+m_{ts}^2 \langle \sqrt{\frac{g_{T\bar T}}{g_{S\bar S}}}\frac{k_T\bar k_T}{\Sigma^2} \rangle-m_{tt}^2 \langle\sqrt{\frac{g_{S\bar S}}{g_{T\bar T}}}\frac{k_T\bar k_S}{\Sigma^2} \rangle \ ,\\
\left[ {{\mathcal{M}_s^{\xi(r)}}} \right]_{13}&=-\beta_{r}\left[ {{\mathcal{M}_s^{\xi(r)}}} \right]^\prime_{11}+\gamma_r \left[ {{\mathcal{M}_s^{\xi(r)}}} \right]_{12}\ ,\\
\left[ {{\mathcal{M}_s^{\xi(r)}}} \right]_{22}&=m_{ss}^2 \langle \frac{g_{T\bar T}}{g_{S\bar S}}\frac{k_T\bar k_T}{\Sigma^2} \rangle -2m_{st}^2 \langle \frac{k_S\bar k_T}{\Sigma^2}\rangle
+m_{tt}^2 \langle \frac{g_{S\bar S}}{g_{T\bar T}}\frac{k_S\bar k_S}{\Sigma^2} \rangle \ ,\\
\left[ {{\mathcal{M}_s^{\xi(r)}}} \right]_{23}&=-\beta_{r} \left[ {{\mathcal{M}_s^{\xi(r)}}} \right]_{12}^2+\gamma_r \left[ {{\mathcal{M}_s^{\xi(r)}}} \right]_{22}\ ,\\
\left[ {{\mathcal{M}_s^{\xi(r)}}} \right]_{33}&=m_1^2-\beta_{r} \left[ {{\mathcal{M}_s^{\xi(r)}}} \right]_{13}+{\gamma_r}\left[ {{\mathcal{M}_s^{\xi(r)}}} \right]_{23}\ .
\end{align}
\endgroup
Since $m_{\text{anom}}\gg m_{\text{soft}}$, $\phi^1$ is actually decoupled from $\phi^2$ and $\phi^3$ in the spectrum. It follows that the $\phi^{i}$  mass terms in the Lagrangian can be written as

\begin{equation}
\mathcal{L}\supset m_{\text{anom}}^2{\phi^1}^2+(\phi^2,\>\phi^3)
\left(  
\begin{matrix}[1.5]
\left[ {{\mathcal{M}_s^{\xi(r)}}} \right]_{22}&\left[ {{\mathcal{M}_s^{\xi(r)}}} \right]_{23}\\
\left[ {{\mathcal{M}_s^{\xi(r)}}} \right]_{23}&\left[ {{\mathcal{M}_s^{\xi(r)}}} \right]_{33}\\
\end{matrix}
\right)\left( \begin{matrix}\phi^2\\ \phi^3\end{matrix}\right) \ .
\end{equation}
The mass matrix is diagonal if and only if
\begin{equation}
\label{align123}
\left[ {{\mathcal{M}_s^{\xi(r)}}} \right]_{23}=0\quad \Rightarrow \quad \gamma_r
=\frac{  \left[{{\mathcal{M}_s^{\xi(r)}}} \right]_{12}    }
{  \left[ {{\mathcal{M}_s^{\xi(r)}}} \right]_{22}   }\beta_{r}\ .
\end{equation}
Therefore, as promised, turning on the non-perturbative effects fixed the remaining angle $\gamma_r$ in the rotation matrix for the real component scalars, that is ${\mathbf{{U}^{(r)}_s}}^{-1}$. The 
$\gamma_r$ angle was fixed such that the scalar perturbations $\phi^2$ and $\phi^3$ - which before turning on any non-perturbative effects were just flat directions orthogonal to $\phi^1$ - are aligned along the new mass eigenstates of the system. The mass of these states are
\begin{align}
\label{eq:second_sc_masses}
m_{\phi^2}^2&=m_{ss}^2 \langle \frac{g_{T\bar T}}{g_{S\bar S}}\frac{k_T\bar k_T}{\Sigma^2} \rangle -2m_{st}^2 \langle \frac{k_S\bar k_T}{\Sigma^2}\rangle
+m_{tt}^2 \langle \frac{g_{S\bar S}}{g_{T\bar T}}\frac{k_S\bar k_S}{\Sigma^2} \rangle\\
&=\mathcal{O}(m_{\text{SUSY}}^2)\nonumber\\
m_{\phi^3}^2&=m_1^2+\beta_{r}^2\frac{\left[ {{\mathcal{M}_s^{\xi(r)}}} \right]_{11}^\prime \left[ {{\mathcal{M}_s^{\xi(r)}}} \right]_{22}-\left[ {{\mathcal{M}_s^{\xi(r)}}} \right]^2_{12}}{\left[ {{\mathcal{M}_s^{\xi(r)}}} \right]_{22}}\nonumber\\
\label{eq:second_sc_masses2}
&=m_1^2+\beta_{r}^2\frac{\left( m_{ss}^2m_{tt}^2-m_{st}^4\right) \left(\langle \frac{g_{S\bar S}}{g_{T\bar T}}\frac{|k_S|^4}{\Sigma^4}\rangle +\langle \frac{g_{T\bar T}}{g_{S\bar S}}\frac{|k_T|^4}{\Sigma^4} \rangle \right)-2m_{st}^4 \langle \frac{|k_S|^2|k_T|^2}{\Sigma^4}\rangle }{m_{ss}^2 \langle \frac{g_{T\bar T}}{g_{S\bar S}}\frac{k_T\bar k_T}{\Sigma^2}\rangle -2m_{st}^2 \langle \frac{k_S\bar k_T}{\Sigma^2} \rangle
+m_{tt}^2 \langle \frac{g_{S\bar S}}{g_{T\bar T}}\frac{k_S\bar k_S}{\Sigma^2} \rangle}\\
&=\mathcal{O}(m_{\text{SUSY}}^2)\nonumber\\
\end{align}

Let us now consider consider the second  term in \eqref{eq:mass_terms_after}--that is, the $\eta^{i}$, $i=1,2,3$ contribution to the Lagrangian. For these axionic components, the conclusions are similar. After decoupling the heavy $\eta^1$ state, the masses of the $\eta^{2}$ and $\eta^{3}$ states left in the low energy spectrum are given by
\begin{align}
\label{eq:second_sc_masses_axion}
m_{\eta^2}^2&=m_{\sigma \sigma}^2 \langle \frac{g_{T\bar T}}{g_{S\bar S}}\frac{k_T\bar k_T}{\Sigma^2} \rangle -2m_{\sigma \chi}^2 \langle \frac{k_S\bar k_T}{\Sigma^2}\rangle
+m_{\chi \chi}^2 \langle \frac{g_{S\bar S}}{g_{T\bar T}}\frac{k_S\bar k_S}{\Sigma^2} \rangle\\
&=\mathcal{O}(m_{\text{SUSY}}^2)\nonumber\\
\label{eq:second_sc_masses_axion2}
m_{\eta^3}^2
&=m_1^2+\beta_{i}^2\frac{\left( m_{\sigma \sigma}^2m_{\chi \chi}^2-m_{\sigma \chi}^4\right) \left(\langle \frac{g_{S\bar S}}{g_{T\bar T}}\frac{|k_S|^4}{\Sigma^4}\rangle +\langle \frac{g_{T\bar T}}{g_{S\bar S}}\frac{|k_T|^4}{\Sigma^4} \rangle \right)-2m_{\sigma \chi}^4 \langle \frac{|k_S|^2|k_T|^2}{\Sigma^4}\rangle }{m_{\sigma \sigma}^2 \langle \frac{g_{T\bar T}}{g_{S\bar S}}\frac{k_T\bar k_T}{\Sigma^2}\rangle -2m_{\sigma \chi}^2 \langle \frac{k_S\bar k_T}{\Sigma^2} \rangle
+m_{\chi \chi}^2 \langle \frac{g_{S\bar S}}{g_{T\bar T}}\frac{k_S\bar k_S}{\Sigma^2} \rangle} \\
&=\mathcal{O}(m_{\text{SUSY}}^2) \nonumber 
\end{align}
where $\beta_{i}=\beta_{f}$ given in \eqref{tf1}.

\subsubsection{Fermion Eigenstates}

After the D-term stabilization process alone, the fermion mass term present in the effective theory is
\begin{equation}
\mathcal{L}\supset  - m_{\text{anom}}  \left(\psi_{\xi}^{1\dagger}\frac{\lambda_{2}^\dag}{\langle g_2\rangle} + \psi_{\xi}^{1} \frac{\lambda_{2}}{\langle g_2\rangle}  \right)
\ ,
\end{equation}
where $\psi_\xi^1$ is given by
\begin{equation}
\psi_\xi^1= \langle g_{S\bar S}\frac{\bar k_S}{\Sigma^\prime} \rangle \psi_S+  \left \langle g_{T\bar T}\frac{\bar k_T}{\Sigma^\prime}
+g_{T\bar C^1}\frac{\bar k_1}{\Sigma^\prime}\right \rangle \psi_T+  \left \langle g_{C^1\bar T}\frac{\bar k_T}{\Sigma^\prime}+g_{C^1\bar C^1}\frac{\bar k_1}{\Sigma^\prime}\right\rangle\psi_1
\end{equation}
and $\lambda$ is the $U(1)$ gaugino.

When non-perturbative effects are turned on, the Lagrangian gets additional fermion mass terms
\begin{align}
\mathcal{L}&\supset
-(\psi_S,\>\psi_T, \>\psi_1){\mathcal{M}_f}
\left( \begin{matrix} \psi_S\\   \psi_T\\  \psi_1 \end{matrix}\right)+h.c.\ ,
\end{align}
where the fermion mass matrix is given by
\begin{equation}
{\mathcal{M}_f}=\left(  
\begin{matrix}
M_{S S}&M_{ST} &0\\
M_{T S}&M_{T T}&0\\
0&0&0
\end{matrix}
\right)\ .
\end{equation}
The mass matrix elements  $M_{S S},\>M_{S T}=M_{T S},\> M_{T T}$ are defined in eq. \eqref{eq:f_mass_1}.
 Next, we rotate the $\psi^A=(\psi_S,\psi_T,\psi_1)$ fermions into the the $\psi_\xi^A=(\psi_\xi^1,\psi_\xi^2,\psi_\xi^3)$ mass eigenstates.
This rotation is achieved using the $\mathbf{U_f}$ matrix, defined in eq. \eqref{eq:fermion_matrix}, which depends on the yet undetermined parameter $\gamma_f$. Assuming the $\psi_\xi^1$ is fixed at the unification scale as discussed above, the angle $\gamma_f$ is set by aligning the linear combinations $\psi_\xi^2,\>\psi_\xi^3$, along the new mass eigenstates of the system.

The conclusions in the case of the fermions are similar to the results for the scalar field components. We find that.

\begin{align}
\mathcal{L}&\supset
= -  m_{\text{anom}}  \left(\psi_{\xi}^{1\dagger}\frac{\lambda_{2}^\dag}{\langle g_2\rangle} + \psi_{\xi}^{1} \frac{\lambda_{2}}{\langle g_2\rangle}  \right)-
\left[(\psi^\dag_S,\>\psi^\dag_T, \>\psi^\dag_1){\mathcal{M}_f}
\left( \begin{matrix} \psi_S\\   \psi_T\\  \psi_1 \end{matrix}\right)+h.c.\right]\ \\
&= -m_{\text{anom}}  \left(\psi_{\xi}^{1\dagger}\frac{\lambda_{2}^\dag}{\langle g_2\rangle} + \psi_{\xi}^{1} \frac{\lambda_{2}}{\langle g_2\rangle}  \right)-\left[
(\psi_\xi^{1},\>\psi_\xi^{2}, \>\psi_\xi^{3})
{\bm{\mathbf{U}}_f^{-1{\text{T}}}}{\mathcal{M}_f}
{\bm{\mathbf{U}}_f^{-1}}\left( \begin{matrix} \psi_\xi^1\\   \psi_\xi^2\\  \psi_\xi^3 \end{matrix}\right)+h.c.\right]
\\
&\equiv -m_{\text{anom}}  \left(\psi_{\xi}^{1\dagger}\frac{\lambda_{2}^\dag}{\langle g_2\rangle} + \psi_{\xi}^{1} \frac{\lambda_{2}}{\langle g_2\rangle}  \right)-\left(\left[{{\mathcal{M}_f^\xi}}\right]_{A B}\psi_\xi^A\psi_\xi^{ B}+h.c.\right)
\end{align}
After decoupling the $\psi_\xi^1$ heavy state, the fermion mass matrix becomes diagonal for
\begin{equation}
 \gamma_f=\frac{\left[{{\mathcal{M}_f}}\right]_{12}}{\left[{{\mathcal{M}_f}}\right]_{22}}\beta_{f} \ ,
\end{equation}
where
\begin{align}
\left[{{\mathcal{M}_f}}\right]_{12}&=M_{S S} \langle \sqrt{\frac{g_{T\bar T}}{g_{S\bar S}}}\frac{k_S\bar k_T}{\Sigma^2}\rangle -M_{ST}\langle \sqrt{\frac{g_{S\bar S}}{g_{T\bar T}}}\frac{k_S\bar k_S}{\Sigma^2} \rangle
\\
&\qquad \qquad+M_{TS} \langle \sqrt{\frac{g_{T\bar T}}{g_{S\bar S}}}\frac{k_T\bar k_T}{\Sigma^2} \rangle -M_{TT} \langle \sqrt{\frac{g_{S\bar S}}{g_{T\bar T}}}\frac{k_T\bar k_S}{\Sigma^2}\rangle\ , \\
\left[{{\mathcal{M}_f}}\right]_{22}&=M_{SS}\langle \frac{g_{T\bar T}}{g_{S\bar S}}\frac{k_T\bar k_T}{\Sigma^2}\rangle -2M_{ST}\langle \frac{k_S\bar k_T}{\Sigma^2}\rangle 
+M_{T T} \langle \frac{g_{S\bar S}}{g_{T\bar T}}\frac{k_S\bar k_S}{\Sigma^2}\rangle 
\end{align}
and $\beta_{f}=\beta_{r}$ is given in \eqref{tf1}.
Just as in the case of the scalars, the $\gamma_f$ angle from the rotation matrix $\mathbf{{U}_f}^{-1}$ was fixed such that the fermion states $\psi_\xi^2$ and $\psi_\xi^3$ - which before turning on any non-perturbative effects were just flat directions orthogonal to $\psi_\xi^1$ - are aligned along the new mass eigenstates of the system. 

Putting everything together, we get two Majorana fermions
\begin{equation}
\Psi_\xi^2= \dbinom{\psi_\xi^2}{\psi_\xi^{2\dag}} \ ,\quad \Psi_\xi^3= \dbinom{\psi_\xi^3}{\psi_\xi^{3\dag}} 
\end{equation}
with non-vanishing mass terms
\begin{equation}
\mathcal{L}\supset -M_{22}\Psi_\xi^2\Psi_\xi^{2\dag}-M_{33}\Psi_\xi^3\Psi_\xi^{3\dag}\ .
\end{equation}
The masses of these Majorana fermions are found to be
\begin{equation}
\begin{split}
\label{eq:second_f_masses}
M_{\psi_{\xi}^2}&=2M_{SS} \langle \frac{g_{T\bar T}}{g_{S\bar S}}\frac{k_T\bar k_T}{\Sigma^2} \rangle -4M_{S T} \langle \frac{k_S\bar k_T}{\Sigma^2} \rangle
+2M_{T T}\langle\frac{g_{S\bar S}}{g_{T\bar T}}\frac{k_S\bar k_S}{\Sigma^2} \rangle \\
&=\mathcal{O}(m_{\text{SUSY}})\\
M_{\psi_\xi^3}
&=2\beta_{f}^2\frac{\left( M_{S S}M_{T T}-M_{ST}^2\right) \left(\langle \frac{g_{S\bar S}}{g_{T\bar T}}\frac{|k_S|^4}{\Sigma^4}\rangle+\langle \frac{g_{T\bar T}}{g_{S\bar S}}\frac{|k_T|^4}{\Sigma^4} \rangle \right)-2M_{ST}^2 \langle \frac{|k_S|^2|k_T|^2}{\Sigma^4}\rangle}{M_{SS} \langle \frac{g_{T\bar T}}{g_{S\bar S}}\frac{k_T\bar k_T}{\Sigma^2}\rangle -2M_{S T}\langle \frac{k_S\bar k_T}{\Sigma^2}\rangle
+M_{T T} \langle \frac{g_{S\bar S}}{g_{T\bar T}}\frac{k_S\bar k_S}{\Sigma^2}\rangle}\\
&=\mathcal{O}(q^2m_{\text{SUSY}})\ .
\end{split}
\end{equation}
where from \eqref{sky1} $q=\frac{v}{M_{U}}=\frac{\langle C^1 \rangle}{M_{U}}$.

\subsubsection{Final Low-Energy States}\label{sec:final_states2}

We conclude this subsection by displaying the Lagrangian for the low-energy spectrum after supersymmetry breaking. Ignoring all interaction terms, as well as the fields from the observable sector, this Lagrangian is given by
\begin{equation}
\begin{split}
\mathcal{L}&= -\partial^{\mu} \phi^{2} \partial_{\mu} \phi^{2}-\partial^{\mu} \eta^{2} \partial_{\mu} \eta^{2}-\partial^{\mu} \phi^{3} \partial_{\mu} \phi^{3}-\partial^{\mu} \eta^{3} \partial_{\mu} \eta^{3}- e^{\kappa_4^2K_T/3}\partial_\mu C^2 \partial^\mu \bar C^2\\
&- m_{\phi^2}^2{\phi^2}^2- m_{\eta^2}^2{\eta^2}^2- m_{\phi^3}^2{\phi^2}^2- m_{\eta^3}^2{\eta^3}^2-m_{1}^2C^2\bar C^2\\
& -i \Psi_{\xi}^{2} \slashed\partial \Psi_{\xi}^{2\dagger}-i \Psi_{\xi}^{3} \slashed\partial \Psi_{\xi}^{3\dagger}-ie^{\kappa_4^2K_T/3}\psi_2 \slashed{\partial} \psi_2^{\dag}-\tfrac{1}{2}M_{\Psi_\xi^2}\Psi_\xi^2\Psi_\xi^{2\dag}-\tfrac{1}{2}M_{\Psi_\xi^3}\Psi_\xi^3\Psi_\xi^{3\dag} \\
 \end{split}
\end{equation}  
where 
\begin{align}
m_{\phi^2}^2&= m_{s s}^2\langle\frac{g_{T\bar T}}{g_{S\bar S}}\frac{k_T\bar k_T}{\Sigma^2}\rangle-2m_{s t}^2\langle\frac{k_S\bar k_T}{\Sigma^2}\rangle
+m_{tt}^2\langle\frac{g_{S\bar S}}{g_{T\bar T}}\frac{k_S\bar k_S}{\Sigma^2} \rangle \  , \\
m_{\eta^2}^2&= m_{\sigma \sigma}^2\langle\frac{g_{T\bar T}}{g_{S\bar S}}\frac{k_T\bar k_T}{\Sigma^2}\rangle-2m_{\sigma\chi}^2\langle\frac{k_S\bar k_T}{\Sigma^2}\rangle
+m_{\chi \chi}^2\langle\frac{g_{S\bar S}}{g_{T\bar T}}\frac{k_S\bar k_S}{\Sigma^2} \rangle \  , \\
M_{\Psi_\xi^2}&= 2M_{S S}\langle\frac{g_{T\bar T}}{g_{S\bar S}}\frac{k_T\bar k_T}{\Sigma^2}\rangle-4M_{ST}\langle\frac{k_S\bar k_T}{\Sigma^2}\rangle
+2M_{T T}\langle\frac{g_{S\bar S}}{g_{T\bar T}}\frac{k_S\bar k_S}{\Sigma^2} \rangle\ \\
m_{\phi^3}^2&=m_1^2+\beta_{r}^2\frac{\left( m_{ss}^2m_{tt}^2-m_{st}^4\right) \left(\langle \frac{g_{S\bar S}}{g_{T\bar T}}\frac{|k_S|^4}{\Sigma^4}\rangle +\langle \frac{g_{T\bar T}}{g_{S\bar S}}\frac{|k_T|^4}{\Sigma^4} \rangle \right)-2m_{st}^4 \langle \frac{|k_S|^2|k_T|^2}{\Sigma^4}\rangle }{m_{ss}^2 \langle \frac{g_{T\bar T}}{g_{S\bar S}}\frac{k_T\bar k_T}{\Sigma^2}\rangle -2m_{st}^2 \langle \frac{k_S\bar k_T}{\Sigma^2} \rangle
+m_{tt}^2 \langle \frac{g_{S\bar S}}{g_{T\bar T}}\frac{k_S\bar k_S}{\Sigma^2} \rangle}\\
m_{\eta^3}^2
&=m_1^2+\beta_{i}^2\frac{\left( m_{\sigma \sigma}^2m_{\chi \chi}^2-m_{\sigma \chi}^4\right) \left(\langle \frac{g_{S\bar S}}{g_{T\bar T}}\frac{|k_S|^4}{\Sigma^4}\rangle +\langle \frac{g_{T\bar T}}{g_{S\bar S}}\frac{|k_T|^4}{\Sigma^4} \rangle \right)-2m_{\sigma \chi}^4 \langle \frac{|k_S|^2|k_T|^2}{\Sigma^4}\rangle }{m_{\sigma \sigma}^2 \langle \frac{g_{T\bar T}}{g_{S\bar S}}\frac{k_T\bar k_T}{\Sigma^2}\rangle -2m_{\sigma \chi}^2 \langle \frac{k_S\bar k_T}{\Sigma^2} \rangle
+m_{\chi \chi}^2 \langle \frac{g_{S\bar S}}{g_{T\bar T}}\frac{k_S\bar k_S}{\Sigma^2} \rangle}\\
M_{\psi_\xi^3}
&=2\beta_{f}^2\frac{\left( M_{S S}M_{T T}-M_{ST}^2\right) \left(\langle \frac{g_{S\bar S}}{g_{T\bar T}}\frac{|k_S|^4}{\Sigma^4}\rangle+\langle \frac{g_{T\bar T}}{g_{S\bar S}}\frac{|k_T|^4}{\Sigma^4} \rangle \right)-2M_{ST}^2 \langle \frac{|k_S|^2|k_T|^2}{\Sigma^4}\rangle}{M_{SS} \langle \frac{g_{T\bar T}}{g_{S\bar S}}\frac{k_T\bar k_T}{\Sigma^2}\rangle -2M_{S T}\langle \frac{k_S\bar k_T}{\Sigma^2}\rangle
+M_{T T} \langle \frac{g_{S\bar S}}{g_{T\bar T}}\frac{k_S\bar k_S}{\Sigma^2}\rangle}\ 
\end{align}
are the moduli masses computed above. The hidden matter fermions $\psi_2$ remain massless.

\section{Coupling of the the Moduli Fields to the Observable Sector}\label{sec:DM}

In Section \ref{sec:Effective theory}, we presented the spectrum for both the observable and hidden sectors of phenomenologically realistic heterotic $M$-theory vacua with an anomalous line bundle on the hidden sector. We specified the associated K\"ahler potentials, the anomalous $U(1)$ transformations of the moduli and hidden sector scalar fields and the generic form of the observable sector and hidden sector perturbative superpotentials. We also briefly discussed possible non-perturbative hidden sector superpotentials. Section 3 was devoted to determining the scalar and fermion mass eigenstates with canonical kinetic energy in the case of a pure $D$-term potential $V_{D}$--both for a vanishing and a non-vanishing Fayet-Iliopoulos term. With the exception of one heavy modulus, which decouples at low energy, all other scalar and fermion masses vanish. In Section 4, we introduced gaugino condensation and the associated non-perterbative superpotential $\hat W_{np}(S,T)$. We showed that the related $F$-terms, and the $V_{F}$ potential generated by them, led to new canonically normalized mass eigenstates. Now, however, in addition to the very massive modulus, most of the other scalars and fermions also had non-vanishing masses--although at a much smaller mass scale. These masses were explicitly computed, both for the $\text{FI}=0$ and $\text{FI} \neq 0$ cases. However, in all cases, interactions of these mass eigenstates were ignored.
In this section, we will derive the interactions between these scalars and fermions for both types of Fayet-Iliopoulos terms. We will, however, limit our discussion to the vertices which directly couple the observable sector fields to the moduli and hidden matter fields. This is motivated by our interest in exploring the possible role of moduli and hidden sector matter as cosmological dark matter candidates.

We continue to use the notation
\begin{equation}
z^A=(S,\>T,\>C^1,\>C^2,\>C_{(o)}^{\cal I})\ ,\quad A=1,\dots 4+\mathcal{N}
\end{equation}
for the scalar component fields in our theory, and 
\begin{equation}
\psi^A=(\psi_S,\>\psi_T,\>\psi_1,\>\psi_2,\>\psi_{(o)}^{\cal I})\ ,\quad A=1,\dots 4+\mathcal{N}
\end{equation}
for the corresponding fermions. We further assume that supersymmetry breaking effects determine a vacuum in which
\begin{equation}
 \left \langle \frac{\partial V}{\partial z^A}\right\rangle=0\ ,\quad  \left \langle \frac{\partial V}{\partial \bar z^A}\right\rangle=0 \ ,
\end{equation}
where $V=V_D+V_F$ is the scalar potential.

In the following, we identify the interactions which are sourced from two distinct terms in the 4D $N=1$ supergravity Lagrangian:

\begin{enumerate}

\item{Kinetic Terms}

We find the following kinetic terms for the matter scalars
\begin{equation}
\begin{split}
\label{eq:kinetic_terms_infl}
\mathcal{L}&\supset  -e^{\kappa_4^2K_T/3}\mathcal{G}_{\cal I\bar {\cal J}} \partial_\mu C_{(o)}^{\cal I}\partial^\mu \bar C_{(o)}^{\bar {\cal J}}
-e^{\kappa_4^2K_T/3} \partial_\mu C^1\partial^\mu \bar C^1-e^{\kappa_4^2K_T/3} \partial_\mu C^2\partial^\mu \bar C^2\ ,
\end{split}
\end{equation}
and for the matter fermions
\begin{equation}
\begin{split}
\mathcal{L}&\supset  -i e^{\kappa_4^2K_T/3}\mathcal{G}_{\cal I\bar {\cal J}}  \psi_{(o)}^{\bar{ \cal J}\dag}\slashed{\mathcal{D}} \psi_{(o)}^{I}
-ie^{\kappa_4^2K_T/3}  \psi_{1}^{\dag}\slashed{\mathcal{D}} \psi_{1}
-ie^{\kappa_4^2K_T/3}  \psi_{2}^{\dag}\slashed{\mathcal{D}} \psi_{2}\ .
\end{split}
\end{equation}
Expanding to linear order in the moduli field scalar perturbations we obtain
\begin{equation}
\begin{split}
\label{eq:int_sc_1}
\mathcal{L}&\supset  - e^{\kappa_4^2\langle K_T/3\rangle}\mathcal{G}_{\cal I\bar {\cal J}}  \partial_\mu C_{(o)}^{\cal I}\partial^\mu \bar C_{(o)}^{\bar{\cal J}}
- e^{\kappa_4^2\langle K_T/3\rangle} \partial_\mu C^1\partial^\mu \bar C^1-e^{\kappa_4^2\langle K_T/3\rangle} \partial_\mu C^2\partial^\mu \bar C^2\ \\
&-\kappa_4^2e^{\kappa_4^2\langle K_T/3\rangle }\left(
\frac{1}{3}\langle\frac{\partial K_T}{\partial T} \rangle\delta T+
\frac{1}{3}\langle \frac{\partial K_T}{\partial T}\rangle \delta \bar T
 \right)\mathcal{G}_{\cal I\bar {\cal J}}\partial_\mu C_{(o)}^{\cal I}\partial^\mu \bar C_{(o)}^{\bar {\cal J}}\\
&-\kappa_4^2e^{\kappa_4^2\langle K_T/3 \rangle}\left(
\frac{1}{3}\langle\frac{\partial K_T}{\partial T}\rangle \delta T+
\frac{1}{3}\langle\frac{\partial K_T}{\partial T}\rangle\delta \bar T
 \right)\left(\partial_\mu C^1\partial^\mu \bar C^1+\partial_\mu C^2\partial^\mu \bar C^2\right)\ \\
 &= - e^{\kappa_4^2\langle K_T/3\rangle}\mathcal{G}_{{\cal I}\bar {\cal J}}  \partial_\mu C_{(o)}^{\cal I}\partial^\mu \bar C_{(o)}^{\bar {\cal J}}
- e^{\kappa_4^2\langle K_T/3\rangle} \partial_\mu C^1\partial^\mu \bar C^1-e^{\kappa_4^2\langle K_T/3\rangle} \partial_\mu C^2\partial^\mu \bar C^2\\
 &+2\frac{e^{\kappa_4^2\langle K_T/3\rangle }}{\langle T+\bar T\rangle }\delta t\mathcal{G}_{{\cal I}\bar {\cal J}}\partial_\mu C_{(o)}^{\cal I}\partial^\mu \bar C_{(o)}^{\bar {\cal J}}
 +2\frac{e^{\kappa_4^2\langle K_T/3\rangle}}{\langle T+\bar T\rangle }\delta t \left(\partial_\mu C^1\partial^\mu \bar C^1+\partial_\mu C^2\partial^\mu \bar C^2\right)
\end{split}
\end{equation}
and
\begin{equation}
\begin{split}
\label{eq:int_sc_2}
\mathcal{L}&\supset  -i e^{\kappa_4^2\langle K_T/3\rangle/3}\mathcal{G}_{{\cal I}\bar {\cal J}}  \psi_{(o)}^{\bar {\cal J}\dag}\slashed{\mathcal{D}} \psi_{(o)}^{{\cal I}}- ie^{\kappa_4^2\langle K_T/3\rangle/3}\psi_{1}^{\dag}\slashed{\mathcal{D}} \psi_{1}-ie^{\kappa_4^2\langle K_T/3\rangle/3}\psi_{2}^{\dag}\slashed{\mathcal{D}} \psi_{2}
\ \\
&-i\kappa_4^2e^{\kappa_4^2\langle K_T/3\rangle/3}\left(
\frac{1}{3}\langle\frac{\partial K_T}{\partial T} \rangle\delta T+
\frac{1}{3}\langle \frac{\partial K_T}{\partial T}\rangle \delta \bar T
 \right)\mathcal{G}_{{\cal I}\bar {\cal J}} \psi_{(o)}^{\bar {\cal J}\dag}\slashed{\mathcal{D}} \psi_{(o)}^{{\cal I}}\\
 &-i\kappa_4^2e^{\kappa_4^2\langle K_T/3\rangle/3}\left(
\frac{1}{3}\langle\frac{\partial K_T}{\partial T} \rangle\delta T+
\frac{1}{3}\langle \frac{\partial K_T}{\partial T}\rangle \delta \bar T
 \right)\left(\psi_{1}^{\dag}\slashed{\mathcal{D}} \psi_{1}+\psi_{2}^{\dag}\slashed{\mathcal{D}} \psi_{2}\right)\\
 &= -i e^{\kappa_4^2\langle K_T/3\rangle/3}\mathcal{G}_{{\cal I}\bar {\cal J}}  \psi_{(o)}^{\bar {\cal J}\dag}\slashed{\mathcal{D}} \psi_{(o)}^{{\cal I}}- ie^{\kappa_4^2\langle K_T/3\rangle/3}\psi_{1}^{\dag}\slashed{\mathcal{D}} \psi_{1}-ie^{\kappa_4^2\langle K_T/3\rangle/3}\psi_{2}^{\dag}\slashed{\mathcal{D}} \psi_{2}
\ \\
&+2i\frac{e^{\kappa_4^2\langle K_T/3\rangle/3}}{\langle T+\bar T\rangle }\delta t\mathcal{G}_{{\cal I}\bar {\cal J}} \psi_{(o)}^{\bar {\cal J}\dag}\slashed{\mathcal{D}} \psi_{(o)}^{I}+2i\frac{e^{\kappa_4^2\langle K_T/3\rangle/3}}{\langle T+\bar T\rangle }\delta t\left(\psi_{1}^{\dag}\slashed{\mathcal{D}} \psi_{1}+\psi_{2}^{\dag}\slashed{\mathcal{D}} \psi_{2}\right)\ .
\end{split}
\end{equation}
These terms represent the couplings of the matter scalars and fermions to the moduli perturbation $\delta t$. The kinetic terms do not source couplings to the axionic components of the moduli. Note however, that in more general models, the K\"ahler metric of the matter fields depends on the dilaton field as well. Such classes of models allow for interactions between the matter fields and the real scalar component $\delta s$.

\item{Fermion Bilinear}

The bilinear fermion terms 
\begin{equation}
\begin{split}
\mathcal{L}&\supset-\frac{1}{2}e^{\kappa_4^2 K_{\text{mod}}/2}\mathcal{D}_A{D}_BW  \psi^{A}\psi^{B}+h.c,
\end{split}
\end{equation}
are a second source of interactions between the scalars and the fermions of the theory. In the expression above, $W$ is the full superpotential of the theory. Expanding these terms around the vacuum state defined above, we obtain couplings of the type
\begin{equation}
\label{shoe2}
\begin{split}
&\mathcal{L}\supset\\
&-{\frac{1}{2}e^{\kappa_4^2 \langle K_{\text{mod}}\rangle/2}\langle \mathcal{D}_A{D}_BW \rangle \psi^{A}\psi^{B}}-{
{\frac{1}{2}\left\langle \partial_C (e^{\kappa_4^2 K_{\text{mod}}/2}\mathcal{D}_A{D}_BW)\right\rangle\delta z^C \psi^{A}\psi^{B}}}\\
&{-\frac{1}{4}\kappa_4^2e^{\kappa_4^2 \langle K_{\text{mod}}\rangle/2} e^{\kappa_4^2\langle K_T\rangle /3}\langle \mathcal{D}_A{D}_BW\rangle  \psi^{A}\psi^{B}
\left(\mathcal{G}_{{\cal I}\bar {\cal J}}C_{(o)}^{{\cal I}}\bar C_{(o)}^{\bar {\cal J}}+C^1\bar C^1+C^2\bar C^2\right)}
+\dots\ 
\\&\hspace{11cm}+h.c\ ,
\end{split}
\end{equation}
where we have also used the expansion
 \begin{equation}
\begin{split}
e^{\kappa_4^2K/2}&\approx e^{\kappa^2_4\hat K_{\text{mod}}/2}\times \\ &\times \left(1+\frac{\kappa^2_4}{2}e^{\kappa_4^2K_T/3}\mathcal{G}_{{\cal I}\bar {\cal J}}C_{(o)}^{{\cal I}}\bar C_{(o)}^{\bar {\cal J}}+\frac{\kappa^2_4}{2}e^{\kappa_4^2K_T/3}C^1\bar C^1+\frac{\kappa^2_4}{2}e^{\kappa_4^2K_T/3}C^2\bar C^2\right)\ .
\end{split}
\end{equation}

As a result, we have recovered the fermion mass terms 
\begin{equation}
-{\frac{1}{2}e^{\kappa_4^2 \langle K_{\text{mod}} \rangle /2} \langle \mathcal{D}_A{D}_BW \rangle \psi^{A}\psi^{B}}+h.c.\ ,
\end{equation}
as well as the interaction terms
\begin{multline}
-{
{\frac{1}{2}\left\langle \partial_C (e^{\kappa_4^2 K_{\text{mod}}/2}\mathcal{D}_A{D}_BW)\right\rangle\delta z^C \psi^{A}\psi^{B}}}\\{-\frac{1}{4}\kappa_4^2e^{\kappa_4^2 \langle K_{\text{mod}}\rangle /2} e^{\kappa_4^2 \langle K_T \rangle/3}\langle \mathcal{D}_A{D}_BW\rangle  \psi^{A}\psi^{B}
\left(\mathcal{G}_{{\cal I}\bar {\cal J}}C_{(o)}^{{\cal I}}\bar C_{(o)}^{\bar {\cal J}}+C^1\bar C^1+C^2\bar C^2\right)}\ .
\end{multline}

The fermion moduli mass matrix has been computed in the previous section, in both the vanishing \text{FI} and non-vanishing \text{FI} scenarios. For the matter fields, the only non-zero contribution is
\begin{equation}
M_{AB}=\mu_{{\cal I}{\cal J}}\ ,
\end{equation}
where $\mu_{{\cal I}{\cal J}}$ occurs in the superpotential \eqref{shoe1} for the observable sector matter fields only, specifically for the Higgs doublet. Indeed, after SUSY is broken by F-terms generated by a non-perturbative superpotential, no soft-SUSY breaking mass terms are generated for any of the matter fermions.

\end{enumerate}

\begin{figure}[t]
   \centering
       \begin{subfigure}[b]{0.49\textwidth}
       \centering
   \includegraphics[width=0.46\textwidth]{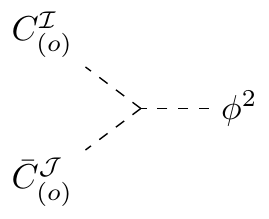}
\caption*{\centerline{(a)}}{{\small$2{p}_\mu^{\cal I}{p}^{\mu\cal J}\left\langle \frac{e^{\kappa_4^2K_T/3}}{T+\bar T}\right\rangle  \mathcal{G}_{{\cal I}\bar {\cal J}} [{\mathbf{U}^{-1}}]^2_2$}}
\label{fig:int_1a}
\end{subfigure}
       \begin{subfigure}[b]{0.49\textwidth}
\centering
 \includegraphics[width=0.46\textwidth]{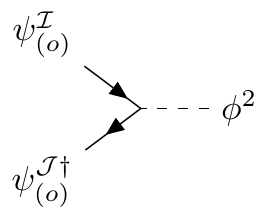}
\caption*{\centerline{(b)}}{$2i\slashed{p}\left\langle \frac{e^{\kappa_4^2K_T/3}}{T+\bar T}\right\rangle \mathcal{G}_{{\cal I}\bar {\cal J}} [{\mathbf{U}^{-1}}]^2_2$}
\label{fig:int_1b}
\end{subfigure}\\
       \begin{subfigure}[b]{0.49\textwidth}
\centering
 \includegraphics[width=0.46\textwidth]{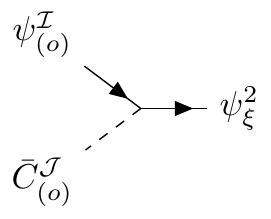}
\caption*{\centerline{(c)}}{${\small-\frac{\kappa_4^2}{2}{\mathcal{G}_{{\cal I}\bar {\cal J}}}e^{\kappa_4^2 \langle K_{\text{mod}}/2+K_T/3\rangle}}\times\newline {\small\left[\langle \hat F_{T} \rangle [{\mathbf{U}^{-1}}]^2_2 +\langle \hat F_{S} \rangle [{\mathbf{U}^{-1}}]^1_2 \right]}$}
\label{fig:int_1c}
\end{subfigure}
       \begin{subfigure}[b]{0.49\textwidth}
\centering
 \includegraphics[width=0.42\textwidth]{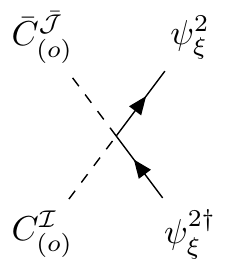}
\caption*{\centerline{(d)}}{$-\frac{1}{2}\kappa_4^2e^{\kappa_4^2\langle K_T\rangle/3}\mathcal{G}_{{\cal I}\bar {\cal J}}\times\newline \left[ {\mathbf{M_f}} \right]_{AB}     [{\mathbf{U}^{-1}}]^A_2 [{\mathbf{U}^{-1}}]^B_2$}
\label{fig:int_1d}
\end{subfigure}

\caption{  Interaction terms when $\text{FI}=0$. The vertices shown in Figures \ref{fig:int_1}(a) and \ref{fig:int_1}(b) originate in the kinetic terms from eq. \eqref{eq:int_sc_1} and \eqref{eq:int_sc_2} respectively, and couple both the scalars and the fermions from the observable sector to the moduli scalar field $\phi^2$. Their associated amplitudes are proportional to the momenta $p^{\cal I},p^{\cal J}$ and $p$, respectively, of the incoming particles. The vertices shown in Figures \ref{fig:int_1}(c) and \ref{fig:int_1}(d) originate the from expression eq. \eqref{shoe2}. Those vertices couple the observable sector fields to the moduli field fermions $\psi_\xi^2$, $\psi_\xi^{2\dag}$. The matrix $\mathbf{M_f}$ is defined in eq. \eqref{mfm}.}
\label{fig:int_1}
\end{figure}

In the following, we will write these interaction terms in the mass eigenstate basis described in Section \ref{mass_spect_sec}, for the two types of vacua we identified.
When the genus-one corrected FI term vanishes, we use the rotations
\begin{equation}
\label{basis1}
\left(\begin{matrix}
\delta s\\\delta t
\end{matrix}\right)
=i\mathbf{U}^{-1}
\left(\begin{matrix}
{\phi^1}\\ {\phi^2}
\end{matrix}\right)\ ,
\quad 
\left(\begin{matrix}
\psi_S\\\psi_T
\end{matrix}
\right)=\mathbf{U}^{-1}\left(\begin{matrix}
\psi_\xi^1\\ \psi_\xi^2
\end{matrix}
\right)\ 
\end{equation}
presented in \eqref{bird1} and \eqref{bird2}  to express equations \eqref{eq:int_sc_1}, \eqref{eq:int_sc_2} and \eqref{shoe2} in the mass eigenstate basis. The $2\times 2$ matrix $\mathbf{U}^{-1}$ was given in eq. \eqref{eq:U22A}. The hidden sector matter multiplets do not mix with the moduli in this case and, therefore, they do not couple directly to the observable sector\footnote{There are higher order terms in the $N=1$ supergravity potential term that do couple the observable and hidden sector directly, but are heavily supressed by powers of $\kappa_4^2$.}.
After the change of basis is performed, we obtain the four vertices shown in Figure \ref{fig:int_1} with their associated amplitudes written underneath.

\begin{figure}[t]
   \centering
       \begin{subfigure}[b]{0.49\textwidth}
       \centering
      \includegraphics[width=0.46\textwidth]{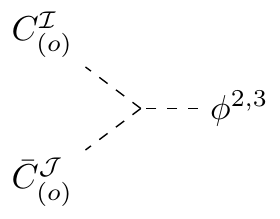}
\caption*{\centerline{(a)}}{{\small$2{p}_\mu^{\cal I}{p}^{\mu\cal J} \left \langle \frac{e^{\kappa_4^2K_T/3}}{T+\bar T}\right\rangle    \mathcal{G}_{{\cal I}\bar {\cal J}} \left[\mathbf{U_s^{(r)}}^{-1}\right]^2_{2,3}$}}
\label{fig:int_2a}
\end{subfigure}
       \begin{subfigure}[b]{0.49\textwidth}
\centering
    \includegraphics[width=0.46\textwidth]{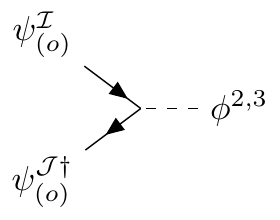}
\caption*{\centerline{(b)}}{$i\slashed{p} \left \langle \frac{e^{\kappa_4^2K_T/3}}{T+\bar T}\right\rangle   \mathcal{G}_{{\cal I}\bar {\cal J}} \left[\mathbf{U_s^{(r)}}^{-1}\right]^2_{2,3}$}
\label{fig:int_2b}
\end{subfigure}\\
       \begin{subfigure}[b]{0.49\textwidth}
\centering
   \includegraphics[width=0.46\textwidth]{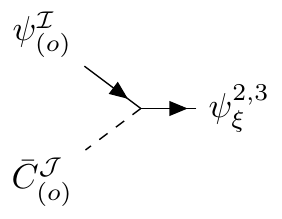}
\caption*{\centerline{(c)}}{${\small-\frac{\kappa_4^2}{2}{\mathcal{G}_{{\cal I}\bar {\cal J}}}e^{\kappa_4^2 \langle K_{\text{mod}}/2+K_T/3\rangle}}\times\newline {\small\left[\langle \hat F_{T} \rangle \left[\mathbf{U_f}^{-1}\right]^2_2 +\langle \hat F_{S} \rangle \left[\mathbf{U_f}^{-1}\right]^1_{2,3} \right]}$}
\label{fig:int_2c}
\end{subfigure}
       \begin{subfigure}[b]{0.49\textwidth}
\centering
   \includegraphics[width=0.42\textwidth]{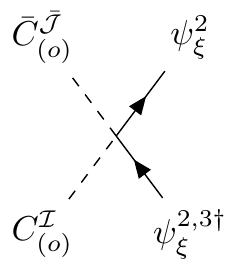}
\caption*{\centerline{(d)}}{$-\frac{1}{2}\kappa_4^2e^{\kappa_4^2\langle K_T\rangle/3}\mathcal{G}_{{\cal I}\bar {\cal J}}\times\newline \left[ {\mathbf{M_f}} \right]_{AB}     \left[\mathbf{U_f}^{-1}\right]^A_2\left[\mathbf{U_f}^{-1}\right]^B_{2,3}$}
\label{fig:int_2d}
\end{subfigure}

\caption{ Interaction terms when $\text{FI}\neq 0$. The vertices shown in Figures \ref{fig:int_2}(a) and \ref{fig:int_2}(b) originate in the kinetic terms from eq. \eqref{eq:int_sc_1} and \eqref{eq:int_sc_2} respectively, and couple both the scalars and the fermions from the observable sector to the scalar fields $\phi^2$ or $\phi^3$. Their associated amplitudes are proportional to the momenta $p^{\cal I},p^{\cal J}$ and $p$ of the incoming particles. The vertices shown in Figures \ref{fig:int_1}(c) and \ref{fig:int_1}(d) originate from the expression in eq. \eqref{shoe2}. Those vertices couple the observable sector fields to the moduli field fermions $\psi_\xi^2$, $\psi_\xi^{2\dag}$, as well as to $\psi_\xi^3$, $\psi_\xi^{3\dag}$. The matrix $\mathbf{M_f}$ is defined in eq. \eqref{mfm}. }
\label{fig:int_2}
\end{figure}

On the other-hand, when the genus-one corrected FI term is non-zero, we use the $3\times 3$ rotations
\begin{equation}
\label{basis2}
\left( \begin{matrix}{\delta s}\\ {\delta t} \\ {\text{Re}(\delta C^1)} \end{matrix}\right)=i\mathbf{U_s^{(r)}}^{-1}
\left( \begin{matrix}{\phi^1}\\ {\phi^2} \\ {\phi^3} \end{matrix}\right)\ ,\quad  \left( \begin{matrix} \psi_S\\   \psi_T\\  \psi_1 \end{matrix}\right)={\bm{\mathbf{U}}_f^{-1}}\left( \begin{matrix} \psi_\xi^1\\   \psi_\xi^2\\  \psi_\xi^3 \end{matrix}\right)\ .
\end{equation}
given in \eqref{fdr3} and \eqref{bird3}  in order to express the interaction terms in the mass eigenstate basis.
The expressions for the matrices  $\mathbf{U_s^{(r)}}^{-1}$ and $\mathbf{U}_f^{-1}$ were derived in Section \ref{mass_spect_sec}. Matrix  $\mathbf{U_s^{(r)}}^{-1}$ was given in \eqref{eq:u_m1_matrix} while  $\mathbf{U}_f^{-1}$ is identical in form but with the parameter $\gamma_{r}$ replaced by $\gamma_{f}$.
After the change of basis is performed, we obtain the four vertices shown in Figure \ref{fig:int_2}, which have the associated amplitudes written underneath. The main difference from the vanishing FI case is that the low-energy spectrum contains an extra scalar-fermion pair of fields ($\phi^3,\psi_\xi^3$). These fields are linear combinations of the moduli fields and the matter fields from the hidden sector. They couple to the observable sector fields as well .

The couplings of the fields $\phi^2$, and $\psi_\xi^2$ to the observable sector are proportional to the matrix elements $[\mathbf{U_s^{(r)}}^{-1}]^2_{2}$ and $[\mathbf{U_f}^{-1}]^2_{2}$. It can be shown that the values of these elements are of order $1$. The couplings of the fields  ($\phi^3,\psi_\xi^3$) to the observable sector, however, are proportional to the matrix elements $[\mathbf{U_s^{(r)}}^{-1}]^2_{3}$ and $[\mathbf{U_f}^{-1}]^2_{3}$. These matrix elements are, in turn, proportional to the the size of the $C^1$ field VEV that is needed to cancel the non-zero FI. It can be shown, therefore, that the values of these matrix elements are of the order
\begin{equation}
\left[\mathbf{U_s^{(r)}}^{-1}\right]^2_{3}\ ,\left[\mathbf{U_f}^{-1}\right]^2_{3}\sim \frac{v}{M_U}=q\ .
\end{equation}
In principle, $q$ can take values as large as $1$. However, within the scenario analyzed in the previous section--in which the matter field VEV that is needed to cancel the non-zero FI term is arbitrarily small--these couplings are relatively suppressed.

\subsection{Some Possible Dark Matter Candidates}

In this section, we propose a scenario in which the inflaton is a linear combination of the observable sector scalar fields
\begin{equation}
\Phi=c_{\cal I}C_{(o)}^{\cal I}\ ,
\end{equation}
such that
\begin{equation}
C_{(o)}^{\cal I}\ =c^{\cal I}\Phi \ .
\end{equation}
Such models have been analyzed in a number of papers~\cite{Deen:2016zfr,Cai:2018ljy,Ibanez:2014swa}.

Let us consider the case in which the FI term is non zero. The conclusions of this section can be easily extended to the vanishing FI case, by taking the limit $v\rightarrow 0$. In our model, the inflaton can produce the fermions $\psi_\xi^{2,3}$ via decay processes of the type

\begin{equation*}
   \includegraphics[width=0.26\textwidth]{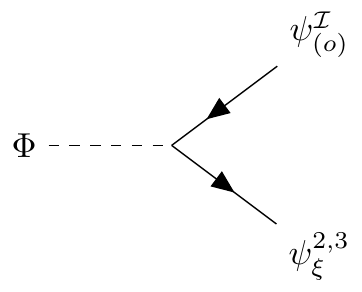}
\end{equation*}

These processes have the associated amplitudes
\begin{equation}
i\mathcal{A}_{2,3}=\frac{\kappa_4^2}{2}{c^J\mathcal{G}_{{\cal I}\bar {\cal J}}}e^{\kappa_4^2 \langle K_{\text{mod}} \rangle/2}e^{\kappa_4^2 \langle K_T \rangle/3}\left[\langle \hat F_{T} \rangle\left[\mathbf{U_f}^{-1}\right]^2_{2,3} +\langle \hat F_{S} \rangle\left[\mathbf{U_f}^{-1}\right]^1_{2,3} \right]\ ,
\end{equation}
where
\begin{equation}
\begin{split}
\left[\mathbf{U_f}^{-1}\right]^1_{2}&=\left\langle\frac{1}{\Sigma}\sqrt{\frac{g_{T\bar T}}{g_{S\bar S}}}k_T\right\rangle\ ,\\
\left[\mathbf{U_f}^{-1}\right]^2_{2}&=- \left\langle\frac{1}{\Sigma}\sqrt{\frac{g_{S\bar S}}{g_{T\bar T}}}k_S\right\rangle\ ,
\end{split}
\end{equation}
and
\begin{equation}
\begin{split}
\left[\mathbf{U_f}^{-1}\right]^1_{3}&=\left\langle\frac{\beta_{f}}{\Sigma} k_S- \frac{\gamma_f}{\Sigma}\sqrt{\frac{g_{T\bar T}}{g_{S\bar S}}}k_T\right\rangle\ ,\\
\left[\mathbf{U_f}^{-1}\right]^2_{3}&=-\left\langle\frac{\beta_{f}}{\Sigma} k_T- \frac{\gamma_f}{\Sigma}\sqrt{\frac{g_{S\bar S}}{g_{T\bar T}}}k_S\right\rangle\ .
\end{split}
\end{equation}
The parameters $\beta_{f}$ and $\gamma_f$ are given by
\begin{equation}
\begin{split}
\beta_{f} (=\beta_{r}=\beta_{i}) &=-\frac{1}{2t^2}\langle C^1\rangle \frac{\bar k_T}{\Sigma}+\frac{1}{2t}\frac{\bar k_1}{\Sigma}\ ,\\
\gamma_f&=\frac{M_{12}}{M_{22}}\beta_{f}\ .
\end{split}
\end{equation}
These expressions are derived in the limit in which $v=\langle C^1\rangle$ is small; that is, such that $q=v/M_U\ll 1$. In this limit, we expect the couplings to be of order
\begin{align}
&\left[\mathbf{U_f}^{-1}\right]^1_{2},\>\left[\mathbf{U_f}^{-1}\right]^2_{2}\sim \mathcal{O}(1)\ ,\\
&\left[\mathbf{U_f}^{-1}\right]^1_{3},\>\left[\mathbf{U_f}^{-1}\right]^2_{3}\sim \mathcal{O}(q)\ .
\end{align}

The low-energy spectrum after supersymmetry breaking, in both the moduli and the hidden sectors, was summarized in Subsection \ref{sec:final_states2}. Including the observable sector fields, we expect the following mass hierarchy during reheating:

\begin{equation}
\begin{split}
\label{mass_hier}
&m_{C_{(o)}^{\cal I}}\sim \mathcal{O}(m_{\text{SUSY}})\ ,M_{\Psi_{(o)}^{\cal I}}=y_{\Phi \psi^{\cal I}\psi^{\cal I}}\sqrt{\langle \Phi^2 \rangle} , \qquad {\cal{I}}=1, \dots, \mathcal{N}  ,\\
&m_{C^2}\sim \mathcal{O}(m_{\text{SUSY}}),\>m_{\Psi_2}=0\ ,\\
&m_{\phi^2},\>m_{\phi^3}\sim \mathcal{O}(m_{\text{SUSY}})\ ,\\
&m_{\eta^2},\>m_{\eta^3}\sim \mathcal{O}(m_{\text{SUSY}})\ ,\\
&M_{\Psi_\xi^2}\sim \mathcal{O}(m_{\text{SUSY}})\ , M_{\Psi_\xi^3}\sim \mathcal{O}(q^2m_{\text{SUSY}})\ .
\end{split}
\end{equation}
In the above equations, the fermion $\Psi^{\cal I}_{C(o)}$ from the observable sector has the mass
\begin{equation}
M_{\psi_{C(o)}^{\cal I}}=y_{\Phi \psi^{\cal I}\psi^{\cal I}}\sqrt{\langle \Phi^2 \rangle}\ ,
\end{equation}
where $\sqrt{\langle \Phi^2 \rangle}$ is the the root mean square value of the inflaton during reheating, and $y_{\Phi \psi^I\psi^I}$ is the Yukawa-like coupling in the inflaton, two Weyl fermion $\psi^{\cal I}$ interaction.

Analyzing the expressions of the amplitudes $\mathcal{A}_{2,3}$, we deduce that
\begin{align}
i\mathcal{A}_2&\sim \kappa_4^3M_U^3\approx 10^{-5}\ ,\\
i\mathcal{A}_3&\sim q\kappa_4^3M_U^3\approx 10^{-5}q\ .
\end{align}
The decay rates associated with the processes $\Phi\rightarrow \Psi^{\cal I}\Psi_\xi^{2,3}$ are
\begin{equation}
\begin{split}
\Gamma_{\Phi\rightarrow \Psi_\xi^2 \Psi^{\cal I}_{(o)}}&=\frac{|\mathcal{A}_2|^2m_\Phi}{8\pi}
\left[1-\left( \frac{M_{\psi_{(o)}^{\cal I}}+M_{\psi_\xi^2}}{m_\Phi} \right)^2 \right]^{3/2}\ ,\\
&\approx 10^{3} ~{\rm GeV }\left[ 1-\left(\frac{y_{\Phi \psi^{\cal I}\psi^{\cal I}}\sqrt{\langle \Phi^2 \rangle}}{m_\Phi}+\frac{M_{\psi_\xi^2}}{m_\Phi}\right)^2\right]^{3/2}\ .
\end{split}
\end{equation}
and
\begin{equation}
\begin{split}
\Gamma_{\Phi\rightarrow \Psi_\xi^3 \Psi^{\cal I}_{(o)}}&=\frac{|\mathcal{A}_3|^2m_\Phi}{8\pi}
\left[1-\left( \frac{M_{\psi_{C(o)}^I}+M_{\psi_\xi^3}}{m_\Phi} \right)^2 \right]^{3/2}\ ,\\
&\approx 10^{3}~\text{GeV}\>q^2\left[ 1-\left(\frac{y_{\Phi \psi^{\cal I}\psi^{\cal I}}\sqrt{\langle \Phi^2 \rangle}}{m_\Phi}+q^2\right)^2\right]^{3/2}\ .
\end{split}
\end{equation}

As $\langle \Phi^2\rangle \rightarrow 0$,
\begin{equation}
\begin{split}
\Gamma_{\Phi\rightarrow \psi_\xi^3 \psi^{\cal I}_{(o)}}\rightarrow 
 10^{3}~\text{GeV}\>q^2\left[ 1-q^2\right]^{3/2}\ .
\end{split}
\end{equation}
The mass of the inflaton is a linear combination of the masses acquired by its component fields after supersymmetry breaking and, therefore, $m_{\Phi}\sim \mathcal{O}(m_{\text{SUSY}})$. Comparing the processes $\Phi\rightarrow \psi_\xi^2 \psi^{\cal I}_{C(o)}$ and $\Phi\rightarrow \psi_\xi^3 \psi^{\cal I}_{C(o)}$, it is not obvious which one is expected to be dominant during reheating. Although the decay rate of $\Phi\rightarrow \psi_\xi^3 \psi^{\cal I}_{C(o)}$ is supressed by $q^2$, the total mass of the decay products is smaller as well, allowing this process to start earlier during reheating. It would be interesting to find out if any of these processes can be responsible, at least partially, for the production of dark matter. The $\psi_\xi^3$ fermions are particularly interesting, because they are relatively light and also stable. However, a proper analysis within the context of realistic inflation and reheating scenarios is beyond the scope of this paper.

 We point out that other mechanisms of producing dark matter have been proposed in literature, in similar contexts. For example, ~\cite{Chowdhury:2018tzw,Dutra:2019nhh}, propose the hidden sector matter fields as possible dark matter candidates. In such scenarios, the moduli fields are produced via processes of the type shown in Figures \ref{fig:int_1}(a) and \ref{fig:int_2}(a), and act as a ``portal'' between the observable and hidden sectors.

\section{SUSY Breaking and Moduli Stabilization $-$ Simple Examples}\label{Sec:Mod_Stab}

In Section \ref{mass_spect_sec}, we introduced non-perturbative effects, such as gaugino condensation, that generate a new potential term $V_{F}$ in the effective Lagrangian. It was assumed that the potential energy $V_{D}+V_{F}$ admits a stable minimum which spontaneously breaks $N=1$ supersymmetry.
The mass spectrum of the moduli and hidden sector matter fields were then explicitly computed in this vacuum. The treatment is completely general and, in principle, can be applied to any specific supersymmetry breaking context.
This section is dedicated to discussing a few simple examples of scalar potentials and stable vacua in which $N=1$ supersymmetry is broken. In each of these examples, we apply the results of Subsections \ref{sec:final_states1} and \ref{sec:final_states2} and compute the resultant low energy mass spectrum.

Turning on non-perturbative effects, such as gaugino condensation, generate a non-vanishing contribution to the superpotential, denoted by $\hat W_{np}(S,T)$. Then, as discussed in Section \ref{mass_spect_sec}, the potential energy now becomes 
\begin{equation}
V=V_{D}+V_{F} \ ,
\label{fly3}
\end{equation}
with the $F$-term potential $V_{F}$ given by
\begin{equation}
V_F=e^{\kappa^2_4K}\left[ g^{A\bar B}(D_A\hat W_{np})(D_{\bar B}\hat W_{np}^*)-3\kappa_4^2{|\hat W_{np}|^2}  \right] 
\label{fly4}
\end{equation}
where
\begin{equation}
D_A\hat W_{np}=\partial_{A}\hat W_{np}+\kappa_4^2K_A\hat W_{np}\ .
\label{fly4A}
\end{equation}
The indices $A,B$ each run over $S,T$. 
We propose a mechanism in which the $\langle s \rangle$ and $\langle t \rangle$ VEVs of the moduli, as well as one of the axion VEVs, which were not completely determined by the D-flatness condition studied in the Section \ref{sec:D-term_s}, can in principle be fixed at the minimum of the potential $V=V_D+V_F$. We analyze scenarios in which $N=1$ supersymmetry is spontaneously broken in the vacuum.

\subsection{Vanishing FI Term}

In this subsection, we offer a useful visualization of the moduli stabilization mechanism, in vacua in which $N=1$ supersymmetry is broken.  We use a simplified setting. That is, we do not consider gauge threshold corrections which appear at order $\kappa_4^{4/3}$ in the gauge kinetic functions. Furthermore, we assume that the complex and bundle moduli have been stabilized, as in \cite{Anderson:2010mh, Anderson:2011cza}. The subject of moduli stabilization in the heterotic theory is a vast one and to the knowledge of the authors, it does not have a clear solution at present. A more realistic analysis of the moduli stabilization mechanism in heterotic vacua in which supersymmetry is broken by non-perturbative effects can be found in \cite{Cicoli:2013rwa}. 

In the vanishing FI term case, the D-term potential $V_D$ is then given in eq. \eqref{late1}, where the matter fields $C^1$ and $C^2$ decouple. The shape of this potential was shown in Figure \ref{fig:D_flat11}. The potential vanishes along a ``D-flat'' direction defined by eq. \eqref{cof1}. Assuming $\frac{\pi\epsilon_S\beta}{3}=1$, this flat direction is along the $s=t$ line. Let us now add the non-perturbative potential $V_{F}$ given in \eqref{fly4}.

\begin{figure}[t]
   \centering
       \begin{subfigure}[b]{0.4\textwidth}
\includegraphics[width=1\textwidth]{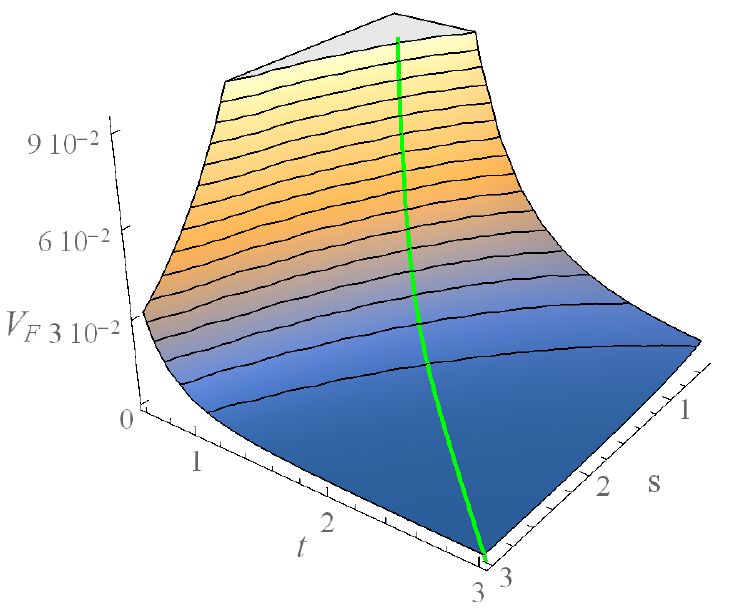}
\caption{}
\label{fig:runaway}
\end{subfigure}
       \begin{subfigure}[b]{0.38\textwidth}
\includegraphics[width=1\textwidth]{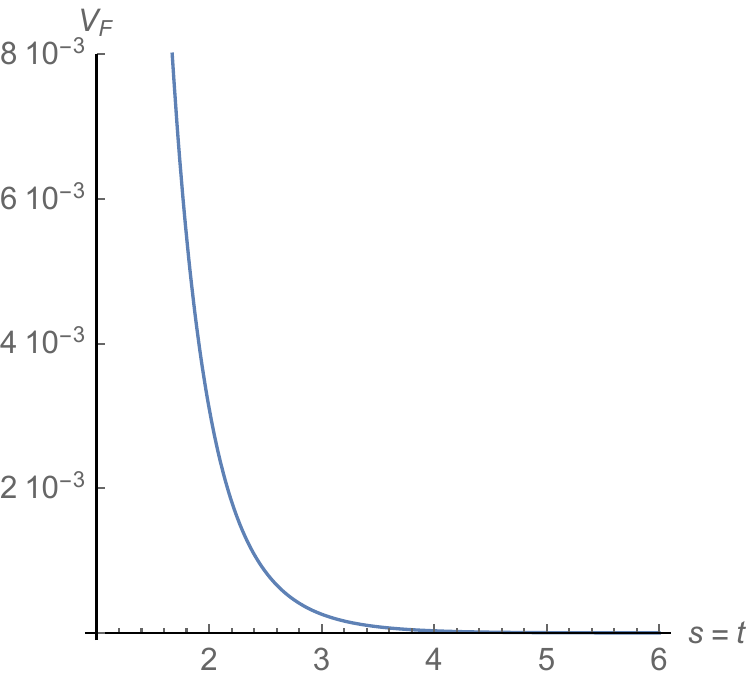}
\caption{}
\label{fig:nonlift}
\end{subfigure}
\\
    \begin{subfigure}[b]{0.4\textwidth}
\includegraphics[width=1\textwidth]{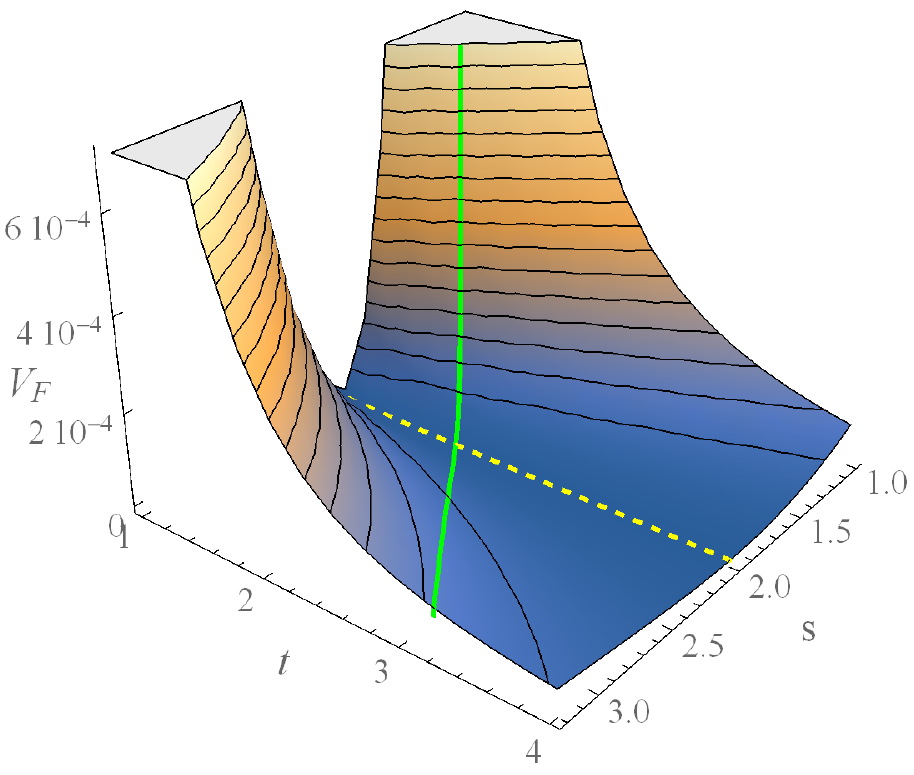}
\caption{}
\label{fig:non-runaway}
\end{subfigure}
  \begin{subfigure}[b]{0.38\textwidth}
\includegraphics[width=1\textwidth]{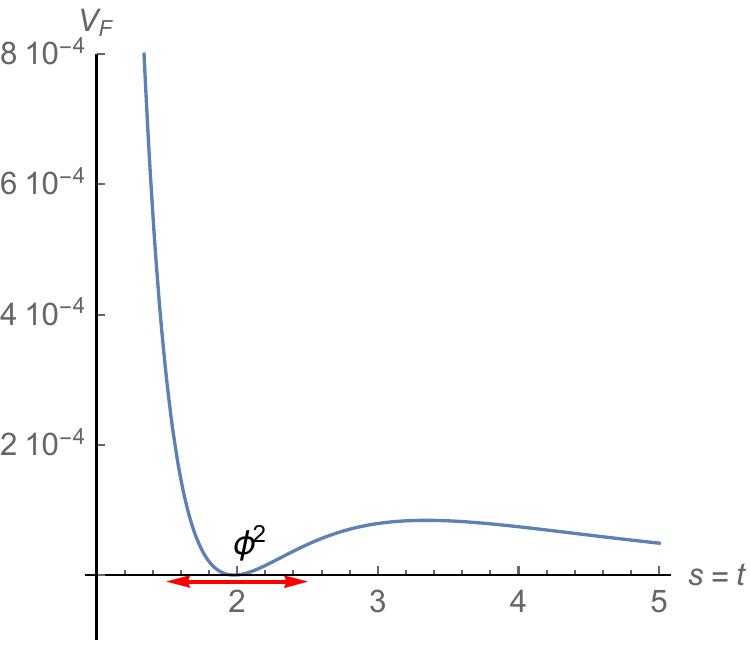}
\caption{}
\label{fig:LiftPotential}
\end{subfigure}
\caption{(a) Runaway F-term potential $V_F$ given in eq. \eqref{eq:runaway_pot} and generated by the superpotential in eq. \eqref{eq:sup1}, for $b=0.75$. The potential $V_F$ depends on the real component moduli fields $s$ and $t$ only. The green line indicates the $V_F$ profile along the D-flat direction $s=t$ in the vanishing FI term case with $\frac{\pi\epsilon_S\beta}{3}=1$. (b) $V_F$ profile along the D-flat direction $s=t$. The $V_F$ potential has no minimum along this direction and therefore, it cannot stabilize the remaining modulus of the theory. (c) Non-runaway F-term potential $V_F^\prime$ given in eq. \eqref{eq:V_F_prime} and generated by the superpotential in eq. \eqref{eq:sup2}, for $b=0.75$ and $c=1$. The potential $V_F^\prime$ depends on the real component moduli fields $s$ and $t$, and also on the value of the dilaton axion $\sigma$. For fixed $s$ and $t$, this potential is always minimized for $\langle \sigma \rangle=\frac{\pi n}{b}$, with $n\in \mathbb{Z}$. Here, as in the rest of the paper, we assume $\langle \sigma \rangle=\frac{\pi}{b}$. The green line indicates the $V_F^\prime$ profile along the D-flat direction $s=t$, in the vanishing FI term case. (d) $V_F^\prime$ profile along the D-flat direction $s=t$. The $V_F^\prime$ potential, calculated for $\sigma$ fixed to its VEV $\langle \sigma \rangle=\frac{\pi}{b}$, has a minimum along the D-flat direction and, therefore, it can stabilize the vacuum completely. The potential minimum is found at $\langle s\rangle=\langle t\rangle\approx 2$, at the intersection between the D-flat line shown with green and the F-flat line, displayed with yellow. 
The field $\phi^{2}$ is the fluctuation around this minimum in the $s=t$ direction. 
All potentials are expressed in multiples of $\kappa_4^2M_U^6$.}
\label{fig:D_flat1}
\end{figure}

One  possible non-perturbative effect is gaugino condensation in a
non-Abelian gauge group in the commutant of the hidden sector anomalous $U(1)$. This configuration has been analyzed in a number of papers~\cite{Barreiro:1998nd,Binetruy:1996uv,Ashmore:2020wwv}. When hidden sector gauginos condense, they produce a moduli dependent non-vanishing superpotental. Including corrections up to order $\kappa_{4}^{2/3}$ only, this superpotential has the expression
\begin{equation}
\label{eq:sup1}
\hat W_{np}=M_U^3e^{-bS}\ ,
\end{equation}
where $M_{U}$ is the unification scale in the hidden sector and $b$ is a positive coefficient associated with the beta-function of the non-Abelian hidden sector gauge coupling. This superpotential leads to the F-term scalar potential
\begin{equation}
\begin{split}
\label{eq:runaway_pot}
V_F&=\kappa_4^2M_U^6\frac{1}{(S+\bar S)(T+\bar T)^3}\left(b(S+\bar S) +1 \right)^2e^{-b\left(S+\bar S\right)} \ ,\\
&=\kappa_4^2M_U^6\frac{1}{16st^3}\left(2bs +1 \right)^2e^{-2bs}\ ,
\end{split}
\end{equation}
which is displayed in Figure \ref{fig:runaway}. However, this is a runaway potential in both $s=\rm{Re}S$ and $t=\rm {Re}T$. Therefore, as shown in Figure \ref{fig:nonlift}, it cannot stabilize these moduli
along the D-flat direction, that is, the dashed green line, in Figure 1.

To do this, we propose another superpotential which can fix these moduli by turning on flux of the non-zero mode of the antisymmetric tensor field in the bulk space. This effect generates a constant superpotential $\hat W_{\text{flux}}$ in the 4D effective theory, proportional to the averaged three-form flux~\cite{Cicoli:2013rwa,Ibanez:2012zz}. The flux quantization condition~\cite{Lukas:1997rb,Gray:2007qy} constraints this constant contribution to be of the form 
\begin{equation}
\label{eq:fluxsup}
\hat W_{\text{flux}} = cM_U^3\ .
\end{equation}
 The value of the dimensionless constant $c$ is quantized such that
 \begin{equation}
  \quad c =\alpha n\ ,\quad n \in \mathbb{Z}\ ,\quad \alpha \sim \mathcal{O}(1)\ .
  \end{equation}
Adding this effect to the gaugino condensate superpotential, we get
\begin{equation}
\label{eq:sup2}
\hat W^\prime_{np}=M_U^3(c+e^{-bS})\ ,
\end{equation}
This new superpotential leads to the F-term potential
\begin{equation}
\begin{split}
\label{eq:V_F_prime}
V_F^\prime&=\kappa_4^2M_U^6\frac{1}{(S+\bar S)(T+\bar T)^3}\Big(\left|c+\left(b(S+\bar S) +1 \right)e^{-bS}\right|^2\Big) \ ,\\
&=\kappa_4^2M_U^6\frac{1}{16st^3}\bigg(\left|c+\left(2bs +1 \right)e^{-b\left(s+i\sigma\right)}\right|^2\bigg) \ ,\\
&=\kappa_4^2M_U^6\frac{1}{16st^3}\left( c^2+(2bs+1)^2e^{-2bs}+2c(2bs+1)e^{-bs}\cos b\sigma\right) \ .
\end{split}
\end{equation}
This potential is a function of the dilaton axion $\sigma$, as well as $s=\rm{Re} S$ and $t=\rm{Re}T$. Note, however, that $V_{F}^{\prime}$ does not contain the K\"ahler axion $\chi$, which we henceforth ignore. The dependence of this potential on the dilaton axion $\sigma$ is periodic, with period $\frac{2\pi}{b}$. For fixed $s=\rm{Re}S$ and $t=\rm{Re}T$, the potential is minimized for 
\begin{equation}
\langle \sigma \rangle =\frac{\pi (2m+1)}{b}\ ,\quad m \in \mathbb{Z}\ .
\end{equation}
We will assume the axion has a fixed VEV in one of these minima, which, for simplicity, we choose to be
\begin{equation}
\langle \sigma\rangle =\frac{\pi}{b}\ .
\end{equation}
When $ \langle \sigma \rangle$ is fixed to $\pi/b$, the potential $V_F^\prime$ given in \eqref{eq:V_F_prime} takes the simple form
\begin{equation}
\begin{split}
\label{eq:V_F_prime2}
V_F^\prime&=\kappa_4^2M_U^6\frac{1}{16st^3}\left( c-(2bs+1)e^{-bs} \right)^2 \ ,
\end{split}
\end{equation}
which is positive definite. The potential vanishes at the unique minumum defined by
\begin{equation}
\label{pap1}
\langle s \rangle=\frac{1}{b}\ln \left(\frac{2b\langle s \rangle+1}{c}\right)\ ,
\end{equation}
This potential stabilizes the dilaton $s$, but leaves the K\"ahler moduli $t$ undetermined along a flat direction. 
It is useful to plot the potential $V_F^\prime$ in \eqref{eq:V_F_prime2} as a function of $s$ and $t$. This is done in Figure \ref{fig:non-runaway} where, for specificity, we choose $c=1$ and take $b=0.75$. Note that for these choices of $c$ and $b$, the dilaton takes the fixed value of $\langle s \rangle \approx 2$, while the value of $t$ remains undetermined. This is shown as the dashed yellow line in Figure \ref{fig:non-runaway}.

Now consider the total potential $V_D+V_F^\prime$. This total potential will have a unique minimum at the intersection of the $V_F^{\prime}=0$  and the $V_{D}=0$ lines. For the values of $c$ and $b$ used in Figure 5,  the potential has a unique minimum in $s,t$ space at $\langle s\rangle=\langle t\rangle\approx2$.
We learn that the D-term stabilization mechanism presented in the previous section, combined with the non-perturbative effects introduced in this section, are sufficient to stabilize the moduli in the theory in a vacuum with vanishing cosmological constant.  The profile of this potential along the D-flat direction is shown in Figure \ref{fig:LiftPotential}. The fluctuation field $\phi^2$ is now to be evaluated at this minimum of the total potential and acquires a non-zero mass.
Furthermore, it can be checked that in this vacuum at $\langle s\rangle=\langle t\rangle\approx2$, the F-term associated with $S$ vanishes, while the F-term associated with the K\"ahler modulus $T$ is non-zero. That is,
\begin{align}
\langle F_S\rangle &=\langle D_S\hat W_{np}^\prime\rangle =0\ ,\\
\langle F_T\rangle&=\langle D_T\hat W_{np}^\prime\rangle\neq0\ .
\end{align}
Therefore, $N=1$ supersymmetry is spontaneously broken. 

The mechanism of supersymmetry breaking via gaugino condensation in the hidden sector is well known. The non-vanishing moduli $F$-terms generate the soft SUSY-breaking Lagrangian in the observable sector via gravitational mediation.
Another consequence of supersymmetry breaking is that the scalar components of the moduli, as well as the corresponding fermions, become massive as well. Since they no longer belong to a supermultiplets, their acquired masses are expected to differ. Note that in scenarios in which moduli are stabilized by turning on a constant flux contribution of the type $\hat W_{\text{flux}}=cM_U^3$, where $c\sim\mathcal{O}(1)$, the scale of the soft-SUSY breaking terms, as well as the masses acquired by the moduli fields, are of the order $\kappa_4^2M_U^3\approx 10^{13}$GeV. Although turning on the flux contributions was a useful tool in stabilizing the vacuum, it has the caveat that it can work in a high-scale SUSY-breaking scenario only.

We can now apply the results derived in Section \eqref{sec:final_states1} to find the low-energy matter spectrum produced after supersymmetry is broken. In the vanishing FI case, after decoupling the heavy $\phi^1$ and $\eta^1$ states, the scalar matter spectrum is composed of two scalar moduli fields, $\phi^2$ and $\eta^2$, and the hidden sector matter fields $C^1$ and $C^2$. The fermion matter spectrum is composed of a Majorana fermion $\Psi_\xi^2$ and the matter fermions $\psi_1$ and $\psi_2$. We now want to compute their masses, using expressions \eqref{eq:mphi2}-\eqref{pro1}, for the vacuum at the minimum of the potential $V_F^{\prime}$ given in eq. \eqref{eq:V_F_prime}, in the case in which the FI term vanishes. We found that when $b=0.75$ and $c=1$, this potential is minimized at $\langle s\rangle=\langle t\rangle\approx2$ and $\langle \sigma\rangle=\pi/b$. The masses of the $\phi^2$ and $\eta^2$ states are given in eq. \eqref{eq:mphi2} and eq. \eqref{eq:meta2} respectively. For the form of the $V_F^{\prime}$ potential that we use, $m^2_{st}=m^2_{ts}=m^2_{tt}=0$, and therefore, we have
\begin{equation}
m^2_{\phi^2}=m_{ss}^2\langle \frac{g_{T\bar T}}{g_{S\bar S}}\frac{k_T\bar k_{\bar T}}{\Sigma^2}\rangle\ 
\end{equation}
and
\begin{equation}
m^2_{\eta^2}=m_{\sigma\sigma}^2\langle\frac{g_{T\bar T}}{g_{S\bar S}}\frac{k_T\bar k_{\bar T}}{\Sigma^2}\rangle\ .
\end{equation}
Using the fact that 
\begin{equation}
\frac{k_S}{\Sigma}=\frac{k_T}{\Sigma}=\frac{\kappa_4}{\sqrt{\tfrac{1}{(S+\bar S)^2}+\tfrac{3}{(T+\bar T)^2}}}\ ,
\end{equation}
we compute the values of these mass terms at $\langle s\rangle=\langle t\rangle\approx2$ and $\langle \sigma\rangle=\pi/b$ and find
\begin{equation}
\begin{split}
m_{\phi^2}=0.5\times 10^{12}\text{GeV}\ ,\quad m_{\eta^2}=2.2\times 10^{12}\text{GeV}\ .
\end{split}
\end{equation}
In our no-scale model, the hidden matter scalars $C^1$ and $C^2$ remain massless,
\begin{equation}
m_{C^1}=m_{C^2}=0\ .
\end{equation}

To compute the mass of the Majorana fermion $\Psi_\xi^2$, we apply eq. \eqref{eq:mPsi2}. However, recall that the vacuum defined by potential $V_F^\prime$ is characterized by $\langle F_S\rangle =0\ ,\langle F_T\rangle\neq 0$.
Therefore, the values of $M_{SS},M_{ST}$ and $M_{TS}$ vanish in these equations. The state ${\psi_\xi^2}$ aquires the mass 
\begin{equation}
M_{\Psi_\xi^2}=2M_{TT}\langle \frac{g_{S\bar S}}{g_{T\bar T}}\frac{k_S\bar k_{S}}{\Sigma^2}\rangle\ .
\end{equation}
 At $\langle s\rangle=\langle t\rangle\approx2$ and $\langle \sigma\rangle=\pi/b$, we find that
\begin{equation}
M_{\Psi_\xi^2}=1.6\times 10^{12}\text{GeV},
\end{equation}
while the matter fermions $\psi_1$ and $\psi_2$ remain massless,
\begin{equation}
M_{\psi_1}=M_{\psi_2}=0\ .
\end{equation}
These masses are indeed of the same order as the ``supersymmetry breaking'' scale, defined as
\begin{equation}
m_{\text{SUSY}}=\kappa_4^2e^{\kappa_4^2 \langle K_{\text{mod} \rangle}/2} \langle |W_{np}^\prime|\rangle=0.8\times 10^{12}\text{GeV}\ .
\end{equation}

\subsection{Non-vanishing FI Term}

\begin{figure}[t]
   \centering
          \begin{subfigure}[b]{0.4\textwidth}
\includegraphics[width=1\textwidth]{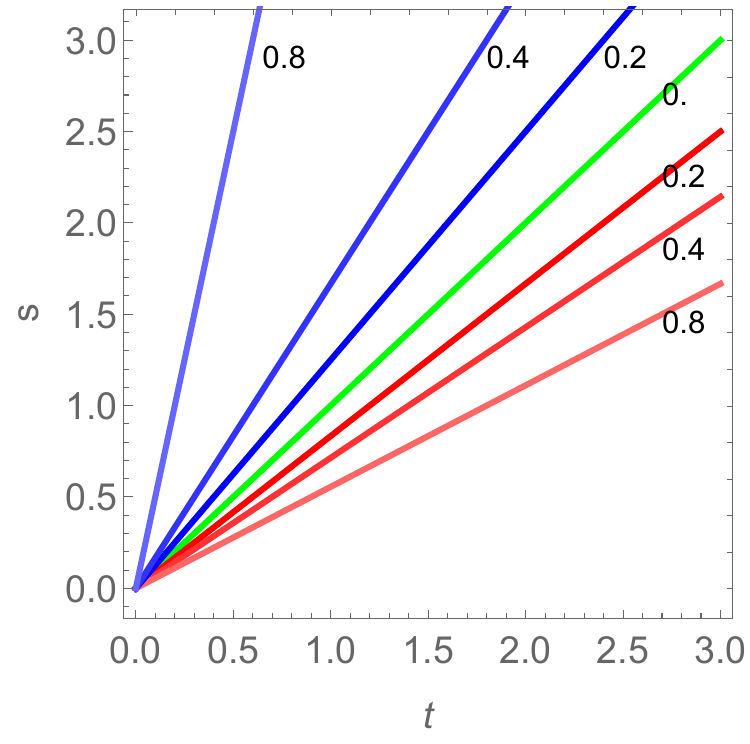}
\caption{}
\label{fig:blue_red1}
\end{subfigure}
       \begin{subfigure}[b]{0.53\textwidth}
\includegraphics[width=1\textwidth]{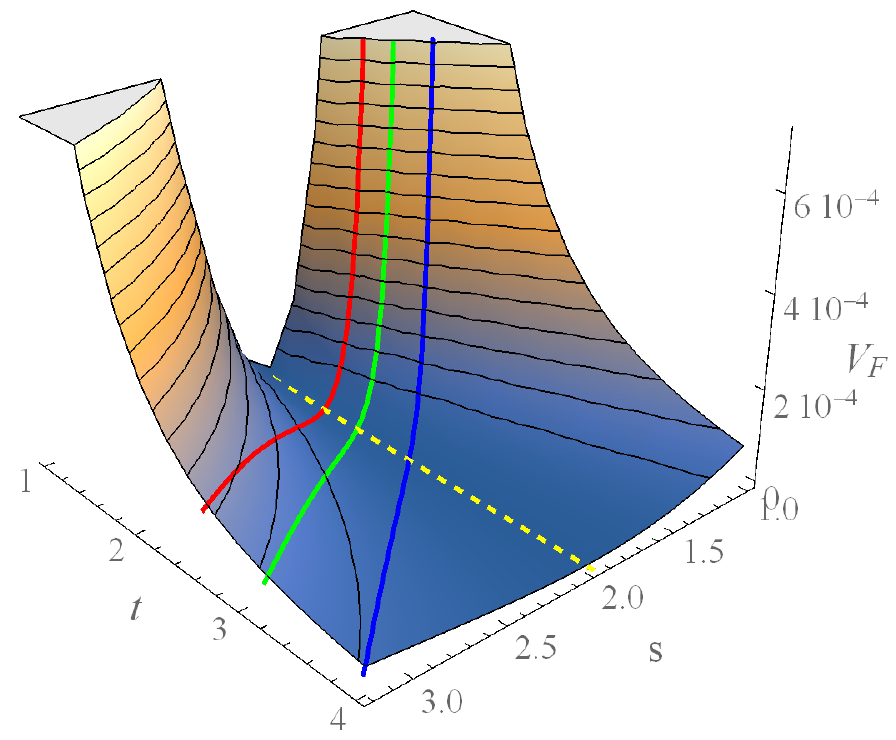}
\caption{}
\label{fig:blue_red2}
\end{subfigure}
\caption{(a) D-flat direction in $(s,t)$ moduli space for different values of $v^2$. The numbers associated with each line represent different values of $\frac{ |\tilde Q_1 |}{l}\frac{v^2}{M_U^2}$. The green line specifies the direction along which the FI term vanishes. With red we represent the D-flat solutions when $l/\tilde Q_1>0$, while with blue we represent the D-flat solutions when $l/\tilde Q_1<0$. (b) Plot of the potential $V_F^\prime$ given in eq. \eqref{eq:V_F_prime2} with $b=0.75$ and $c=1$.  Its minimum, defined in \eqref{pap1}, is shown by the dashed yellow line.  Turning on VEVs for  $C^1$ defines D-flat lines in the $s$ and $t$  moduli space, displayed in red, green and blue--as given in \ref{fig:blue_red1}. 
The vacuum state for any choice of $v^2$ lies at the intersection between the F-flat dashed yellow-line and the D-flat (red, green or blue) line.}
\label{fig:blue_red3}
\end{figure}

In Section \ref{sec:D-term_s}, we showed that the D-term potential imposes a relation between the VEVs of the moduli fields $s$ and $t$ and the matter scalars $C^1$ and $C^2$. That is, the D-flatness condition is satisfied at compactification if and only if
\begin{equation}
\label{umb33}
-\frac{ a\epsilon_S\epsilon_R^2}{\kappa^{2}_{4}}
\left( -\frac{1}{\langle s \rangle}\pi\beta \epsilon_Sl+\frac{3l}{\langle t \rangle} \right) - e^{\kappa_{4}^{2}\langle K_{T} \rangle /3} \left(\langle \tilde Q_1 \rangle  \langle C^1\rangle \langle\bar C^1\rangle
+\langle \tilde Q_2 \rangle  \langle C^2\rangle \langle\bar C^2\rangle \right)  = 0 \ .
\end{equation}
Let us now assume, as we did in  Section \ref{sec:D-term_s}, that of the matter scalars only $C^1$ can obtain a non-vanishing VEV; that is
\begin{equation}
\langle C^1\rangle =v\ ,
\end{equation}
where $v$ can always be taken to be real. Let us continue to work under the assumption that $\pi\beta\epsilon_S/3=1$. Furthermore, based on our results in \cite{Ashmore:2020ocb}, the magnitude of the $\text{FI}$-term is expected to be of the order of the compactification scale squared. Therefore, for simplicity, we will take
\begin{equation}
\label{sc1}
\frac{ 3a\epsilon_S\epsilon_R^2}{\kappa^{2}_{4}}= \frac{M_{U}^{2}}{2} \ .
\end{equation}
It then follows from \eqref{cup1} that
\begin{equation}
\text{FI}=-\frac{M_U^2l}{2}\left(\frac{1}{\langle s\rangle}-\frac{1}{\langle t\rangle}\right)\ .
\end{equation}
Under these assumptions, and using expression \eqref{bl1} for the K\"ahler potential $K_{T}$, the D-flatness condition becomes
\begin{equation}
\label{eq:blabla}
v^2=M_U^2\frac{l}{\tilde Q_1}\left(\frac{\langle t\rangle }{\langle s\rangle}-1\right)\ .
\end{equation}
There are two cases to be discussed: 
\begin{enumerate}
\item{$\frac{l}{\tilde Q_1}>0$}: In this case, the D-flatness condition has solutions only for $\langle s\rangle <\langle t \rangle$. For fixed $v=qM_U$, it follows from \eqref{eq:blabla} that the D-flat direction in $(s,t)$ moduli space is given by
\begin{equation}
\label{eq:st_Dflat1}
\langle t\rangle=\langle s\rangle \big(1+\frac{\tilde Q_1}{l}\frac{v^2}{M_U^2}\big) .
\end{equation}

\item{$\frac{l}{\tilde Q_1}<0$}: The D-flatness condition has solutions only for $\langle s\rangle >\langle t\rangle$. For fixed $v=qM_U$, the D-flat direction in $(s,t)$ moduli space is given by
\begin{equation}
\label{eq:st_Dflat2}
\langle t\rangle=\langle s\rangle \big(1-\big|\frac{\tilde Q_1}{l}\frac{v^2}{M_U^2}\big| \big) .
\end{equation}
\end{enumerate}
Note that when $v=0$, that is, when $FI=0$, both expressions reduce to 
\begin{equation}
\label{wire1}
\langle s \rangle = \langle t \rangle \ ,
\end{equation}
as given in \eqref{cof1}. These solutions are represented in Figure \ref{fig:blue_red1} where for specificity, and to be consistent with the $FI=0$ case above, we choose $b=0.75$ and $c=1$.

In the previous subsection we analyzed the vanishing FI case in which the hidden sector matter fields $C^1$ and $C^2$ decouple and have vanishing VEVs. Hence, the associated vacua must sit along the $FI=0$ line in the $s$ and $t$ moduli space. In this subsection, we extended our analysis to the more complicated case in which the VEVs of the hidden sector matter scalars can be non-zero. We explicitly look at vacua for which only the hidden sector matter field $C^1$ potentially has a non-zero VEV. The main consequence is that the D-flat line no-longer has a fixed direction in the $(s,t)$ moduli space. Instead, this direction can be shifted in moduli space by turning on a larger or smaller $\langle C^1 \rangle=v$ VEV, as seen in Figure \ref{fig:blue_red1}. As discussed above, and shown in Figure \ref{fig:blue_red2}, having chosen a value for $v$, the vacuum values for both $\langle s \rangle$ and $\langle t \rangle$ are fixed to be the intersection point of the red, green or blue D-flat line with the dashed yellow F-flat line of the $V_{F}^{\prime}$ potential. That is, turning on the non-perturbative potential $V_{F}^{\prime}$ given in \eqref{eq:V_F_prime2}, stabilizes both the $s$ and $t$ moduli completely. However, can it also stabilize the $C^1$ matter scalar? After a careful analysis, we have shown that $V_{F}^{\prime}$ in \eqref{eq:V_F_prime2} cannot by itself stabilize $C^1$ and, hence, fix the value of $\langle C^1 \rangle=v$. This would require additional non-perturbative contributions--which are beyond the scope of the present paper. Here, we will simply treat $v$ as a free parameter and, having chosen it, compute the associated values of $\langle s \rangle$ and $\langle t \rangle$.

We can now apply the results derived in Section \eqref{sec:final_states2} to find the low-energy matter spectrum produced after supersymmetry is broken. 
In the non-vanishing FI case, the matter field $C^1$ mixes with the two moduli $S$ and $T$. The final low energy spectrum contains two sets of scalar moduli fields, ($\phi^2$, $\eta^2$) and ($\phi^3$, $\eta^3$), and the hidden sector matter field $C^2$, which does not mix with the rest of the states. In Section 3.2.1, we computed the masses acquired by these scalar mass eigenstates for the case in which the VEV of the matter field $C^1$ needed to cancel the non-zero FI is small compared to the unification scale; that is, $v=qM_U$, $q\ll 1$. 
Let us again consider the vacuum state produced at the minimum of the potential $V_F^{\prime}$ given in eq. \eqref{eq:V_F_prime}, but for the case in which the FI term does not vanish. We asssume that the non-zero FI term can be cancelled by the VEV $v=qM_U$ of $C^1$, and that $q$ is relatively small. For specificity, we take
\begin{equation}
\left|\tfrac{\tilde Q_1}{l}\right|q^2=0.1\ .
\end{equation}
We have shown that when $b=0.75$ and $c=1$, the $V_F^\prime$ potential is minimized along the $\langle s\rangle\approx2$ direction and that $\langle \sigma\rangle=\pi/b$. Furthermore, when the $C^1$ field VEV is turned on, the D-flatness conditions shown in eq. \eqref{eq:st_Dflat1} and eq. \eqref{eq:st_Dflat2}, determine the value of the $t$ modulus. Using these expressions,  we find that 1) for $\tilde Q_1/l>0$ we get $\langle t\rangle\approx2.2$, while 2) for $\tilde Q_1/l<0$ we find $\langle t\rangle\approx1.8$.

The masses of the $\phi^2$ and $\eta^2$ states are given in eq. \eqref{eq:second_sc_masses} and \eqref{eq:second_sc_masses_axion}, respectively. For the form of the $V_F^{\prime}$ potential that we use, $m^2_{st}=m^2_{ts}=m^2_{tt}=0$ and, therefore, these expressions simplify to
\begin{equation}
m^2_{\phi^2}=m_{ss}^2\langle\frac{g_{T\bar T}}{g_{S\bar S}}\frac{k_T\bar k_{\bar T}}{\Sigma^2}\rangle\ 
\end{equation}
and
\begin{equation}
m^2_{\eta^2}=m_{\sigma\sigma}^2\langle\frac{g_{T\bar T}}{g_{S\bar S}}\frac{k_T\bar k_{\bar T}}{\Sigma^2}\rangle\ .
\end{equation}
For ``branch'' 1), where the moduli values are fixed to $\langle s\rangle\approx2, \>\langle t\rangle\approx2.2$ and $\langle \sigma\rangle=\pi/b$, we find that these masses are given by
\begin{equation}
\begin{split}
m_{\phi^2}=0.42\times 10^{12}\text{GeV}\ ,\quad m_{\eta^2}=1.9\times 10^{12}\text{GeV}\ .
\end{split}
\end{equation}
On the other hand, if we move to ``branch'' 2), with the moduli values $\langle s\rangle\approx2$ and $\langle t\rangle\approx1.8$, we find that
\begin{equation}
\begin{split}
m_{\phi^2}=0.58\times 10^{12}\text{GeV}\ ,\quad m_{\eta^2}=2.6\times 10^{12}\text{GeV}\ .
\end{split}
\end{equation}
In the no-scale model, the hidden matter scalar $C^2$ remains massless,
\begin{equation}
m_{C^2}=0\ .
\end{equation}
The masses of the $\phi^3$ and $\eta^3$ states are given in eq. \eqref{eq:second_sc_masses2} and \eqref{eq:second_sc_masses_axion2}, respectively. We find that for our choice of $V_{F}^{\prime}$, for which $m^2_{st}=m^2_{ts}=m^2_{tt}$ as well as $m_1^2$ vanish, these scalars eigenstates remain massless as well. That is,
\begin{equation}
m_{\phi^3}=m_{\eta^3}=0\ .
\end{equation}

The analysis of the fermion mass spectrum is similar. The spectrum contains a state $\psi_\xi^1$ which forms a Dirac fermion with the $U(1)$ gaugino and becomes part of a massive vector multiplet which decouples below the compactification scale.
To compute the rest of the fermion mass spectrum, we need to calculate the F-terms associated with the moduli fields.
For the family of vacua defined along the $t$-flat line, the F-term associated with $S$ vanishes, while the F-term associated with the K\"ahler modulus $T$ is non-zero. Hence, the only non-zero element in the fermion mass matrix defined in Appendix \ref{AppendixB2} is $M_{TT}$. Therefore, the mass of the $\Psi_\xi^2$ fermion is given by the expression
\begin{equation}
M_{\Psi_\xi^2}=2M_{TT}\langle \frac{g_{S\bar S}}{g_{T\bar T}}\frac{k_S\bar k_{S}}{\Sigma^2}\rangle\ .
\end{equation}
 At ``branch'' 1), where $\langle s\rangle\approx2$,$\langle t\rangle\approx2.2$ and $\langle \sigma\rangle=\pi/b$, we compute that
\begin{equation}
M_{\Psi_\xi^2}=0.9\times 10^{12}\text{GeV},
\end{equation}
while at ``branch'' 2), where $\langle s\rangle\approx2$,$\langle t\rangle\approx1.8$ and $\langle \sigma\rangle=\pi/b$, we find
\begin{equation}
M_{\Psi_\xi^2}=2.54\times 10^{12}\text{GeV}.
\end{equation}
The moduli fermion field $\psi_\xi^3$, as well as the matter fermion $\psi_2$, remain massless.

The fact that the masses of the $\phi^3$ and $\eta^3$ scalars vanish can be attributed to the fact that we have not managed to stabilize all of the moduli in this simple model. While $\langle s\rangle $ is fixed at a definite value, the values of $\langle t\rangle$ and $v$ can vary as long as the D-flatness constraint is satisfied. More specifically, the value of $\langle t \rangle$ is uniquely determined once the value of $v$ is chosen. However, for the non-perturbative potential $V_{F}^{\prime}$ we have chosen, the value of $v$ is not determined. In other words, there is no limit on how small or how large the FI term can be in this model, or the matter field VEV $v$ needed to cancel it. However, we should specify that features of this model, in particular the masslessness of the states $\phi^3,\eta^3$ and $\Psi_{\xi}^3$, are not expected to remain true in more general and realistic string theory models; which include, for example, gauge threshold corrections present at order $\kappa^{4/3}$. If stable vacuum states can be proven to exist in such more realistic models, then the mass hierarchy proposed in eq. \eqref{mass_hier} is expected to be valid.

It is interesting, however, to point out that even in our simple models, the masses of the fermions $\Psi_\xi^3$ can receive some non-zero contributions. These are sourced by terms that we have thus far neglected, such as those given in eq. \eqref{non-zerobla}. These terms are suppressed by $q^2M_U^2/M_P^2$ and, therefore, do not impact the mass matrix diagonalization conditions derived in Section \ref{mass_spect_sec}. They are in general negligible relative to the non-zero masses obtained by the $\Psi_\xi^3$ fermions in more realistic models, as discussed in the paragraph above. However, in our simple model they provide a lower bound for the masses that these fermions can take. That is,
\begin{equation}
M_{\Psi_\xi^3}\geq e^{\kappa_4^2 \langle K_{\text{mod}}/2+2K_T/3\rangle }\kappa_4^4v^2 \langle \hat W_{np}\rangle \approx q^2\times 5\times 10^7\text{GeV}\ .
\end{equation}

We conclude this section by pointing that we have chosen this simple model to serve as a useful visualization of how moduli can be stabilized. Note that there are many other non-perturbave superpotentials that can arise in string theory, such as in vacua with multiple gaugino condensates and so on. Other non-perturbative effects are generated by five-brane instantons~\cite{Carlevaro:2005bk,Gray:2003vw,Lima:2001nh}, for example. It is possible that when additional non-perturbative effects are included, as well as higher order corrections, the remaining unfixed VEV in our theory, namely, $\langle C^1 \rangle=v$, will also be stabilized. The subject of moduli stabilization is a vast one. Attempts have been made in the context of the heterotic string, with various degrees of succes~\cite{Choi:1998nx,Gukov:2003cy,PaccettiCorreia:2007ret,Anderson:2011cza,Cicoli:2013rwa,Dundee:2010sb,Parameswaran:2010ec}. To the knowledge of the authors, however, within the
context of phenomenologically realistic models, there are no models that fix all moduli.

Although the results in this paper were derived within the context of a simple $h^{1,1}=1$ model,  we believe they will remain valid within the context of more generic, and more complicated, string vacua with $h^{1,1} \geq 1$, provided such stable vacua can be found.

\subsubsection*{Acknowledgements}

We would like to thank Anthony Ashmore for many helpful conversations. Sebastian Dumitru is supported in part by research grant DOE No.~DESC0007901. Burt Ovrut is supported in part by both the research grant DOE No.~DESC0007901 and SAS Account 020-0188-2-010202-6603-0338.\\

\appendix

\section{ The Effective Supergravity Lagrangian}

In this paper, we work in the 4D $N=1$ supergravity formalism to compute the masses of the moduli fields and the hidden sector matter fields, as well all the relevant couplings, after non-perturbative effects are turned on. In this Appendix we reproduce the $N=1$ supergravity Lagrangian (in Weyl spinor notation) that we use~\cite{Wess:1992cp,Freedman:2012zz}:
\begingroup
\allowdisplaybreaks
\begin{align}
\nonumber \frac{1}{e}\mathcal{L}&=-\frac{M_P^2}{2}R-g_{A\bar B}\mathcal{ D}_\mu z^{A}\mathcal{D}^\mu \bar z^{\bar B}-ig_{A\bar B}\psi^{B\dag}\bar\sigma^\mu\mathcal{D}_\mu\psi^{A}\\
\nonumber&+\epsilon^{\mu \nu \rho \sigma} {\tilde G}^\dag_{L \mu}\sigma_\nu  \mathcal{D}_\rho {\tilde G}_{L \sigma}-\tfrac{1}{4}(\text{Re}f_{ab})F_{\mu \nu }^{a}F^{b\>\mu \nu}
+\tfrac{1}{8}(\text{Im}f_{ab})F_{\mu \nu}^{a}\tilde F^{b\>{\mu \nu}}\\
\nonumber&+\tfrac{i}{2}\text{Re}f_{ab} \lambda^{a}_{(\text{s})}\sigma^\mu\mathcal{D}_\mu\lambda^{ \dag b}-\tfrac{1}{2e}\text{Im}f_{ab}
\mathcal{\tilde D}_\mu\left[e \lambda^{a}\sigma^\mu \lambda^{ b\dag}\right]\\
\nonumber&+\left(\tfrac{i}{16}\sqrt{2}g_{A\bar B}\partial^{A}f_{ab}\lambda^{a\dag}[\gamma^\mu,\gamma^\nu]\hat F_{\mu \nu}\psi^{\bar B\dag}+\text{h.c}\right)\\
\nonumber&+\left[\sqrt{2}g_ag_{A\bar B}\partial^{A}D_{a} \lambda^{ a\dag}\psi^{\bar B\dag}
-i\tfrac{\sqrt{2}}{4}g_{a}[\text{Re}f^{-1}]^{ab}g_{A\bar B}\partial^A f_{bc}D_{a} \lambda^{c\dag}\psi^{\bar B\dag}+\text{h.c}\right]\\
&+\tfrac{1}{M_P}\bigg[-\tfrac{1}{2}(\text{Re}f_{ab})D^{a} {\tilde G}_\mu\sigma^\mu\lambda^{b\dag}+
\tfrac{1}{2}(\text{Re}f_{ab})D^{a} {\tilde G}^\dag_\mu\bar \sigma^\mu\lambda^{b}-\left(\tfrac{1}{\sqrt{2}}g_{A\bar B}
\mathcal{D}_\nu \bar z^{\bar B} \psi^{A}\sigma^\mu  \bar \sigma^\nu{\tilde G}_\mu + \text{h.c}\right) \\
\nonumber&+\tfrac{i}{4}(\text{Re}f_{ab})\left[{\tilde G}_\mu\sigma^{\rho \xi}\sigma^\mu\lambda^{a\dag}
+{\tilde G}^\dag_\mu\bar\sigma^{\rho \xi}\bar\sigma^\mu\lambda^{a}\right](F_{\rho \xi}^{b}+\hat F^{b}_{\rho\xi})\bigg]\\
\nonumber&-e^{\kappa_4^2K/2}\Bigg[  \tfrac{1}{4M_P^2}W^* {\tilde G}_\mu\sigma^\mu \sigma^\nu{\tilde G}_\nu+  \tfrac{1}{4M_P^2}W {\tilde G}^\dag_\mu\sigma^\mu \sigma^\nu{\tilde G}^\dag_\nu\\
&\qquad+\tfrac{i}{2M_P}\sqrt{2}D_AW{\tilde G^\dag}_\mu\sigma^\mu \psi^{A}+\tfrac{i}{2M_P}\sqrt{2}D_{\bar A}W^*{\tilde G}_\mu\bar\sigma^\mu \psi^{\dag A}\\
\nonumber&+\tfrac{1}{2}\mathcal{D}_A\mathcal{D}_BW\psi^{A}\psi^B+\tfrac{1}{2}\mathcal{D}_{\bar A}\mathcal{D}_{\bar B}W^*\psi^{A}\psi^B+\left(\tfrac{1}{4}g^{A\bar B}
D_{\bar B}W^*\partial_Af_{ab} \lambda^{a}\lambda^{b}+\text{h.c.}\right)\Bigg]\\
\nonumber&-\frac{1}{2}[(\text{Re}f_{ab})^{-1}]^{ab}D^aD^b-e^{\kappa_4^2K}\left[ g^{A\bar B}(D_AW)(D_{\bar B}W^*)-3\kappa^2_4|W|^2 \right]+L_{4f}\ ,
\end{align}
\endgroup

where
\begin{equation}
\begin{split}
&D_AW=W_A+\kappa_4^2K_AW\ ,\\
&\mathcal{D}_A{D}_BW=W_{AB}+\kappa_4^2(K_{AB}W+K_AD_BW+K_BD_AW-K_AK_BW)\\
&\hspace{5cm}-\Gamma^C_{AB}D_CW+\mathcal{O}(M_P^{-3})\ .
\end{split}
\end{equation}

\section{Mass matrices}

In this appendix, we show how the scalars and the fermions of the theory acquire masses in the heterotic vacuum in which 4D $N=1$ supersymmetry was spontaneously broken by low-energy non-perturbative effects.

\subsection{Scalars}\label{AppendixB1}

The total F-term scalar potential is given by
\begin{equation}
V=e^{\kappa^2_4K}\left[ g^{A\bar B}(D_AW)(D_{\bar B}W^*)-3\kappa_4^2{| W|^2}  \right] 
\label{fly222}
\end{equation}
where we consider the superpotential:
\begin{equation}
\begin{split}
\label{eq:app_super}
W&=W^{\text{(hid)}}+W^{\text{(mod)}}\\
&=\hat W_{np}(S,T)+ m^{LM}C_{L}C_{M}+\lambda^{KLM}C_KC_LC_M \ .
\end{split}
\end{equation}
and the K{a}hler potential
\begin{equation}
\begin{split}
\label{eq:app_K\"ahler}
K&=K_S+K_T+K_{\text{matter}}^{\text{(hid)}}\\
&=\hat K_{\text{mod}}+G_{L\bar M}C^L\bar C^{\bar M} \\
&=-\kappa_4^{-2}\ln(S+\bar S)-3\kappa_4^{-2}\ln(T+\bar T)+e^{\kappa_4^2K_T/3}C^1\bar C^1+e^{\kappa_4^2K_T/3}C^2\bar C^2\ .
\end{split}
\end{equation}

The first derivatives of the K\"ahler potential are
\begin{align}
\partial_S K&=\partial_S K_S\ ,\\
\partial_T K&=\partial_T K_T+\partial_T G_{L\bar M}C^L\bar C^{\bar M}\ ,\\
\partial_{C^L}K&=G_{L\bar M}\bar C^{\bar M}\ .
\end{align}
while the K\"ahler metrics and their inverses are given by
\begin{align}
g_{S\bar S}&=\partial_S\partial_{\bar S}K_S\ ,\\
g_{T\bar T}&=\partial_T\partial_{\bar T}K_T+\partial_T\partial_{\bar T}G_{L\bar M}C^L\bar C^{\bar M}\ ,\\
g_{T\bar C^{\bar M}}&=\partial_{ T}G_{L\bar M}C^L\ ,\\
g_{C^L\bar C^{\bar M}}&=G^{L\bar M}\ ,
\end{align}
and 
\begin{align}
g^{S\bar S}&=\frac{1}{\partial_S\partial_{\bar S}K_S}\ ,\\
g^{T\bar T}&=\frac{1}{\partial_T\partial_{\bar T}K_T}\left(1-\frac{\partial_T\partial_{\bar T}G_{L\bar M}C^L\bar C^{\bar M}}{\partial_T\partial_{\bar T}K_T}\right)\ ,\\
g^{T\bar C^{\bar M}}&=\frac{1}{\partial_T\partial_{\bar T}K_T}G^{L\bar M}\partial_{ \bar T}G_{L\bar P}\bar C^{\bar P}\ ,\\
g^{C^L\bar C^{\bar M}}&=G^{L\bar M}\ ,
\end{align}
respectively.

The covariant derivatives of the superpotential are
\begin{align}
D_SW&=\partial_S\hat W_{np}+\kappa_4^{2}\partial_SK_S\hat W_{np}+\kappa_4^{2}\partial_SK_S(m_{LM}C^{L}C^{M}+\lambda_{KLM}C^KC^LC^M)\ ,\\
\nonumber D_TW&=\partial_T\hat W_{np}+\kappa_4^{2}\partial_TK_T\hat W_{np}+\kappa_4^{2}\partial_TK_T(m_{LM}C^{L}C^{M}+\lambda_{KLM}C^KC^LC^M)\\
&+\kappa_4^{2}\partial_T G_{L\bar M}C^L\bar C^{\bar M}(m_{PR}C^{P}C^{R}+\lambda_{PQR}C^PC^QC^R)\\
&+\kappa_4^{2}\partial_T G_{L\bar M}C^L\bar C^{\bar M}\hat W_{np}\nonumber\\
\nonumber D_{C^L}W&=m_{LM}C^{M}+\lambda_{KLM}C^KC^M+\kappa_4^{2}G_{L\bar M}\bar C^{\bar M}\hat W_{np}\\
&+\kappa_4^{2}G_{L\bar M}\bar C^{\bar M}(m_{PR}C^{P}C^{R}+\lambda_{PQR}C^PC^QC^R)\ .
\end{align}

Let assume 
\begin{equation}
m_{LM}=\lambda_{KLM}=0\ ,
\end{equation}
 first, and define 
\begin{align}
F_S&=\partial_S\hat W_{np}+\kappa_4^{2}\partial_SK_S\hat W_{np}\ ,\\
F_T&=\partial_T\hat W_{np}+\kappa_4^{2}\partial_TK_T\hat W_{np}\ .
\end{align}

With these choices, we find the following expressions for the terms in the F-term potential
\begin{equation}
\begin{split}
&g^{S\bar S}D_{S}W D_{\bar S}W^*=\frac{1}{\partial_S\partial_{\bar S}K_S}F_S\bar F_{\bar S}\ ,\\
&g^{T\bar T}D_{T}W D_{\bar T}W^*=\frac{1}{\partial_T\partial_{\bar T}K_T}F_T\bar F_{\bar T}\\
&\qquad+\frac{\kappa_4^{2}}{\partial_T\partial_{\bar T}K_T}(\partial_T G_{L\bar M}\bar F_{\bar T}\hat W_{np}+\partial_{\bar T}G_{L\bar M}F_T\hat W_{np}^*)C^L\bar C^{\bar M}\\
&\qquad+\frac{\kappa_4^{4}}{\partial_T\partial_{\bar T}K_T}\partial_TG_{L\bar M}\partial_{\bar T}G_{P\bar Q}|\hat W_{np}|^2C^L\bar C^{\bar M}C^P\bar C^{\bar Q}-\frac{\partial_T\partial_{\bar T}G_{L\bar M}}{(\partial_T\partial_{\bar T}K_T)^2}F_T\bar F_{\bar T}C^L\bar C^{\bar M}
\\&\qquad-\frac{\partial_T\partial_{\bar T}G_{R\bar N}}{(\partial_T\partial_{\bar T}K_T)^2}\kappa_4^{4}\partial_TG_{L\bar M}\partial_{\bar T}G_{P\bar Q}|\hat W_{np}|^2C^L\bar C^{\bar M}C^P\bar C^{\bar Q}C^R\bar C^{\bar N}
\\&\qquad-\frac{\partial_T\partial_{\bar T}G_{P\bar Q}}{(\partial_T\partial_{\bar T}K_T)^2}\kappa_4^{2}(\partial_T G_{L\bar M}\bar F_{\bar T}\hat W_{np}+\partial_{\bar T}G_{L\bar M}F_T\hat W_{np}^*)C^L\bar C^{\bar M}C^P\bar C^{\bar Q}\\
&g^{T\bar C^{\bar M}}D_{T}W D_{\bar C^{\bar M}}W^*=\\
&\qquad=\frac{\kappa_4^{2}}{\partial_T\partial_{\bar T}K_T}G^{L\bar M}\partial_{ \bar T}G_{L\bar P}\bar C^{\bar P}(F_T+\kappa_4^{2}\partial_TG_{P\bar Q}C^P\bar C^{\bar Q}\hat W_{np})G_{N\bar M}C^N\hat W_{np}^*\\
&g^{C^L\bar C^{\bar M}}D_{C^L}W D_{\bar C^{\bar M}}W^*=\kappa_4^{4}G^{L\bar M}G_{L\bar P}\bar C^{\bar P}G_{Q\bar M}C^Q|\hat W_{np}|^2=\kappa_4^4G_{L\bar M}C^L\bar C^{\bar M}|\hat W_{np}|^2\ .
\end{split}
\end{equation}

To quartic order in the $C^L$ fields, the total F-term scalar potential is given by
\begin{equation}
\begin{split}
V&=e^{\kappa^2_4K}\left[ g^{A\bar B}(D_AW)(D_{\bar B}W^*)-3\kappa_4^2{| W|^2}  \right] \\
&=V_F(S,T)+m^2_{L\bar M}(S,T)C^L\bar C^{\bar M}+\Lambda_{L\bar M P\bar Q}(S,T)C^L\bar C^{\bar M}C^P\bar C^{\bar Q}+\dots\ ,
\end{split}
\label{fly2222}
\end{equation}
where 
\begin{equation}
V_F(S,T)=e^{\kappa^2_4\hat K_{\text{mod}}}\left[\frac{1}{\partial_S\partial_{\bar S}K_S}F_S\bar F_{\bar S} + \frac{1}{\partial_T\partial_{\bar T}K_T}F_T\bar F_{\bar T} -3\kappa_4^2|\hat W_{np}|^2\right]\ ,
\end{equation}
\begin{equation}
\begin{split}
\label{eq:msquaredSc}
m_{L\bar M}^2&=\kappa_4^2\langle V_F\rangle G_{L\bar M}+ e^{\kappa_4^2\hat K_{\text{mod}}}\kappa^4_4|\hat W_{np}|^2G_{L\bar M}-e^{\kappa_4^2\hat K_{\text{mod}}}\frac{\partial_T\partial_{\bar T}G_{L\bar M}}{(\partial_T\partial_{\bar T}K_T)^2}F_T\bar F_{\bar T}\\
&+\frac{e^{\kappa_4^2\hat K_{\text{mod}}}}{\partial_T\partial_{\bar T}K_T}\kappa_4^{2}(\partial_T G_{L\bar M}\bar F_{\bar T}\hat W_{np}+\partial_{\bar T}G_{L\bar M}F_T\hat W_{np}^*)\\
&+\frac{e^{\kappa_4^2\hat K_{\text{mod}}}}{\partial_T\partial_{\bar T}K_T}G^{N\bar P}\partial_{ \bar T}G_{N\bar M}G_{L\bar P}\kappa_4^{2}F_T \hat W_{np}^*+\frac{e^{\kappa_4^2\hat K_{\text{mod}}}}{\partial_T\partial_{\bar T}K_T}G^{N\bar P}\partial_{ T}G_{N\bar M}G_{L\bar P}\kappa_4^{2}\bar F_T \hat W_{np}\ ,\\
\end{split}
\end{equation}
and
\begin{equation}
\begin{split}
\Lambda_{L\bar M P\bar Q}&=\kappa_4^2m^2_{L\bar M}G_{P\bar Q}+
\frac{e^{\kappa_4^2\hat K_{\text{mod}}}}{\partial_T\partial_{\bar T}K_T}\kappa_4^{4}\partial_TG_{L\bar M}\partial_{\bar T}G_{P\bar Q}|\hat W_{np}|^2\\
&-e^{\kappa_4^2\hat K_{\text{mod}}}\frac{\partial_T\partial_{\bar T}G_{P\bar Q}}{(\partial_T\partial_{\bar T}K_T)^2}\kappa_4^{2}(\partial_T G_{L\bar M}\bar F_{\bar T}\hat W_{np}+\partial_{\bar T}G_{L\bar M}F_T\hat W_{np}^*)\\
&+\frac{e^{\kappa_4^2\hat K_{\text{mod}}}}{\partial_T\partial_{\bar T}K_T}\kappa_4^{4}G_{L\bar R}G^{N\bar R}\partial_{ \bar T}G_{N\bar M}\partial_TG_{P\bar Q}|\hat W_{np}|^2
\end{split}
\end{equation}

Note that to obtain the expressions above, we made use of the expansion
\begin{equation}
\begin{split}
e^{\kappa_4^2K}&\approx e^{\kappa^2_4\hat K_{\text{mod}}}\left(1+\kappa_4^2G_{L\bar M}C^L\bar C^{\bar M}\right)\ .
\end{split}
\end{equation}

For our hidden sector model, $G_{L\bar M}=e^{\kappa_4^2K_T/3}\delta_{L\bar M}$, for $L,M=1,2$. This corresponds to no-scale supegravity models, for which
\begin{equation}
m^2_{L\bar M}=0\ .
\end{equation}
Therefore, neglecting the quartic order, the F-term potential is given by
\begin{equation}
V=V_F(S,T)\ .
\end{equation}
The scalar mass matrix has the form
\begin{equation}
\begin{pmatrix}
m^2_{AB}&m^2_{A\bar B}\\
m^2_{\bar A B}&m^2_{\bar A\bar B}
\end{pmatrix}=
\begin{pmatrix}
\partial_S\partial_S V_F& \partial_S\partial_T V_F&\partial_S\partial_{\bar S} V_F&\partial_S\partial_{\bar T} V_F\\
\partial_T\partial_S V_F& \partial_T\partial_T V_F&\partial_T\partial_{\bar S} V_F&\partial_T\partial_{\bar T} V_F\\
\partial_{\bar S }\partial_S V_F& \partial_{\bar S }\partial_T V_F&\partial_{\bar S }\partial_{\bar S} V_F&\partial_{\bar S }\partial_{\bar T} V_F\\
\partial_{\bar T }\partial_S V_F& \partial_{\bar T }\partial_T V_F&\partial_{\bar T }\partial_{\bar S} V_F&\partial_{\bar T}\partial_{\bar T} V_F
\end{pmatrix}\ , \quad A,B=S,T\ ,
\end{equation}
while all other elements are set to zero. 

The no-scale model does not survive the inclusion of genus-one corrections. When these are included, the K\"ahler metric associated with the hidden matter fields becomes~\cite{Brandle:2003uya}
\begin{equation}
G_{L\bar M}=e^{\kappa_4^2K_T/3}\delta_{L\bar M}\mapsto (e^{\kappa_4^2K_T/3}+\beta e^{\kappa_4^2K_S} )\delta_{L\bar M}\ .
\end{equation}
In this case, the scalar squared mass matrix $m^2_{L\bar M}$ has the form
\begin{equation}
m^2_{L\bar M}=\begin{pmatrix}
m_1^2&0\\
0&m_2^2
\end{pmatrix}\ ,
\end{equation}
with 
\begin{equation}
m_1^2=m_2^2\sim \beta e^{\kappa_4^2\hat K_{\text{mod}}}\kappa_4^4|\hat W_{np}^2|\ .
\end{equation}

Hence, the mass terms $m^2_{L\bar M}(S,T)C^L\bar C^{\bar M}$ are generically non-zero, and therefore, the scalar mass matrix contains additional terms of the type 
\begin{equation}
\begin{split}
\label{eq:non-diagonal}
&\partial_{C^L}\partial_{\bar C^{\bar M}}V=m^2_{L\bar M}\ ,\\
&\partial_{S}\partial_{\bar C^{\bar M}}V=\frac{\partial m^2_{L\bar M}}{\partial S}C^L\ ,\\
&\partial_{T}\partial_{\bar C^{\bar M}}V=\frac{\partial m^2_{L\bar M}}{\partial T}C^L\ .
\end{split}
\end{equation}
In the vanishing FI term case, the scalar mass matrix is block diagonal, composed of the moduli mass matrix shown above, and the mass terms $m^2_{L\bar M}$ for the hidden matter fields. Restoring the moduli fields with their natural mass units, that is
\begin{equation}
S\rightarrow \kappa_4^{-1/2}S\ ,\quad T\rightarrow \kappa_4^{-1/2}T\ ,
\end{equation}
we are able to obtain estimates for the mass scales of these fields. We learn the mass matrix elements associated with the moduli fields $S$ and $T$ alone, are of the order of  the``SUSY-breaking scale'', defined as
\begin{equation}
m_{\text{SUSY}}\sim e^{\kappa_4^2\hat K_{\text{mod}}/2}\kappa_4^2|\hat W_{np}|\ ,
\end{equation}  
In this language, when we consider the string one-loop corrections, the squared mass matrix elements associated with the hidden scalar fields are of the order
\begin{equation}
m_1=m_2\sim \sqrt{\beta} m_{\text{SUSY}}\ .
\end{equation}

In the non-vanishing FI case, the scalar mass matrix is modified by the inclusion of the non-diagonal terms of the type $\partial_{S}\partial_{\bar C^{\bar M}}V$ and $\partial_{T}\partial_{\bar C^{\bar M}}V$, defined in \eqref{eq:non-diagonal}, which become proportional to the VEV of the scalar field that is turned on to cancel the FI term. However, assuming that the sizes of these matter fields VEVs are no larger than the unification scale, that is, $v=qM_U$, such that $q<1$, one can check that these extra contributions are suppressed by powers of 
$q\frac{M_U}{M_P}$. Explicitly, both the moduli field masses and the matter field masses discussed above get corrections suppressed by $q^2M_U^2/M_P^2$, while the non-diagonal terms of the type $\partial_{S}\partial_{\bar C^{\bar M}}V$ and $\partial_{T}\partial_{\bar C^{\bar M}}V$ are suppressed by $qM_U/M_P$. In our analysis we will neglect these extra corrections, and assume that the scalar mass matrix has the same expressions in both the vanishing FI term and non-vanishing FI term cases.

The scalar mass matrices are most easily expressed in a basis composed of the real scalar components and the corresponding imaginary component fields. Including the contributions from the $V_D$ potential, which has the form $V_D=m_{\text{anom}}^2{\phi^1}^2+m_{\text{anom}}^2{\eta^1}^2$ around the D-flat vacuum, we have
\begin{equation}
\begin{split}
\label{xyz}
\mathcal{L}\supset
= -m_{\text{anom}}^2{\phi^1}^2-&m_{\text{anom}}^2{\eta^1}^2-\left( \delta s,\>\delta t, \>\text{Re}(\delta  C^1),\>\text{Re}(\delta C^2)\right)
{\mathcal{M}_s^{(r)}}
\left( \begin{matrix}\delta s\\\delta  t\\ \text{Re}(\delta C^1) \\\text{Re}(\delta C^2)\end{matrix}\right)\\
&-\left( \delta \sigma,\>\delta \chi, \>\text{Im}(\delta  C^1),\>\text{Im}(\delta C^2)\right)
{\mathcal{M}_s^{(i)}}
\left( \begin{matrix}\delta \sigma \\\delta  \chi\\\text{Im}(\delta C^1)\\\text{Im}(\delta C^2) \end{matrix}\right) \ .
	\end{split}
\end{equation}
 In the above equation, the $4\times 4$ scalar squared mass matrices are given by
\begin{equation}
\label{eq:xyz}
{\mathcal{M}^{(r)}_s}=\left(  
\begin{matrix}
m^2_{ss}&m^2_{st} &0\\
m^2_{ts}&m^2_{tt}&0\\
0&0&m^2_{L\bar M}
\end{matrix}
\right)\ ,\quad 
{\mathcal{M}^{(i)}_s}=\left(  
\begin{matrix}
m^2_{\sigma \sigma}&m^2_{\sigma \chi} &0\\
m^2_{\chi \sigma}&m^2_{\chi \chi}&0\\
0&0&m^2_{L\bar M}
\end{matrix}
\right)\ ,\quad L,M=1,2\ .
\end{equation}
Note that we assumed that the real scalar and imaginary/axionic degrees of freedom separate. For the examples of F-term potentials we give in Section \ref{Sec:Mod_Stab}, this is indeed the case. In the above expressions, the moduli mass terms are obtained by doubly differentiating the $F$-term potential with respect to $s$ and $t$, $\sigma$ or $\chi$. For example
\begin{equation}
\begin{split}
\left[{\mathbf{M_s^{(r)}}}   \right]_{11}&\equiv m_{\sigma \sigma}^2=\frac{1}{2}\left\langle\frac{\partial^2 V_F}{\partial \sigma\partial \sigma}\right\rangle\ .
\end{split}
\end{equation}

 The matrix elements $m^2_{L\bar M}$ have been given in eq. \eqref{eq:msquaredSc}. If the ``no-scale'' model, with $G_{L\bar M}=e^{\kappa_4^2K_T/3}$, these elements vanish. At one-loop order, however, $m^2_{L\bar M}=\text{diag}(m_1^2,m_2^2)$, as explained above.

\subsection{Fermions}\label{AppendixB2}

The fermion mass terms originate in the fermion bilinear
\begin{equation}
-\tfrac{1}{2}e^{\kappa_4^2K/2}\mathcal{D}_A\mathcal{D}_BW\psi^{A}\psi^B+h.c.\ ,
\end{equation}
where
\begin{equation}
\begin{split}
D_AW&=W_A+\kappa_4^2K_AW\ ,\\
\mathcal{D}_A{D}_BW&=W_{AB}+\kappa_4^2(K_{AB}W+K_AD_BW+K_BD_AW-\kappa_4^2K_AK_BW)\\
&-\kappa_4^4K^{C\bar D}K_{AB\bar D}D_CW+\mathcal{O}(M_P^{-3})\\
&=W_{AB}+\kappa_4^2(K_{AB}W+K_AW_B+K_BW_A+\kappa_4^2K_AK_BW)\\
&-\kappa_4^4K^{C\bar D}K_{AB\bar D}D_CW+\mathcal{O}(M_P^{-3})
\end{split}
\end{equation}
and the superpotential and the K\"ahler potential were given in eq. \eqref{eq:app_K\"ahler} and \eqref{eq:app_K\"ahler}, respectively. 

When SUSY is broken, such that $F_S=D_SW\neq 0$ and $F_T=D_TW\neq 0$, we obtain the following fermion mass matrix

\begin{align}
\mathcal{L}&\supset
-(\psi_S,\>\psi_T, \>\psi_1,\>\psi_2) [\mathbf{M}]_{AB}
\left( \begin{matrix} \psi_S\\   \psi_T\\  \psi_1\\ \psi_2 \end{matrix}\right)+h.c.\ ,
\end{align}

The matrix elements are given by
\begin{equation}
\begin{split}
\label{eq:f_mass_1}
&M_{SS}=\tfrac{1}{2} e^{\kappa_4^2 \langle K_{\text{mod}}\rangle /2} \langle \partial^2_S\hat {W}_{np}+\kappa_4^2(\partial^2_SK_S {\hat W}_{np}\\
&\quad\qquad+2\partial_SK_S{ \partial_S\hat W}_{np}+\kappa_4^2(\partial_SK_S)^2\hat W_{np})+\Gamma_{SS}^AD_A\hat W_{np} \rangle \ ,\\
&M_{S T},\>M_{T S}=\tfrac{1}{2} e^{\kappa_4^2 \langle K_{\text{mod}}\rangle/2}\langle \partial_S\partial_{T}{\hat W_{np}}+\\
&\qquad \qquad+\kappa_4^2(\partial_SK_S   \partial_{T}{\hat W_{np}}+\partial_{T}K_T \partial_S{\hat W_{np}}+\partial_TG_{LM}C^L\bar C^{\bar M}\partial_S{\hat W_{np}} \\
&\qquad \qquad+
\kappa_4^2\partial_SK_S(\partial_TK_T+\partial_TG_{L\bar M}C^L\bar C^{\bar M})\hat W_{np})+\Gamma^A_{ST}D_A\hat W_{np}\rangle\ , \\
&M_{T T}=\tfrac{1}{2} e^{\kappa_4^2 \langle K_{\text{mod}}\rangle/2}\langle \partial^2_{T}{\hat W_{np}}\\
&\qquad +\kappa_4^2(\partial^2_{T}K_T \hat W_{np}+\partial^2_TG_{L\bar M}C^L\bar C^{\bar M}\hat W_{np}+2\partial_{T}K_T \partial_{T}{\hat W_{np}}\\
&\qquad+2\partial_{T}G_{L\bar M}C^L\bar C^{\bar M}\partial_T{\hat W_{np}}+\kappa_4^2
(\partial_TK_T+\partial_TG_{L\bar M}C^L\bar C^{\bar M})^2\hat W_{np})+\Gamma^A_{TT}D_A\hat W_{np}\rangle \ ,\\
&M_{C^LT}=\tfrac{1}{2} e^{\kappa_4^2 \langle K_{\text{mod}}\rangle /2}\kappa_4^2\langle \partial_TG_{L\bar M}\bar C^{\bar M}\hat W_{np}+G_{L\bar M}\bar C^{\bar M}\partial_T\hat W_{np}\\
&\qquad \qquad+\kappa_4^2(\partial_TK_T+\partial_TG_{P\bar Q}C^P\bar C^{\bar Q})G_{L\bar M}\bar C^{\bar M}\hat W_{np}+\Gamma^A_{TC}D_A\hat W_{np}\rangle\ ,\\
&M_{C^LC^M}=\tfrac{1}{2} e^{\kappa_4^2 \langle K_{\text{mod}}\rangle /2}\kappa_4^4\langle G_{L\bar P }G_{M\bar Q}\bar C^{\bar P }\bar C^{\bar Q} \hat W_{np}\rangle\ .
\end{split}
\end{equation}

For our hidden sector model, $G_{L\bar M}=e^{\kappa_4^2K_T/3}\delta_{L\bar M}$, for $L,M=1,2$. The fermion mass matrix has different forms, depending weather the fields $C^1$ or $C^2$ aquired VEVs during the D-term stabilization process. There are two cases we consider in this paper. In the vanishing FI term case, neither of the fields obtain a VEV and therfore, the fermion mass matrix has the form
\begin{equation}
{\mathbf{M_f}}=\left(
\begin{matrix}
M_{S S}& M_{ST}&0&0\\
M_{T S}& M_{TT}&0&0\\
0&0&0&0\\
0&0&0&0\\
\end{matrix}
\right)\ ,
\end{equation}
where any terms proportional to $C^1,C^2$ have been set to zero

On the other hand, when the $C^1$ field get a VEV $v$, the fermion mass matrix becomes

\begin{equation}
{\mathcal{M}_f}=\left(  
\begin{matrix}
M_{S S}&M_{ST} &0&0\\
M_{T S}&M_{T T}&M_{T C^1}&0\\
0&M_{C^1 T}&M_{C^1 C^1}&0\\
0&0&0&0
\end{matrix}
\right)\ .
\end{equation}
Just as in the case of the scalars, the moduli mass terms such as $M_{SS}$ obtain corrections suppressed by $q^2M_U^2/M_P^2$, where we wrote the matter field VEV as $v=\langle C^1\rangle=qM_U$. The matrix elements $M_{C^1T}$ and $M_{C^1C^1}$ are, up to quadratic order in $v$,
\begin{equation}
\begin{split}
\label{non-zerobla}
M_{C^1T}&=\tfrac{1}{2}e^{\kappa_4^2\langle K_{\text{mod}}/2+K_T/3\rangle}\kappa_4^2v\langle F_T\rangle+\mathcal{O}(v^3)\ ,\\
M_{C^1C^1}&=\tfrac{1}{2} e^{\kappa_4^2 \langle K_{\text{mod}}/2+2K_T/3\rangle }\kappa_4^4v^2 \langle \hat W_{np}\rangle\ .
\end{split}
\end{equation}
The non-diagonal terms $M_{C^1T}$ are suppressed by $qM_U/M_P$, while the masses of the matter field fermions are suppressed by  $q^2M_U^2/M_P^2$. We therefore neglect all these extra contributions and assume that the fermion mass matrices have the same expressions in both the non-vanishing and vanishing FI cases,
\begin{equation}
{\mathcal{M}_f}={\mathbf{M_f}}\ .
\end{equation}


\begin{thebibliography}{90}
\bibitem{Horava:1995qa}
P.~Horava and E.~Witten,
``Heterotic and type I string dynamics from eleven-dimensions,''
Nucl. Phys. B \textbf{460}, 506-524 (1996)
doi:10.1016/0550-3213(95)00621-4
[arXiv:hep-th/9510209 [hep-th]].

\bibitem{Horava:1996ma}
P.~Horava and E.~Witten,
``Eleven-dimensional supergravity on a manifold with boundary'',
Nucl. Phys. B \textbf{475}, 94-114 (1996)
doi:10.1016/0550-3213(96)00308-2
[arXiv:hep-th/9603142 [hep-th]].

\bibitem{Lukas:1997fg}
A.~Lukas, B.~A.~Ovrut and D.~Waldram,
``On the four-dimensional effective action of strongly coupled heterotic string theory'',
Nucl. Phys. B \textbf{532}, 43-82 (1998)
doi:10.1016/S0550-3213(98)00463-5
[arXiv:hep-th/9710208 [hep-th]].

\bibitem{Lukas:1998yy}
A.~Lukas, B.~A.~Ovrut, K.~S.~Stelle and D.~Waldram,
``The Universe as a domain wall'',
Phys. Rev. D \textbf{59}, 086001 (1999)
doi:10.1103/PhysRevD.59.086001
[arXiv:hep-th/9803235 [hep-th]].

\bibitem{Lukas:1998tt}
A.~Lukas, B.~A.~Ovrut, K.~S.~Stelle and D.~Waldram,
``Heterotic M theory in five-dimensions'',
Nucl. Phys. B \textbf{552}, 246-290 (1999)
doi:10.1016/S0550-3213(99)00196-0
[arXiv:hep-th/9806051 [hep-th]].

\bibitem{Donagi:1998xe}
R.~Donagi, A.~Lukas, B.~A.~Ovrut and D.~Waldram,
``Nonperturbative vacua and particle physics in M theory'',
JHEP \textbf{05}, 018 (1999)
doi:10.1088/1126-6708/1999/05/018
[arXiv:hep-th/9811168 [hep-th]].

\bibitem{Ovrut:2000bi}
B.~A.~Ovrut,
``The universe as a three-brane'',
Fortsch. Phys. \textbf{48}, 183-190 (2000)

\bibitem{Donagi:1999gc}
R.~Donagi, A.~Lukas, B.~A.~Ovrut and D.~Waldram,
``Holomorphic vector bundles and nonperturbative vacua in M theory'',
JHEP \textbf{06}, 034 (1999)
doi:10.1088/1126-6708/1999/06/034
[arXiv:hep-th/9901009 [hep-th]].

\bibitem{Braun:2005nv}
V.~Braun, Y.~H.~He, B.~A.~Ovrut and T.~Pantev,
``The Exact MSSM spectrum from string theory'',
JHEP \textbf{05}, 043 (2006)
doi:10.1088/1126-6708/2006/05/043
[arXiv:hep-th/0512177 [hep-th]].

\bibitem{Braun:2005bw}
V.~Braun, Y.~H.~He, B.~A.~Ovrut and T.~Pantev,
``A Standard model from the E(8) x E(8) heterotic superstring'',
JHEP \textbf{06}, 039 (2005)
doi:10.1088/1126-6708/2005/06/039
[arXiv:hep-th/0502155 [hep-th]].

\bibitem{Braun:2005ux}
V.~Braun, Y.~H.~He, B.~A.~Ovrut and T.~Pantev,
``A Heterotic standard model'',
Phys. Lett. B \textbf{618}, 252-258 (2005)
doi:10.1016/j.physletb.2005.05.007
[arXiv:hep-th/0501070 [hep-th]].

\bibitem{Bouchard:2005ag}
V.~Bouchard and R.~Donagi,
``An SU(5) heterotic standard model'',
Phys. Lett. B \textbf{633}, 783-791 (2006)
doi:10.1016/j.physletb.2005.12.042
[arXiv:hep-th/0512149 [hep-th]].

\bibitem{Anderson:2009mh}
L.~B.~Anderson, J.~Gray, Y.~H.~He and A.~Lukas,
``Exploring Positive Monad Bundles And A New Heterotic Standard Model'',
JHEP \textbf{02}, 054 (2010)
doi:10.1007/JHEP02(2010)054
[arXiv:0911.1569 [hep-th]].

\bibitem{Braun:2011ni}
V.~Braun, P.~Candelas, R.~Davies and R.~Donagi,
``The MSSM Spectrum from (0,2)-Deformations of the Heterotic Standard Embedding'',
JHEP \textbf{05}, 127 (2012)
doi:10.1007/JHEP05(2012)127
[arXiv:1112.1097 [hep-th]].

\bibitem{Anderson:2011ns}
L.~B.~Anderson, J.~Gray, A.~Lukas and E.~Palti,
``Two Hundred Heterotic Standard Models on Smooth Calabi-Yau Threefolds'',
Phys. Rev. D \textbf{84}, 106005 (2011)
doi:10.1103/PhysRevD.84.106005
[arXiv:1106.4804 [hep-th]].

\bibitem{Anderson:2012yf}
L.~B.~Anderson, J.~Gray, A.~Lukas and E.~Palti,
``Heterotic Line Bundle Standard Models'',
JHEP \textbf{06}, 113 (2012)
doi:10.1007/JHEP06(2012)113
[arXiv:1202.1757 [hep-th]].

\bibitem{Anderson:2013xka}
L.~B.~Anderson, A.~Constantin, J.~Gray, A.~Lukas and E.~Palti,
``A Comprehensive Scan for Heterotic SU(5) GUT models'',
JHEP \textbf{01}, 047 (2014)
doi:10.1007/JHEP01(2014)047
[arXiv:1307.4787 [hep-th]].

\bibitem{Nibbelink:2015ixa}
S.~Groot Nibbelink, O.~Loukas, F.~Ruehle and P.~K.~S.~Vaudrevange,
``Infinite number of MSSMs from heterotic line bundles?'',
Phys. Rev. D \textbf{92}, no.4, 046002 (2015)
doi:10.1103/PhysRevD.92.046002
[arXiv:1506.00879 [hep-th]].

\bibitem{Nibbelink:2015vha}
S.~Groot Nibbelink, O.~Loukas and F.~Ruehle,
``(MS)SM-like models on smooth Calabi-Yau manifolds from all three heterotic string theories'',
Fortsch. Phys. \textbf{63}, 609-632 (2015)
doi:10.1002/prop.201500041
[arXiv:1507.07559 [hep-th]].

\bibitem{Braun:2006ae}
V.~Braun, Y.~H.~He and B.~A.~Ovrut,
``Stability of the minimal heterotic standard model bundle'',
JHEP \textbf{06}, 032 (2006)
doi:10.1088/1126-6708/2006/06/032
[arXiv:hep-th/0602073 [hep-th]].

\bibitem{Blaszczyk:2010db}
M.~Blaszczyk, S.~Groot Nibbelink, F.~Ruehle, M.~Trapletti and P.~K.~S.~Vaudrevange,
``Heterotic MSSM on a Resolved Orbifold'',
JHEP \textbf{09}, 065 (2010)
doi:10.1007/JHEP09(2010)065
[arXiv:1007.0203 [hep-th]].

\bibitem{Andreas:1999ty}
B.~Andreas, G.~Curio and A.~Klemm,
``Towards the Standard Model spectrum from elliptic Calabi-Yau''.
Int. J. Mod. Phys. A \textbf{19}, 1987 (2004)
doi:10.1142/S0217751X04018087
[arXiv:hep-th/9903052 [hep-th]].

\bibitem{Curio:2004pf}
G.~Curio,
``Standard model bundles of the heterotic string'',
Int. J. Mod. Phys. A \textbf{21}, 1261-1282 (2006)
doi:10.1142/S0217751X06025109
[arXiv:hep-th/0412182 [hep-th]].

\bibitem{Braun:2013wr}
V.~Braun, Y.~H.~He and B.~A.~Ovrut,
``Supersymmetric Hidden Sectors for Heterotic Standard Models'',
JHEP \textbf{09}, 008 (2013)
doi:10.1007/JHEP09(2013)008
[arXiv:1301.6767 [hep-th]].

\bibitem{Ovrut:2018qog}
B.~A.~Ovrut,
``Vacuum Constraints for Realistic Strongly Coupled Heterotic M-Theories'',
Symmetry \textbf{10}, no.12, 723 (2018)
doi:10.3390/sym10120723
[arXiv:1811.08892 [hep-th]].

\bibitem{Ashmore:2020ocb}
A.~Ashmore, S.~Dumitru and B.~A.~Ovrut,
``Line Bundle Hidden Sectors for Strongly Coupled Heterotic Standard Models'',
Fortsch. Phys. \textbf{69}, no.7, 2100052 (2021)
doi:10.1002/prop.202100052
[arXiv:2003.05455 [hep-th]].

\bibitem{Lukas:1999nh}
A.~Lukas and K.~S.~Stelle,
``Heterotic anomaly cancellation in five-dimensions'',
JHEP \textbf{01}, 010 (2000)
doi:10.1088/1126-6708/2000/01/010
[arXiv:hep-th/9911156 [hep-th]].

\bibitem{Witten:1996mz}
E.~Witten,
``Strong coupling expansion of Calabi-Yau compactification'',
Nucl. Phys. B \textbf{471}, 135-158 (1996)
doi:10.1016/0550-3213(96)00190-3
[arXiv:hep-th/9602070 [hep-th]].

\bibitem{Banks:1996ss}
T.~Banks and M.~Dine,
``Couplings and scales in strongly coupled heterotic string theory'',
Nucl. Phys. B \textbf{479}, 173-196 (1996)
doi:10.1016/0550-3213(96)00457-9
[arXiv:hep-th/9605136 [hep-th]].

\bibitem{Banks:1996rr}
T.~Banks and M.~Dine,
``Phenomenology of strongly coupled heterotic string theory'',
[arXiv:hep-th/9609046 [hep-th]].

\bibitem{Ashmore:2021xdm}
A.~Ashmore, S.~Dumitru and B.~A.~Ovrut,
``Hidden Sectors from Multiple Line Bundles for the $B-L$ MSSM'',
[arXiv:2106.09087 [hep-th]].

\bibitem{Dine:1987xk}
M.~Dine, N.~Seiberg and E.~Witten,
``Fayet-Iliopoulos Terms in String Theory'',
Nucl. Phys. B \textbf{289}, 589-598 (1987)
doi:10.1016/0550-3213(87)90395-6

\bibitem{Dine:1987gj}
M.~Dine, I.~Ichinose and N.~Seiberg,
``F Terms and d Terms in String Theory'',
Nucl. Phys. B \textbf{293}, 253-265 (1987)
doi:10.1016/0550-3213(87)90072-1

\bibitem{Anastasopoulos:2006cz}
P.~Anastasopoulos, M.~Bianchi, E.~Dudas and E.~Kiritsis,
``Anomalies, anomalous U(1)'s and generalized Chern-Simons terms'',
JHEP \textbf{11}, 057 (2006)
doi:10.1088/1126-6708/2006/11/057
[arXiv:hep-th/0605225 [hep-th]].

\bibitem{Blumenhagen:2005ga}
R.~Blumenhagen, G.~Honecker and T.~Weigand,
``Loop-corrected compactifications of the heterotic string with line bundles'',
JHEP \textbf{06}, 020 (2005)
doi:10.1088/1126-6708/2005/06/020
[arXiv:hep-th/0504232 [hep-th]].

\bibitem{Ashmore:2020wwv}
A.~Ashmore, S.~Dumitru and B.~A.~Ovrut,
``Explicit soft supersymmetry breaking in the heterotic M-theory B \ensuremath{-} L MSSM,''
JHEP \textbf{08}, 033 (2021)
doi:10.1007/JHEP08(2021)033
[arXiv:2012.11029 [hep-th]].

\bibitem{Choi:1997cm}
K.~Choi, H.~B.~Kim and C.~Munoz,
``Four-dimensional effective supergravity and soft terms in M theory,''
Phys. Rev. D \textbf{57}, 7521-7528 (1998)
doi:10.1103/PhysRevD.57.7521
[arXiv:hep-th/9711158 [hep-th]].

\bibitem{Kaplunovsky:1993rd}
V.~S.~Kaplunovsky and J.~Louis,
``Model independent analysis of soft terms in effective supergravity and in string theory,''
Phys. Lett. B \textbf{306}, 269-275 (1993)
doi:10.1016/0370-2693(93)90078-V
[arXiv:hep-th/9303040 [hep-th]].

\bibitem{Horava:1996vs}
P.~Horava,
``Gluino condensation in strongly coupled heterotic string theory,''
Phys. Rev. D \textbf{54}, 7561-7569 (1996)
doi:10.1103/PhysRevD.54.7561
[arXiv:hep-th/9608019 [hep-th]].

\bibitem{Lukas:1997rb}
A.~Lukas, B.~A.~Ovrut and D.~Waldram,
``Gaugino condensation in M theory on s**1 / Z(2),''
Phys. Rev. D \textbf{57}, 7529-7538 (1998)
doi:10.1103/PhysRevD.57.7529
[arXiv:hep-th/9711197 [hep-th]].

\bibitem{Nilles:1998sx}
H.~P.~Nilles, M.~Olechowski and M.~Yamaguchi,
``Supersymmetry breakdown at a hidden wall,''
Nucl. Phys. B \textbf{530}, 43-72 (1998)
doi:10.1016/S0550-3213(98)00418-0
[arXiv:hep-th/9801030 [hep-th]].

\bibitem{Binetruy:1996xja}
P.~Binetruy, M.~K.~Gaillard and Y.~Y.~Wu,
``Dilaton stabilization in the context of dynamical supersymmetry breaking through gaugino condensation,''
Nucl. Phys. B \textbf{481}, 109-128 (1996)
doi:10.1016/S0550-3213(96)90125-X
[arXiv:hep-th/9605170 [hep-th]].

\bibitem{Antoniadis:1997xk}
I.~Antoniadis and M.~Quiros,
``On the M theory description of gaugino condensation,''
Phys. Lett. B \textbf{416}, 327-333 (1998)
doi:10.1016/S0370-2693(97)01246-X
[arXiv:hep-th/9707208 [hep-th]].

\bibitem{Minasian:2017eur}
R.~Minasian, M.~Petrini and E.~E.~Svanes,
``On Heterotic Vacua with Fermionic Expectation Values,''
Fortsch. Phys. \textbf{65}, no.3-4, 1700010 (2017)
doi:10.1002/prop.201700010
[arXiv:1702.01156 [hep-th]].

\bibitem{Gray:2007qy}
J.~Gray, A.~Lukas and B.~Ovrut,
``Flux, gaugino condensation and anti-branes in heterotic M-theory,''
Phys. Rev. D \textbf{76}, 126012 (2007)
doi:10.1103/PhysRevD.76.126012
[arXiv:0709.2914 [hep-th]].

\bibitem{Lukas:1999kt}
A.~Lukas, B.~A.~Ovrut and D.~Waldram,
``Five-branes and supersymmetry breaking in M theory,''
JHEP \textbf{04}, 009 (1999)
doi:10.1088/1126-6708/1999/04/009
[arXiv:hep-th/9901017 [hep-th]].

\bibitem{Font:1990nt}
A.~Font, L.~E.~Ibanez, D.~Lust and F.~Quevedo,
``Supersymmetry Breaking From Duality Invariant Gaugino Condensation,''
Phys. Lett. B \textbf{245}, 401-408 (1990)
doi:10.1016/0370-2693(90)90665-S

\bibitem{Dumitru:2021jlh}
S.~Dumitru and B.~A.~Ovrut,
``Heterotic $M$-Theory Hidden Sectors with an Anomalous $U(1)$ Gauge Symmetry,''
[arXiv:2109.13781 [hep-th]].

\bibitem{Ibanez:2001nd}
L.~E.~Ibanez, F.~Marchesano and R.~Rabadan,
``Getting just the standard model at intersecting branes,''
JHEP \textbf{11}, 002 (2001)
doi:10.1088/1126-6708/2001/11/002
[arXiv:hep-th/0105155 [hep-th]].

\bibitem{Aldazabal:2000dg}
G.~Aldazabal, S.~Franco, L.~E.~Ibanez, R.~Rabadan and A.~M.~Uranga,
``D = 4 chiral string compactifications from intersecting branes,''
J. Math. Phys. \textbf{42}, 3103-3126 (2001)
doi:10.1063/1.1376157
[arXiv:hep-th/0011073 [hep-th]].

\bibitem{Blumenhagen:2006ux}
R.~Blumenhagen, S.~Moster and T.~Weigand,
``Heterotic GUT and standard model vacua from simply connected Calabi-Yau manifolds,''
Nucl. Phys. B \textbf{751}, 186-221 (2006)
doi:10.1016/j.nuclphysb.2006.06.005
[arXiv:hep-th/0603015 [hep-th]].

\bibitem{Weigand:2006yj}
T.~Weigand,
``Compactifications of the heterotic string with unitary bundles,''
Fortsch. Phys. \textbf{54}, 963-1077 (2006)
doi:10.1002/prop.200610327

\bibitem{Anderson:2009nt}
L.~B.~Anderson, J.~Gray, A.~Lukas and B.~Ovrut,
``Stability Walls in Heterotic Theories,''
JHEP \textbf{09}, 026 (2009)
doi:10.1088/1126-6708/2009/09/026
[arXiv:0905.1748 [hep-th]].

\bibitem{Anderson:2010mh}
L.~B.~Anderson, J.~Gray, A.~Lukas and B.~Ovrut,
``Stabilizing the Complex Structure in Heterotic Calabi-Yau Vacua,''
JHEP \textbf{02}, 088 (2011)
doi:10.1007/JHEP02(2011)088
[arXiv:1010.0255 [hep-th]].

\bibitem{Binetruy:1996uv}
P.~Binetruy and E.~Dudas,
``Gaugino condensation and the anomalous U(1),''
Phys. Lett. B \textbf{389}, 503-509 (1996)
doi:10.1016/S0370-2693(96)01305-6
[arXiv:hep-th/9607172 [hep-th]].

\bibitem{Ambroso:2009jd}
M.~Ambroso and B.~Ovrut,
``The B-L/Electroweak Hierarchy in Heterotic String and M-Theory,''
JHEP \textbf{10}, 011 (2009)
doi:10.1088/1126-6708/2009/10/011
[arXiv:0904.4509 [hep-th]].

\bibitem{Marshall:2014kea}
Z.~Marshall, B.~A.~Ovrut, A.~Purves and S.~Spinner,
``Spontaneous $R$-Parity Breaking, Stop LSP Decays and the Neutrino Mass Hierarchy,''
Phys. Lett. B \textbf{732}, 325-329 (2014)
doi:10.1016/j.physletb.2014.03.052
[arXiv:1401.7989 [hep-ph]].

\bibitem{Marshall:2014cwa}
Z.~Marshall, B.~A.~Ovrut, A.~Purves and S.~Spinner,
``LSP Squark Decays at the LHC and the Neutrino Mass Hierarchy,''
Phys. Rev. D \textbf{90}, no.1, 015034 (2014)
doi:10.1103/PhysRevD.90.015034
[arXiv:1402.5434 [hep-ph]].

\bibitem{Ovrut:2012wg}
B.~A.~Ovrut, A.~Purves and S.~Spinner,
``Wilson Lines and a Canonical Basis of SU(4) Heterotic Standard Models,''
JHEP \textbf{11}, 026 (2012)
doi:10.1007/JHEP11(2012)026
[arXiv:1203.1325 [hep-th]].

\bibitem{Ovrut:2014rba}
B.~A.~Ovrut, A.~Purves and S.~Spinner,
``A statistical analysis of the minimal SUSY B\textendash{}L theory,''
Mod. Phys. Lett. A \textbf{30}, no.18, 1550085 (2015)
doi:10.1142/S0217732315500856
[arXiv:1412.6103 [hep-ph]].

\bibitem{Barger:2008wn}
V.~Barger, P.~Fileviez Perez and S.~Spinner,
``Minimal gauged U(1)(B-L) model with spontaneous R-parity violation,''
Phys. Rev. Lett. \textbf{102}, 181802 (2009)
doi:10.1103/PhysRevLett.102.181802
[arXiv:0812.3661 [hep-ph]].

\bibitem{FileviezPerez:2009gr}
P.~Fileviez Perez and S.~Spinner,
``Spontaneous R-Parity Breaking in SUSY Models,''
Phys. Rev. D \textbf{80}, 015004 (2009)
doi:10.1103/PhysRevD.80.015004
[arXiv:0904.2213 [hep-ph]].

\bibitem{Anderson:2011cza}
L.~B.~Anderson, J.~Gray, A.~Lukas and B.~Ovrut,
``Stabilizing All Geometric Moduli in Heterotic Calabi-Yau Vacua,''
Phys. Rev. D \textbf{83}, 106011 (2011)
doi:10.1103/PhysRevD.83.106011
[arXiv:1102.0011 [hep-th]].

\bibitem{Anderson:2011ty}
L.~B.~Anderson, J.~Gray, A.~Lukas and B.~Ovrut,
``The Atiyah Class and Complex Structure Stabilization in Heterotic Calabi-Yau Compactifications,''
JHEP \textbf{10}, 032 (2011)
doi:10.1007/JHEP10(2011)032
[arXiv:1107.5076 [hep-th]].

\bibitem{Cicoli:2013rwa}
M.~Cicoli, S.~de Alwis and A.~Westphal,
``Heterotic Moduli Stabilisation,''
JHEP \textbf{10}, 199 (2013)
doi:10.1007/JHEP10(2013)199
[arXiv:1304.1809 [hep-th]].

\bibitem{Brandle:2003uya}
M.~Brandle,
``Aspects of branes in (heterotic) M-theory,''

\bibitem{Freedman:2012zz}
D.~Z.~Freedman and A.~Van Proeyen,
``Supergravity,''

\bibitem{Soni:1983rm}
S.~K.~Soni and H.~A.~Weldon,
``Analysis of the Supersymmetry Breaking Induced by N=1 Supergravity Theories'',
Phys. Lett. B \textbf{126}, 215-219 (1983)
doi:10.1016/0370-2693(83)90593-2

\bibitem{Brignole:1997wnc}
A.~Brignole, L.~E.~Ibanez and C.~Munoz,
``Soft supersymmetry breaking terms from supergravity and superstring models'',
Adv. Ser. Direct. High Energy Phys. \textbf{18}, 125-148 (1998)
doi:10.1142/9789812839657\_0003
[arXiv:hep-ph/9707209 [hep-ph]].

\bibitem{Martin:1997ns}
S.~P.~Martin,
``A Supersymmetry primer,''
Adv. Ser. Direct. High Energy Phys. \textbf{18}, 1-98 (1998)
doi:10.1142/9789812839657\_0001
[arXiv:hep-ph/9709356 [hep-ph]].

\bibitem{Deen:2016zfr}
R.~Deen, B.~A.~Ovrut and A.~Purves,
``Supersymmetric Sneutrino-Higgs Inflation,''
Phys. Lett. B \textbf{762}, 441-446 (2016)
doi:10.1016/j.physletb.2016.09.059
[arXiv:1606.00431 [hep-ph]].

\bibitem{Cai:2018ljy}
Y.~Cai, R.~Deen, B.~A.~Ovrut and A.~Purves,
``Perturbative reheating in Sneutrino-Higgs cosmology'',
JHEP \textbf{09}, 001 (2018)
doi:10.1007/JHEP09(2018)001
[arXiv:1804.07848 [hep-th]].

\bibitem{Ibanez:2014swa}
L.~E.~Ibanez, F.~Marchesano and I.~Valenzuela,
``Higgs-otic Inflation and String Theory'',
JHEP \textbf{01}, 128 (2015)
doi:10.1007/JHEP01(2015)128
[arXiv:1411.5380 [hep-th]].

\bibitem{Chowdhury:2018tzw}
D.~Chowdhury, E.~Dudas, M.~Dutra and Y.~Mambrini,
``Moduli Portal Dark Matter,''
Phys. Rev. D \textbf{99}, no.9, 095028 (2019)
doi:10.1103/PhysRevD.99.095028
[arXiv:1811.01947 [hep-ph]].

\bibitem{Dutra:2019nhh}
M.~Dutra,
``The moduli portal to dark matter particles,''
doi:10.1007/978-3-030-55777-5\_39
[arXiv:1911.11862 [hep-ph]].

\bibitem{Barreiro:1998nd}
T.~Barreiro, B.~de Carlos, J.~A.~Casas and J.~M.~Moreno,
``Anomalous U(1), gaugino condensation and supergravity'',
Phys. Lett. B \textbf{445}, 82-93 (1998)
doi:10.1016/S0370-2693(98)01402-6
[arXiv:hep-ph/9808244 [hep-ph]].

\bibitem{Ibanez:2012zz}
L.~E.~Ibanez and A.~M.~Uranga,
``String theory and particle physics: An introduction to string phenomenology'',

\bibitem{Carlevaro:2005bk}
L.~Carlevaro and J.~P.~Derendinger,
Nucl. Phys. B \textbf{736}, 1-33 (2006)
doi:10.1016/j.nuclphysb.2005.11.019
[arXiv:hep-th/0502225 [hep-th]].

\bibitem{Gray:2003vw}
J.~Gray and A.~Lukas,
``Gauge five-brane moduli in four-dimensional heterotic models'',
Phys. Rev. D \textbf{70}, 086003 (2004)
doi:10.1103/PhysRevD.70.086003
[arXiv:hep-th/0309096 [hep-th]].

\bibitem{Lima:2001nh}
E.~Lima, B.~A.~Ovrut and J.~Park,
``Five-brane superpotentials in heterotic M theory'',
Nucl. Phys. B \textbf{626}, 113-164 (2002)
doi:10.1016/S0550-3213(02)00030-5
[arXiv:hep-th/0102046 [hep-th]].

\bibitem{Choi:1998nx}
K.~Choi, H.~B.~Kim and H.~D.~Kim,
``Moduli stabilization in heterotic M theory'',
Mod. Phys. Lett. A \textbf{14}, 125-134 (1999)
doi:10.1142/S021773239900016X
[arXiv:hep-th/9808122 [hep-th]].

\bibitem{Gukov:2003cy}
S.~Gukov, S.~Kachru, X.~Liu and L.~McAllister,
``Heterotic moduli stabilization with fractional Chern-Simons invariants'',
Phys. Rev. D \textbf{69}, 086008 (2004)
doi:10.1103/PhysRevD.69.086008
[arXiv:hep-th/0310159 [hep-th]].

\bibitem{PaccettiCorreia:2007ret}
F.~Paccetti Correia and M.~G.~Schmidt,
``Moduli stabilization in heterotic M-theory,''
Nucl. Phys. B \textbf{797}, 243-267 (2008)
doi:10.1016/j.nuclphysb.2008.01.005
[arXiv:0708.3805 [hep-th]].

\bibitem{Dundee:2010sb}
B.~Dundee, S.~Raby and A.~Westphal,
``Moduli stabilization and SUSY breaking in heterotic orbifold string models'',
Phys. Rev. D \textbf{82}, 126002 (2010)
doi:10.1103/PhysRevD.82.126002
[arXiv:1002.1081 [hep-th]].

\bibitem{Parameswaran:2010ec}
S.~L.~Parameswaran, S.~Ramos-Sanchez and I.~Zavala,
``On Moduli Stabilisation and de Sitter Vacua in MSSM Heterotic Orbifolds'',
JHEP \textbf{01}, 071 (2011)
doi:10.1007/JHEP01(2011)071
[arXiv:1009.3931 [hep-th]].

\bibitem{Wess:1992cp}
J.~Wess and J.~Bagger,
``Supersymmetry and supergravity,''
\end{thebibliography}
\end{document}